\documentclass[useAMS,usenatbib]{mn2e}
\usepackage{subfloat}
\usepackage{epsf}
\usepackage{amssymb}
\usepackage[usenames]{color}
\def\plotone#1{\centering \leavevmode
\epsfxsize=\columnwidth \epsfbox{#1}}
\def\plotwide#1{\centering \leavevmode
\epsfxsize=1.4\columnwidth \epsfbox{#1}}
\def\plottwo#1#2{\centering \leavevmode
\epsfxsize=.99\columnwidth \epsfbox{#1} \hfil
\epsfxsize=.99\columnwidth \epsfbox{#2}}
\def\plottwoo#1#2{\centering \leavevmode
\epsfxsize=.8\columnwidth \epsfbox{#1} \hfil
\epsfxsize=.8\columnwidth \epsfbox{#2}}
\def\plottwoq#1#2{\centering \leavevmode
\epsfxsize=.75\columnwidth \epsfbox{#1} \hfil
\epsfxsize=.75\columnwidth \epsfbox{#2}}
\def\plotthree#1#2#3{\centering \leavevmode
\epsfysize=.8\columnwidth \epsfbox{#1} \hfil
\epsfysize=.8\columnwidth \epsfbox{#2} \hfil
\epsfysize=.8\columnwidth \epsfbox{#3}
}
\def\plotfour#1#2#3#4{\centering \leavevmode
\epsfxsize=0.75\columnwidth \epsfbox{#1}
\epsfxsize=0.75\columnwidth \epsfbox{#2}\vspace{1mm}\hfil
\epsfxsize=0.75\columnwidth \epsfbox{#3}
\epsfxsize=0.75\columnwidth \epsfbox{#4}}
\def\plotfive#1#2#3#4#5{\centering \leavevmode
\epsfxsize=.805\columnwidth \epsfbox{#1}
\epsfxsize=.805\columnwidth \epsfbox{#2}\vspace{1mm}\hfil
\epsfxsize=.805\columnwidth \epsfbox{#3}
\epsfxsize=.805\columnwidth \epsfbox{#4}
\epsfxsize=.805\columnwidth \epsfbox{#5}}
\def\plotfiveq#1#2#3#4#5{\centering \leavevmode
\epsfxsize=.691\columnwidth \epsfbox{#1} \hfil
\epsfxsize=.691\columnwidth \epsfbox{#2} \hfil
\epsfxsize=.691\columnwidth \epsfbox{#3} \vspace{4mm}\hfil
\epsfxsize=.691\columnwidth \epsfbox{#4} 
\epsfxsize=.691\columnwidth \epsfbox{#5} 
}
 
\def\kms{\,{\rm km~s}^{-1}}

\newcommand{\be}{\begin{equation}}
\newcommand{\ee}{\end{equation}}

\def\disp {\displaystyle}
\newcommand{\appropto}{\mathrel{\vcenter{\offinterlineskip\halign{\hfil$##$\cr\propto\cr\noalign{\kern2pt}\sim\cr\noalign{\kern-2pt}}}}}

\title[X-ray bright elliptical galaxies]{Stellar kinematics of X-ray bright massive elliptical galaxies
\thanks{Based on observations obtained with the 6-m telescope of the Special Astrophysical Observatory of the Russian Academy of Sciences. The observations were carried out with the financial support of the Ministry of Education and Science of Russian Federation (contracts no. 16.518.11.7073 and 14.518.11.7070).}
}
\author[Lyskova et al.]{N.~Lyskova,$^{1,2}$
E.~Churazov,$^{1,2}$ A.~Moiseev,$^{3,4}$
O.~Sil'chenko,$^{4,5}$ I.~Zhuravleva$^{6,7}$  
\newauthor \\
$^1$ Max-Planck-Institut f\"ur Astrophysik, Karl-Schwarzschild-Strasse 1, 85741
Garching, Germany\\
$^2$ Space Research Institute (IKI), Profsoyuznaya 84/32, Moscow 117997, Russia\\
$^3$ Special Astrophysical Observatory, Russian Academy of Sciences, Nizhnii Arkhyz, Karachaevo-Cherkesskaya Republic, 369167 Russia \\
$^4$ Sternberg Astronomical Institute, M.~V. Lomonosov Moscow State University, Moscow, 119992 Russia \\
$^5$ Isaac Newton Institute of Chile, Moscow Branch \\
$^6$ Kavli Institute for Particle Astrophysics and Cosmology, Stanford University, 452 Lomita Mall, Stanford, CA 94305-4085, USA  \\
$^7$ Department of Physics, Stanford University, 382 Via Pueblo Mall, Stanford, CA 94305-4060, USA \\
}

\begin{document}
 \pagerange{\pageref{firstpage}--\pageref{lastpage}}
\pubyear{2013}

\maketitle

\label{firstpage}
\begin{abstract}
  
We discuss a simple and fast method for estimating masses of early-type galaxies from optical data and compare the results with X-ray derived masses. The optical method relies only on the most basic observables such as the surface brightness $I(R)$ and the line-of-sight velocity dispersion $\sigma_p(R)$ profiles and provides an anisotropy-independent estimate of the galaxy circular speed $V_c$. The mass-anisotropy degeneracy is effectively overcome by evaluating $V_c$ at a characteristic radius $R_{\rm sweet}$ defined from {\it local} properties of observed profiles. The sweet radius $R_{\rm sweet}$ is expected to lie close to $R_2$, where $I(R) \propto R^{-2}$, and not far from the effective radius $R_{\rm eff}$.  We apply the method to a sample of five X-ray bright elliptical galaxies observed with the 6-m telescope BTA-6 in Russia. We then compare the optical $V_c$-estimate with the X-ray derived value, and discuss possible constraints on the non-thermal pressure in the hot gas and configuration of stellar orbits. We find that the average ratio of the optical $V_c$-estimate to the X-ray one is equal to $\approx 0.98$ with $11 \%$ scatter, i.e. there is no evidence for the large non-thermal pressure contribution in the gas at $\sim R_{\rm sweet}$. From analysis of the Lick indices H$\beta$, Mgb, Fe5270 and Fe5335, we calculate the mass of the stellar component within the sweet radius. We conclude that a typical dark matter fraction inside $R_{\rm sweet}$ in the sample galaxies is $\sim 60\%$ for the Salpeter IMF and $\sim 75 \%$ for the Kroupa IMF.

\end{abstract}


%
%________________________________________________________________

\sloppypar

\section{Introduction}

Being the most massive galaxies in the local Universe, giant elliptical galaxies provide a natural laboratory to study galaxy formation, assembly and evolution processes. The current paradigm of galaxy formation is the hierarchical scenario which suggests that early-type galaxies have complex merging histories of assembling most of the mass through accretion of small galaxies with rare major merger events \citep[e.g.][]{de.Lucia.Blaizot.2007, Naab.et.al.2007}.  Accurate mass determinations and disentangling a luminous and dark matter components at different redshifts are the key steps towards a consistent theory for elliptical galaxies formation. 

Determining the mass profile of early-type galaxies is a notoriously difficult problem as there are no dynamical tracers with the known intrinsic shape and structure of orbits, so that circular velocity curves of elliptical galaxies cannot be measured directly. A number of methods are in use for constraining the mass of early-type galaxies and the shape of dark matter halos, each having its own set of assumptions and limitations. Comparison of the mass profiles obtained from different independent techniques is necessary to get reliable estimates. It also helps to control the systematic uncertainties, inherent in all methods, as well as leads to interesting constraints on properties of elliptical galaxies. 

One of the mass estimation techniques comes from X-ray observations of extended hot X-ray-emitting coronae of massive elliptical galaxies. It is a powerful tool to probe the mass distribution over several decades in radius: from $\sim0.1 R_{\rm eff}$ out to $\sim10R_{\rm eff}$. In this approach spherical symmetry of a galaxy and hydrostatic equilibrium of the gas are commonly assumed. While the spherical symmetry approximation introduces only a small bias, if any \citep[e.g.][]{Piffaretti.et.al.2003, Boute.Humphrey.2012c}, validity of the hydrostatic equilibrium assumption is the subject of debate. When one is able to quantify deviations from hydrostatic equilibrium, it allows to estimate (although indirectly) pressure of the non-thermal gas motions. Most simulations suggest that in relaxed systems hydrostatic approximation works well, with non-thermal support at the level of 5\% to 35\% of the total gas pressure \citep[e.g.][]{Nagai.Vikhlinin.Kravtsov.2007}. When  X-ray observations are combined with, for instance, optical data on the stellar kinematics, then comparison between the X-ray gravitating mass profile and the optical mass  allows one to estimate the magnitude of the non-thermal motions of the hot gas, to constrain the mass-to-light ratio, to disentagle stellar and dark matter contributions to the total gravitating mass profile and to characterize the distribution of stellar orbits. 
 
Although elliptical galaxies suffer from a lack of `ideal' traces like disc rotation curves in spiral galaxies and there is an inherent degeneracy between anisotropy and mass, studies of stellar kinematics and dynamics provide the tools for measuring the gravitating mass profile with sufficient accuracy (up to $\sim 15 \%$, \citealt{Thomas.et.al.2005}). Methods based on the Schwarschild modeling of stellar orbits in axisymmetric (or even triaxial) potentials are considered to be the state-of-the-art techniques in this field. The most sophisticated approaches operate with full information on the line-of-sight velocity distribution including  higher-order moments. The orbit-based methods allow to infer not only the total mass profile, but also to measure the dark matter content, derive mass-to-light ratios and get the distribution function of stellar orbits. Among the drawbacks of these methods are the high computational cost and the necessity to have high-quality observational data. So only nearby elliptical galaxies can be studied by means of Schwarschild modelling, and for a large sample of objects, especially with noisy photometric and/or kinematical data, such an approach is not justified.

In this paper we discuss a simple approach for estimating the mass from the stellar kinematics \citep{2010MNRAS.404.1165C, Lyskova.et.al.2012} that relies only on the most basic observables such as the surface brightness and line-of-sight velocity dispersion profiles. By design   
the method is simple and fast and has a modest scatter ($\disp \Delta V_{c} / V_{c} \sim 5-10 \% $, \citealt{Lyskova.et.al.2012}). This makes it suitable for  large samples of elliptical galaxies even with limited and/or noisy observational data. Of course, the method is not intended to replace a thorough investigation of each indvidual galaxy. 
  
We apply the method to a small and rather arbitrarily selected sample of massive elliptical galaxies located at the centers of groups and clusters, and bright in X-rays. The surface brightness and projected velocity dispersion profiles up to several effective radii have been measured with optical long-slit spectroscopic facilities on the 6-m telescope of the Special Astrophysical Observatory of the Russian Academy of Sciences (SAO RAS).  Using publicly available Chandra data we also derive the X-ray mass profile and compare it with simple optical estimates. 

The paper is organized as follows.  In Section~\ref{sec:method}, we provide a brief description of the method. We apply it to real elliptical galaxies in Section~\ref{sec:analysis}, starting with the illustration of the method on the example of the extensively studied giant elliptical galaxy M87 in Section~\ref{subsec:M87}.
Details on the observations of the sample of early-type galaxies are presented in Section \ref{subsec:observations}. We derive circular speed estimates from optical and X-ray analyses and estimate  stellar contrubutions in Section \ref{subsec:rotcurve}. Results are summarized in Section \ref{sec:discussion}, and  Section \ref{sec:conclusion} contains conclusions.

%________________________________________________________________

\section{Description and justification of the method}

\label{sec:method}

Recent studies of massive elliptical galaxies based on different approaches (stellar dynamical methods, weak and strong lensing, hydrostatic mass modeling, and their combinations) suggest that the gravitational potential $\Phi(r)$ is close to  isothermal \citep[e.g.][]{Gerhard.et.al.2001, Treu.et.al.2006, Koopmans.et.al.2006, Fukazawa.et.al.2006, 2010MNRAS.404.1165C}. For a singular isothermal sphere the gravitational potential can be written as $\disp \Phi(r)=V_c^2\ln(r)$, the circular velocity curve is flat, $\disp V_c(r) =  const$, and the mass profile scales as $\disp M(r) \propto r$. So if the total gravitational potential of the galaxy  is indeed isothermal, it can be characterized with a single parameter - the circular speed $V_c$. Therefore to the first approximation  it is sufficient to determine $V_c$ at any radius and the task is to identify the radius at which the circular speed can be measured most accurately.

Stars in early-type galaxies can be considered as a collisionless system immersed in a gravitational field. Let us consider a spherical galaxy in the equilibrium state. Stars in such systems obey the Jeans equations which in the spherical coordinates ($r$, $\theta$, $\phi$) can be simply written as

\be
{d\over dr}j\sigma_r^2+2\frac{\beta}{r}j\sigma_r^2=-j{d\Phi\over dr}=-j{GM(r) \over r},
\label{eq:jeans}
\ee
where $j(r)$\footnote{Throughout this paper we denote a projected 2D radius as $R$ and a 3D radius as $r$.} is the stellar number  density, $\disp \sigma_r(r)$ is the radial velocity dispersion (weighted by luminosity), $\disp \beta(r)~=~1~-~\frac{\sigma_{\phi}^2+\sigma_{\theta}^2}{2\sigma_{r}^2}$ is the stellar anisotropy parameter, $M(r)$ is the total mass profile\footnote{Here we treat stars as test particles in the gravitational field $\disp \Phi(r)$.}. The anisotropy $\beta(r)$ reflects the distribution of stellar orbits. If all stars in a galaxy are on circular orbits, then $\beta \rightarrow -\infty$, for pure radial orbits $\beta = 1$, and $\beta = 0$ for isotropic distribution of orbits.

We can link the unobservable quantites $j(r)$, $\beta(r)$ and $\sigma_r(r)$ with observable ones - a surface brightness $I(R)$ and a line-of-sight velocity dispersion $\sigma_p(R)$ profiles - via the following equations: 

\be
I(R)=2\int_R^\infty\!\! {j(r)r\,dr\over\sqrt{r^2-R^2}},
\label{eq:sb}
\ee

\be
\sigma_p^2(R)\cdot I(R)=2\int_R^\infty\!\! j(r)\sigma_r^2(r)\left(1-\frac{R^2}{r^2}\beta(r)\right){r\,dr\over\sqrt{r^2-R^2}}.
\label{eq:disp}
\ee

Nevertheless, the set of equations (\ref{eq:jeans})-(\ref{eq:disp}) is not closed. An inherent mass-anisotropy degeneracy does not allow us to solve it for the mass $M(r)$ and anisotropy $\beta(r)$ profiles simultaneously. Traditionally the degeneracy is overcome by assuming some parametric form of the mass or anisotropy profiles and fitting the resulting models to the observed $I(R)$ and $\sigma_p(R)$. However, both $M(r)$ and $\beta(r)$ are still poorly constrained from observational data alone (i.e., $I(R)$ and $\sigma_p(R)$) without resorting to the state-of-the-art modeling or adding detailed information on the line profiles. Here we discuss a technique that allows to estimate the mass of a galaxy without apriori parametrization of $M(r)$ and/or $\beta(r)$. 

Assuming the logarithmic (isothermal) form of the gravitational potential $\disp \Phi(r)=V_c^2\ln(r)+ const$ one can solve analytically the spherical Jeans equation coupled with equations (\ref{eq:sb})-(\ref{eq:disp}) for three types of tracers' orbits - isotropic ($\beta=0$), radial ($\beta=1$) and circular ($\disp \beta\rightarrow -\infty$). Note that for a typical stellar distribution $j(r)$ the projected velocity dispersion profile $\sigma_p(R)$ behaves differently depending on the value of $\beta$. In case of pure radial stellar orbits $\sigma_p(R)$ rapidly declines with the projected radius $R$, for the isotropic distribution of orbits $\sigma_p(R)$ declines much slower and, finally, $\sigma_p(R)$ increases with $R$ for the circular orbits \citep[e.g.][Figure 3]{Richstone&Tremaine1984, 2010MNRAS.404.1165C}. So there is an `optimal radius' where the projected velocity dispersion profiles for different values of anisotropy (almost) intersect each other. The existence of such a radius is 
discussed in \cite{Richstone&Tremaine1984}, \cite{Gerhard1993}. The method based on this observation is presented in detail in \cite{2010MNRAS.404.1165C} and \cite{Lyskova.et.al.2012}.    

It is practical to express the circular speed in terms of the observable surface brightness and line-of-sight velocity dispersion profiles. 
For the logarithmic form of the gravitational potential $\disp \Phi(r)=V_c^2\ln(r)$ the circular velocity $V_c$ profiles for isotropic, radial and circular orbits are given by \cite{2010MNRAS.404.1165C}: 

\[
V^{\rm iso}_c=\sigma_p(R)  \sqrt{1+\alpha+\gamma} 
\]
\be
V^{\rm circ}_c=\sigma_p(R)  \sqrt{2 \frac{1+\alpha+\gamma}{\alpha}}
\label{eq:main} 
\ee
\[
V^{\rm rad}_c=\sigma_p(R)  \sqrt{\left(\alpha+\gamma\right
  )^2+\delta-1}, 
\]
where 
\be
\alpha\equiv-\frac{d\ln I(R)}{d\ln R}, \ \ \gamma\equiv -\frac{d\ln
  \sigma_p^2}{d\ln R},\ \ \delta\equiv \frac{d^2\ln[I(R)\sigma_p^2]}{d
  (\ln R)^2}.
\label{eq:agd}
\ee 

Often the subdominant terms $\gamma$ and $\delta$ can be neglected, i.e. the dispersion profile is assumed to be flat and the curvature of $I(R)$ to be small, and equations (\ref{eq:main}) are simplified to:

\[
V^{\rm iso}_c=\sigma_p  \sqrt{\alpha+1} 
\]
\be
V^{\rm circ}_c=\sigma_p   \sqrt{2\frac{\alpha+1}{\alpha}}
\label{eq:agd_simple} 
\ee
\[
V^{\rm rad}_c=\sigma_p  \sqrt{\alpha^2-1}. \\
\]

Let us call a `sweet spot' $R_{\rm sweet}$ the radius at which all three curves $\disp V^{\rm iso}_c(R), V^{\rm circ}_c(R)$ and $\disp V^{\rm rad}_c(R)$ are very close to each other.  At this radius the circular speed uncertainty due to the unknown stellar anisotropy is minimal.  From the equations (\ref{eq:agd_simple}) it is clear that for $\alpha=2$ the relation between $V_c$ and $\sigma_p$ is the same for all types of orbits. So in the general case the sweet spot is expected to be located not far from the radius $R_{2}$ where the surface brightness declines as $R^{-2}$ which is in turn close to the half-light radius $R_{\rm eff}$ (see also \citealt{2010MNRAS.406.1220W}). If $I(R) \propto R^{-2}$ over some range of radii $[R_1, R_2]$, then the estimates based on equations (\ref{eq:main}) or (\ref{eq:agd_simple}) should work well over the whole range $[R_1, R_2]$. 

While the derivation of equations (\ref{eq:main}), (\ref{eq:agd_simple}) relies on the assumption of a flat circular velocity profile, it works well even in case of slowly varying $V_c(r)$. \cite{Lyskova.et.al.2012} have tested the method on a sample of cosmological simulations of elliptical galaxies from \cite{2010ApJ...725.23120} and have shown that the circular speed can be recovered to a reasonable accuracy. The rms-scatter in the circular velocity estimate has been found to be 5.4\% for present-day simulated massive elliptical galaxies without signs of significant rotation, while the sample averaged bias is less than 1\%.

\subsection{Rotation of galaxies.}

\label{subsec:rotation}
\begin{figure*}
\plotone{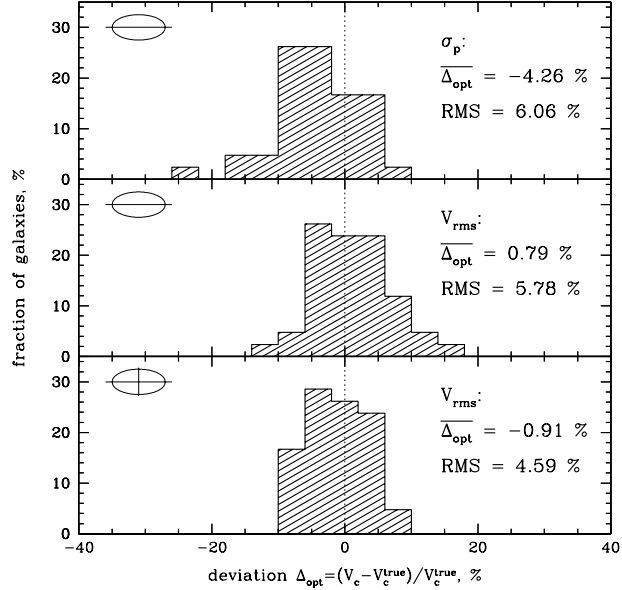}
\caption{The fraction of galaxies (in \%) as a function of deviation $\disp \Delta_{opt}=\left(V_c^{\rm iso}-V_c^{true}\right)/V_c^{true}$ for the sample of simulated galaxies with $\disp \sigma_p(R_{\rm eff})>150$ $\kms$. Each galaxy is analysed for a set of random projections. For each projection the values of $I(R)$, $\sigma_p(R)$ and $V_{\rm rot}(R)$ are calculated. The resulting $V_c$-estimates for all possible inclination angles for each galaxy are averaged. The black histogram in the upper panel results from `traditional' analysis of $I(R)$ and $\sigma_p(R)$ profiles measured along a slit that is aligned with a major axis of a galaxy without taking into account $\disp V_{\rm rot}(R)$. In this case the $V_c$-estimate when averaged over the sample is biased low. The bias could be compensated when considering $\disp V_{\rm rms}(R)=\sqrt{\sigma_p(R)+V_{\rm rot}(R)}$ instead of $\sigma_p(R)$ in equations (\ref{eq:main}), which is shown in the middle panel.
The histogram in the lower panel shows $V_c$-estimates, when profiles along major and minor axes are available and $V_{\rm rms}$ is used. The averaged over the sample estimate is almost unbiased, the distribution looks roughly symmetric and the rms-scatter is moderate.  
\label{fig:histo}
}
\end{figure*}

Elliptical galaxies can be divided into two broad families: (1) normal ellipticals, which show significant rotation, tend to be flattened and have an oblate-spheroidal shape; and (2) giant ellipticals, which are almost non-rotating, less flattened and tend to be triaxial \citep[e.g.][]{Kormendy.et.al.2009, Emsellem.et.al.2007}. Strictly speaking, the method in its original form is applicable only to non-rotating spherical galaxies, i.e. to no real ellipticals. Nevertheless, as tests on simulated galaxies show, the method still allows to recover the circular speed for massive elliptical galaxies without signs of significant rotation. For galaxies with rotational support the value of $V_c$ derived from the observed $\sigma_p(R)$ using equations (\ref{eq:main}) or (\ref{eq:agd_simple}) will be likely underestimated. Can we reduce a bias arising from rotation to extend the method on fast-rotating elliptical galaxies?

Let us consider a disk rotating with the velocity $\tilde{V}_{\rm rot}(R)$. When observed at an inclination angle $i$, where $i = 0^{\circ}$ corresponds to a face-on projection, the observed rotation velocity along an apparent major axis is simply $\disp V_{\rm rot}(R) = \tilde{V}_{\rm rot}(R)\sin i $. After averaging over different inclination angles $0 \le i \le \pi /2$  we get

\be
\langle V^2_{\rm rot} \rangle = \int_0^{\pi /2} V^2_{\rm rot}\cos i\,\mathrm{d}i=\int_0^{\pi /2} \tilde{V}^2_{\rm rot}\sin^2 i \cos i \,\mathrm{d}i=\frac{1}{3} \tilde{V}^2_{\rm rot}.
\ee

Thus the true rotation velocity is $\sqrt{3}$ times larger than the sample averaged observed velocity. This relation is similar to the relation between the simple $V_c$-estimate and the observed projected velocity dispersion near the sweet point (eq. \ref{eq:agd_simple}, $\alpha=2$). As the conversion coefficient at the sweet point does not (strongly) depend on the unknown configuration of stellar orbits, one can use the quantity $\disp V^2_{\rm rms}(R)=\sigma_p^2(R)+V_{\rm rot}^2(R)$ (rms-speed), where $\disp V_{\rm rot}(R)$ is the observed rotation velocity, instead of $\disp \sigma_p(R)$ in equation (\ref{eq:main}) or (\ref{eq:agd_simple}) to estimate the circular speed of a sample of galaxies that includes also fast rotators.  It is clear that for oblate rotating galaxies the $V_c$ inferred from $V_{\rm rms}$ is overestimated for the edge-on view and underestimated when the disk is viewed face-on. But after averaging over different inclination angles the bias disappears. The conjecture on using $V_{\rm rms}$ instead of $\sigma_p(R)$ in equation (\ref{eq:main}) or (\ref{eq:agd_simple}) has been further tested on a sample of resimulated galaxies from the high-resolution cosmological simulations of \cite{2010ApJ...725.23120}. The sample includes both fast and slow rotators in a proportion that is generally consistent with findings of ATLAS$^{3d}$ project \citep{Emsellem.et.al.2007, Emsellem.et.al.2011, Naab.et.al.2013}.

First, for each simulated galaxy in the sample we measure the surface brighness, projected velocity dispersion and rotational velocity profiles along the apparent major axis of the galaxy, mimicing long-slit observations. Then we estimate the circular speed in two ways: 1) using information about $\sigma_p(R)$ (eq. \ref{eq:main}) and 2) using $\disp V_{\rm rms}(R)=\sqrt{\sigma_p^2(R)+V_{\rm rot}^2(R)}$ instead of $\sigma_p(R)$. As a next step we calculate the average deviation $\disp \Delta_{opt}$ of the estimated circular speed from the true one $\disp V_c^{true}(r)=\sqrt{GM(<r)/r}$, after averaging over all possible inclination angles. We consider only galaxies with the value of the projected velocity dispersion at the effective radius $\sigma_p(R_{\rm eff})$ greater than 150 $\kms$ (when the galaxy is viewed edge-on). The sample consists of 26 objects. The results of the analysis are presented in the form of histograms (fraction of galaxies versus deviation of the $V_c$-estimate from the true value) in Figure \ref{fig:histo}. In the upper panel of Figure \ref{fig:histo} we show the histogram for the case when rotation is neglected. On average $V_c$ is underestimated by  $\disp \overline{\Delta_{opt}}=-4.3 \%$. If we substitute $\sigma_p(R)$ with $\disp V_{\rm rms}(R)=\sqrt{\sigma_p^2(R)+V_{\rm rot}^2(R)}$ then we get almost unbiased (within statistical errors) estimate of the circular speed with rms-scatter of $\approx 6\%$ (the middle panel of Figure \ref{fig:histo}). While for oblate ellipticals observations along the major axis carry all information needed for simple mass estimation, for triaxial galaxies rotation along the apparent minor axis might be significant. In a case when information is available along major and minor axes of a galaxy, using $\disp V_{\rm rms}^2=\frac{I_1V_{\rm {rms,} 1}^2+I_2V_{\rm {rms,} 2}^2}{I_1+I_2}$ makes the distribution of $\disp \Delta_{opt}=(V_c-V_c^{true})/V_c^{true}$ more symmetric than for the `one slit' case and reduces the rms-scatter down to $4.6 \%$ (lower panel of Figure \ref{fig:histo}). Note, that for the sample consisting of oblate rotating galaxies only there is no sense to use the weighted rms-speed $\disp V_{\rm rms}^2=\frac{I_1V_{\rm {rms,} 1}^2+I_2V_{\rm {rms,} 2}^2}{I_1+I_2}$ as it leads to the underestimated value of $V_c$ (compare the averaged deviations in the middle and lower panels). But for the sample containing also triaxial halos this approach helps to reduce the scatter and does not strongly bias the $V_c$-estimate. At least, for our sample of 26 simulated objects the bias is not significant, i.e., $\disp \overline{\Delta_{opt}} < RMS/\sqrt{N}$.

\subsection{An algorithm for estimating $V_c$}
\label{subsec:algorithm}

Based on the results of \cite{Lyskova.et.al.2012} and the arguments presented in the previous section, the following algorithm has been developed:

\begin{enumerate}
\item Calculate the logarithmic derivatives $\alpha$, $\gamma$ and $\delta$ from the observed surface brighness $I(R)$ and line-of-sight velocity dispersion $\sigma_p(R)$ profiles using equations (\ref{eq:agd}).

\item Calculate the circular speed $V_c(R)$ for isotropic, radial and circular stellar orbits using equations (\ref{eq:main}) in the case of reliable data (full analysis) or equations (\ref{eq:agd_simple}) in the case of poor or noisy observational data (simplified analysis). For rotating galaxies use $\disp V_{\rm rms}(R)$ instead of $\disp \sigma_p(R)$ in equations (\ref{eq:main}) or (\ref{eq:agd_simple}).

\item Estimate $V_c$ as $\disp V_c^{\rm iso}(R_{\rm sweet})$ at the sweet spot $R_{\rm sweet}$ - the radius at which all three curves $\disp V^{\rm iso}_c(R), V^{\rm circ}_c(R)$ and $\disp V^{\rm rad}_c(R)$ are maximally close to each other. At $R_{\rm sweet}$ the sensitivity of the method to the anisotropy parameter $\beta$ is believed to be minimal so the estimation of the circular speed at this particular point is not affected much by the unknown distribution of stellar orbits.
\end{enumerate}

%________________________________________________________________

\section{Analysis}
\label{sec:analysis}

\subsection{M87, revisited. Illustration of the Method}
\label{subsec:M87}

In this section we illustrate all the steps of the described algorithm on one massive galaxy - M87 (NGC4486). M87 is a nearby (16.1 Mpc) giant elliptical galaxy, luminous in X-rays. It's mass profile has been investigated in detail by a variety of methods. A technique analogous to the one described in Section \ref{subsec:algorithm} has already been applied to M87 \citep{2010MNRAS.404.1165C}. However, new data on stellar kinematics warrant a reanalysis of the data.

Since the 1960s M87 has been extensively explored and there is a large amount of observational data. We focus here on a recent work by \cite{Murphy.et.al.2011} (hereafter `M11'), who estimated  M87's mass profile from axisymmetric orbit-based modeling and compared the resulting total enclosed mass profile with other mass estimates available in the literature. 

Schwarzschild modeling \citep{Schwarzschild1979} is considered to be a state-of-the-art method for dynamical investigation of nearby galaxies  which allows one to recover masses and orbital anisotropies with $\sim 15\%$ accuracy \citep{Thomas.et.al.2005}. This technique consists in analysis of the three-dimensional orbital structure of a stellar system in an assumed gravitational potential and representation of the observed photometric and kinematic data by a superposition of constructed orbits. The system is assumed to be in dynamical equilibrium and as a rule to be viewed edge-on \citep[e.g.][]{Gebhardt.et.al.2000, Gebhardt.et.al.2003, Thomas.et.al.2004, Thomas.et.al.2005}. 

We assume that the mass and circular speed profiles for M87 derived in  M11 are accurate and unbiased (as significant inherent systematic uncertainties are not expected to be in dynamical models \citep[e.g.][]{Thomas.et.al.2007}) and compare our simple estimates with these curves.  

\begin{figure*}
\plotone{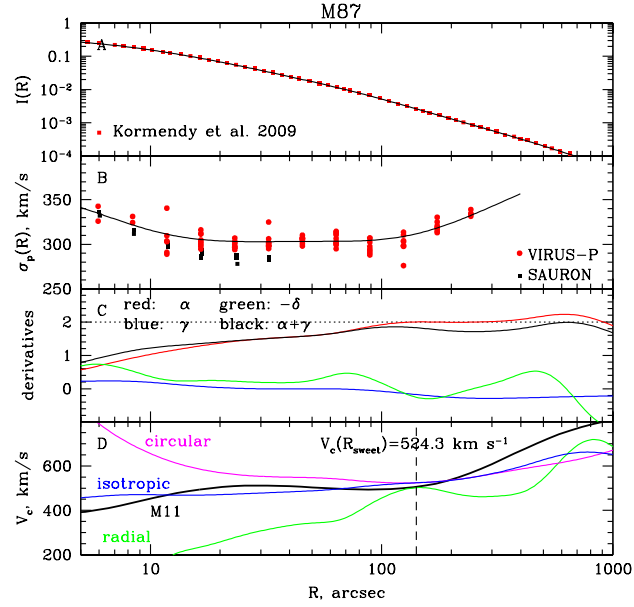}
\caption{Circular speed estimate of M87. The stellar surface brightness and the line-of-sight velocity dispersion profiles are shown in panels (A) and (B) correspondingly. Data are represented as points and smoothed curves used to compute the auxilary coefficients $\alpha, \gamma, \delta$  as black solid lines. The logarithmic derivatives $\alpha, \gamma, -\delta$ and $\alpha+\gamma$ (eq. \ref{eq:agd}) are shown in panel (C) in red, blue, green and black, respectively. Circular velocity curves for isotropic orbits of stars (in blue), pure radial (green) and pure circular (magenta) orbits as well as the circular speed (in black) derived in M11 are presented in panel (D).       
\label{fig:M87a}
}
\end{figure*}

\begin{figure*}
\plotone{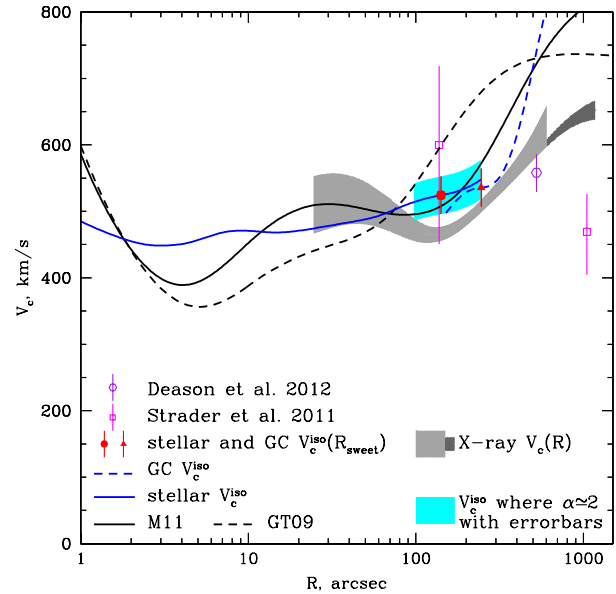}
\caption{Comparison of the simple circular speed estimates for M87 with $V_c$-profiles inferred from the state-of-the-art modeling. The solid line shows the best-fitting model of M11, the dashed line -  the $V_c$-profile from GT09. $V_c^{\rm iso}$ resulting from the same set of data (mainly, stellar kinematic and photometry) as used in M11 is shown as a blue solid line. The dashed blue line is $V_c^{\rm iso}$-profile derived from GC data as used in GT09. The red circle and the red triangle represent our simple $V_c$-estimates derived on the basis of same data sets used in M11 and in GT09, respectively. Shown in grey are X-ray $V_c$-profiles from Chandra and XMM data with errorbars. $V_c$-estimates from recent works (Deason et al. 2012 and Strader et al. 2011) are shown as purple and magenta dots with errorbars.  
\label{fig:M87all}
}
\end{figure*}

For the analysis we use the same set of data as in M11. Namely, the stellar surface brightness profile comes from \cite{Kormendy.et.al.2009}, the stellar kinematic data come from the publicly available SAURON data set \citep{Emsellem.et.al.2004} and from M11 (VIRUS-P instrument).

We first compute the auxiliary coefficients $\alpha$, $\gamma$ and $\delta$ (eq. \ref{eq:agd}) from  the smoothed $I(R)$ and $\sigma_p(R)$ profiles (the smoothing procedure is described in \citealt{2010MNRAS.404.1165C}). The profiles $I(R)$ and $\sigma_p(R)$ for M87 are shown in Figure \ref{fig:M87a}, panels (A) and (B). The stellar surface density from \cite{Kormendy.et.al.2009} is shown as red squares. The velocity dispersion measurements from SAURON and VIRUS-P are shown as red circles and black squares, respectively. When smoothing the projected velocity dispersion profile, we use only SAURON data in the radial range $R \lesssim 8''$, both SAURON and VIRUS-P between $8''\leq R \leq 16''$ and only VIRUS-P for $R \geq 16''$ as is done in M11. The logarithmic derivatives are shown in panel (C). Note that $\alpha \approx 2$ in the range  $ 100'' \leq R \leq 300''$. In this radial range we expect a weak dependence on the anisotropy parameter $\beta$ and the isotropic curcular velocity curve to be a good representation of the true circular speed profile. Using equations (\ref{eq:main}) we calculate $V_c(R)$ for 
an isotropic distibution of stellar orbits (shown in blue), for pure radial and pure circular orbits (in green and magenta, correspondingly). These three curves intersect each other at the sweet spot $R_{\rm sweet}\approx 141''$ where we estimate $\disp V_c^{\rm opt}(R_{\rm sweet})\equiv V_c^{\rm iso}(R_{\rm sweet})\approx 524.3$ $\kms$. The relative error of this estimate at $R_{\rm sweet}$ with respect to the circular speed $V_c^M(R_{\rm sweet})$ from M11 (Figure \ref{fig:M87a}, panel (D), solid black line) is equal to $\disp \Delta=\left(V_c^{\rm opt}- V_c^M \right)/ V_c^M = 3.3\%$. Within expected uncertainties\footnote{$5.4\%$ was found in \cite{Lyskova.et.al.2012} for a sample of simulated massive slowly rotating simulated galaxies} our simple estimate at the sweet point agrees well with the circular speed obtained from Schwarzschild modeling.

The method under consideration is not only simple and fast in implementation, the resulting estimate does not strongly depend on the quality of observational data. To demonstrate this we apply the analysis to the set of data used in the state-of-the-art modeling of M87 by \cite{Gebhardt.Thomas.2009} (hereafter `GT09'). GT09 as well as  M11 derive the mass profile of M87 employing basically the same axisymmetric orbit-based dynamical models. The only difference is the observational data used. GT09 use stellar kinematic data from SAURON \citep{Emsellem.et.al.2004} and from \cite {van.der.Marel.1994} at $R \lesssim 40''$. At larger radii GT09 use the surface density profile from \cite{McLaughlin1999} for the globular clusters (GC) and individual globular cluster velocities reported in \cite{Cote.et.al.2001}. For this data set our analysis results in  $V_c = 535.6 \kms$ at $R_{\rm sweet}=245.5''$. This estimate is basically based only on the GC data. It is about 21\% smaller than the circular speed derived by GT09 but it agrees well ($\disp \Delta=6.4\%$) with the rotation curve from M11. Figure \ref{fig:M87all} shows the circular speed profiles resulting from different methods. The profiles derived from detailed dynamical modeling of M11 and GT09 are shown as black solid and dashed lines, respectively. The simple $V_c$-estimates derived in this work are shown as a red circle (based on the set of data used in M11 ) and a red triangle (from data used in GT09). 
$V_c^{\rm iso}$ in the radial range where the log-slope of the surface brightness $\alpha \simeq 2$ with expected uncertainties ($\pm 5.4 \%$) is shown in cyan. 
 The grey shaded region represents the result of the X-ray analysis of available archive XMM and Chandra data. The X-ray data are deprojected assuming spherical symmetry to derive gas density and temperature profiles (see Section \ref{subsubsec:xray} for more detail). Using derived profiles and the hydrostatic equilibrium equation the mass profile is derived. The width of the shaded area is determined mainly by systematic deviations of $V_c(R)$ derived under an assumption of a fixed or free metal abundances (see details in Section \ref{subsubsec:xray}) rather than statistical variations. The dark grey shaded region is based on XMM data only. The discrepancy between the profiles from dynamical modeling and from X-ray analysis can be explained by the contribution of a gas non-thermal pressure to the total pressure \citep{2008MNRAS.388.1062C}. A wiggle in the X-ray-based $V_c(R)$ at $R \sim 200''$ is due to a quasi spherical shock generated by the supermassive black hole at the center of M87.

If we compare the optical circular speed estimate for M87 at the sweet spot with the X-ray based one at the same radius, we get $\disp \frac{V_c^{opt}}{V_c^X} \approx 1.12$, what implies $\sim 25 \%$ non-thermal pressure support. It should be noted, however, that the sweet radius happens to lie in the vicinity of the shock front \citep{Forman.et.al.2007} and  the X-ray circular speed might be underestimated in the region of a `dip'. 
Previously, the comparison of the X-ray data on M87 with the analysis of the optical data in \cite{Romanowsky.Kochanek.2001} and GT09 has been done in 
\cite{2008MNRAS.388.1062C} and \cite{2010MNRAS.404.1165C} respectively.  In the first case (\citealt{Romanowsky.Kochanek.2001} + X-Rays) no evidence for non-thermal pressure in excess of $\sim$10\% of the thermal pressure was found, while the comparison with GT09 results yielded a large non-thermal component of order 50\%. This discrepancy can be traced to the difference in the optical data. With new data and analysis of M11 this discrepancy largely goes away.
Note also, that in \cite{2008MNRAS.388.1062C,2010MNRAS.404.1165C} the circular speed estimate $\sim 440$  $\kms$ was derived from X-rays via fitting the gravitational potential by $\disp \Phi(R) = V_c^2\ln r + const$ in the broad radial range from $0.1'$ to $5'$.

\begin{table*}
\centering
\caption{Sample of elliptical galaxies. \label{tab:sample} The columns
are: (1) - common name of the galaxy; (2) - redshift from the NASA/IPAC
Extragalactic Database; (3) - adopted distance; (4) - central velocity dispersion from HyperLeda; (5) - hydrogen column density from \citep{Dickey.Lockman.1990}.}
\begin{tabular}{lccccccrll}
\hline
Name   & $z$ & $D$, Mpc & $\sigma_c$, km s$^{-1}$ &  $N_H$, $10^{20}$ cm$^{-2}$ \\
(1)   & (2) & (3) & (4) & (5)  \\
\hline

NGC 708        &0.016195 & 68.3 & $229.8 \pm 9.7$ & 5.37 \\
NGC 1129        &0.017325 & 73.1 & $329.5 \pm 15.1$ & 9.81\\
NGC 1550        &0.012389 & 52.1 & $308.0 \pm 6.2$ & 11.5 \\
NGC 4125        &0.004523 & 23.9 & $226.8 \pm 6.9$ & 1.84 \\
UGC 3957        &0.034120 & 145.9& $331.1 \pm 35.1$ & 4.63  \\

\hline
\end{tabular}
\end{table*}

%%%%%%%%%%%%%%%%%%%%%%%%%%
\subsection{Observations and data reduction}
\label{subsec:observations}

\begin{table*}
\caption{Log of the observations}\label{tab_obs}
\begin{tabular}{lrrccccc}
\hline
Galaxy    & Slit $PA$ & Date                 & Slit width & Sp. range         &  Sp. resol. & Exp. time    &  Seeing  \\
             &   (deg)&                                &   (arcsec)  &   (\AA)         &   (\AA)   &  (min)          &   (arcsec)\\
\hline
NGC 708   & -4   & 05.10.2011   &   1.0         & 4840--5610 &    2.2     & $180$         &   1.1--1.2\\ % Scorp-1 2300G 1''
                 &  215         & 03.10.2011   &   1.0         & 4840--5610 &    2.2    & $185$         &   1.4--3.5\\  % Scorp-1 2300G 1"
NGC 1129  &  166 & 21.10.2012  &   0.5         & 4080--5810 &    3.0     & $180$         &   1.2--1.4\\ % Scorp-1 1200G 0.5''
                 &  256      & 15.10.2012   &   0.5         & 4080--5810 &    3.0    & $180$         &   1.4--1.5\\  % Scorp-1 1200G  0.5''
NGC 1550  &  116 & 09.12.2012  &   1.0         &3700--7200&     5.1     & $180$         &   1.5--1.6\\ % Scorp-2 1200@540 1''
                 &  206    & 16-17.10.2012 & 0.5         &   4080--5810 &    3.0  & $180$         &   1.3--1.6\\  % Scorp-1 1200G  0.5'
NGC 4125  &  175 & 14.04.2013   &   1.0        & 4840--5610 &    2.2    & $160$        &   1.2--1.3\\ % Scorp-1 2300G  1''
                 &  265      & 19.10.2012   &    0.5         &4080--5810 &    3.0    & $200$         &   1.6--1.7\\  % Scorp-1 1200G 0.5''
UGC 3957  &  287 & 03.11.2010  &    1.0          & 4415--6015 &    2.2     & $180$         &   1.1--1.5\\ % Scorp-2   2300@520 1''
   \hline
\end{tabular}
\end{table*}

The spectroscopic observations at the prime focus of the SAO RAS 6-m telescope  were  made with the multi-mode focal reducer SCORPIO  \citep{scorpio} and  its new version SCORPIO-2 \citep{scorpio2}. When operated in the long-slit mode, both devices have same slit  6 arcmin in height  with a scale of 0.36 arcsec per pixel. However, with a similar spectral resolution SCORPIO-2 provides twice larger spectral range. The CCDs employed were an EEV 42-40 in the SCORPIO and E2V 42-90  in the SCORPIO-2.

Table \ref{tab:sample} lists the target galaxies and Table \ref{tab_obs} gives the log of observations:  the position angles of the spectrograph slit for each galaxy, the observing date, the slit width, spectral range, spectral resolution (estimated by the mean FWHM of air glow lines), total exposure $T_{\rm exp}$, and seeing. Usually we observed targets with two slit positions: along photometric major and minor axes. The exceptions are NGC 708 (the second slit was placed along the dust lane crossed galaxy nucleus) and UGC~3957 where  observations only along major axis have been performed.

\begin{figure*}
\plotthree{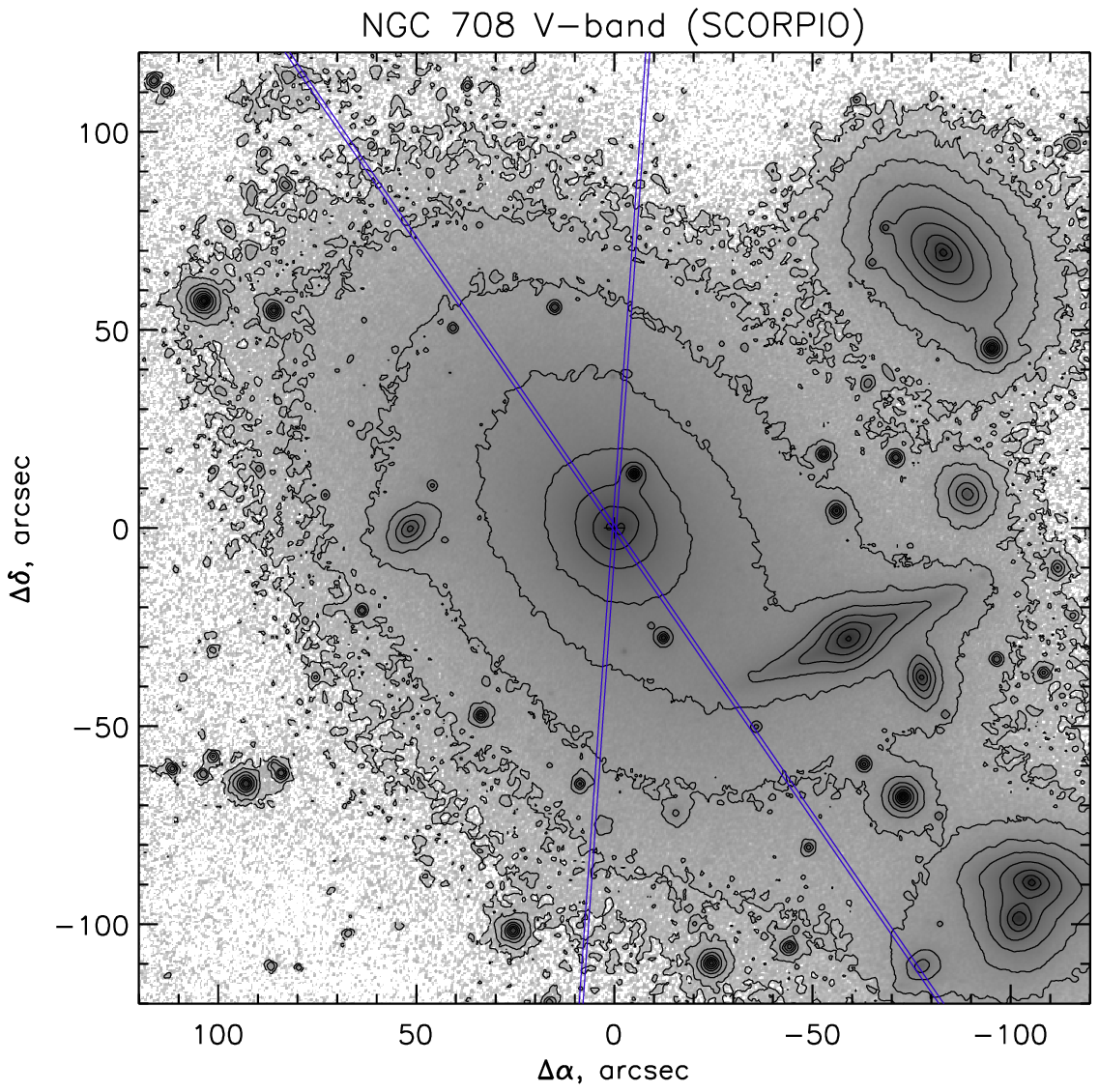}{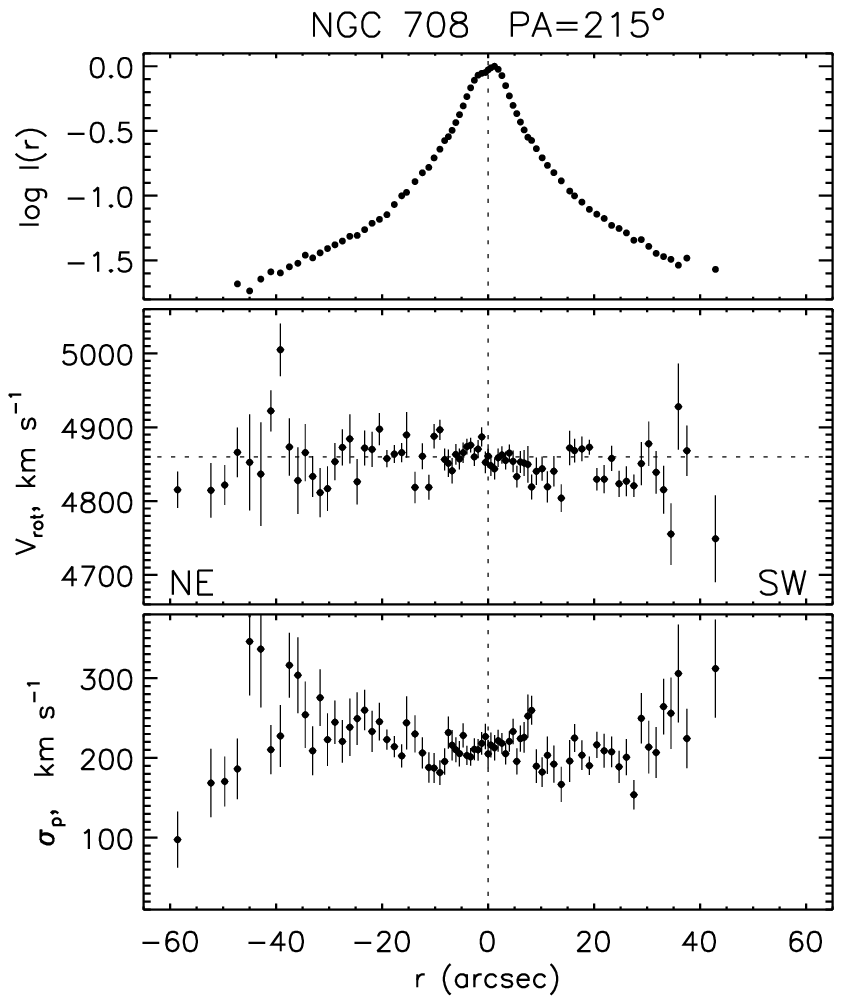}{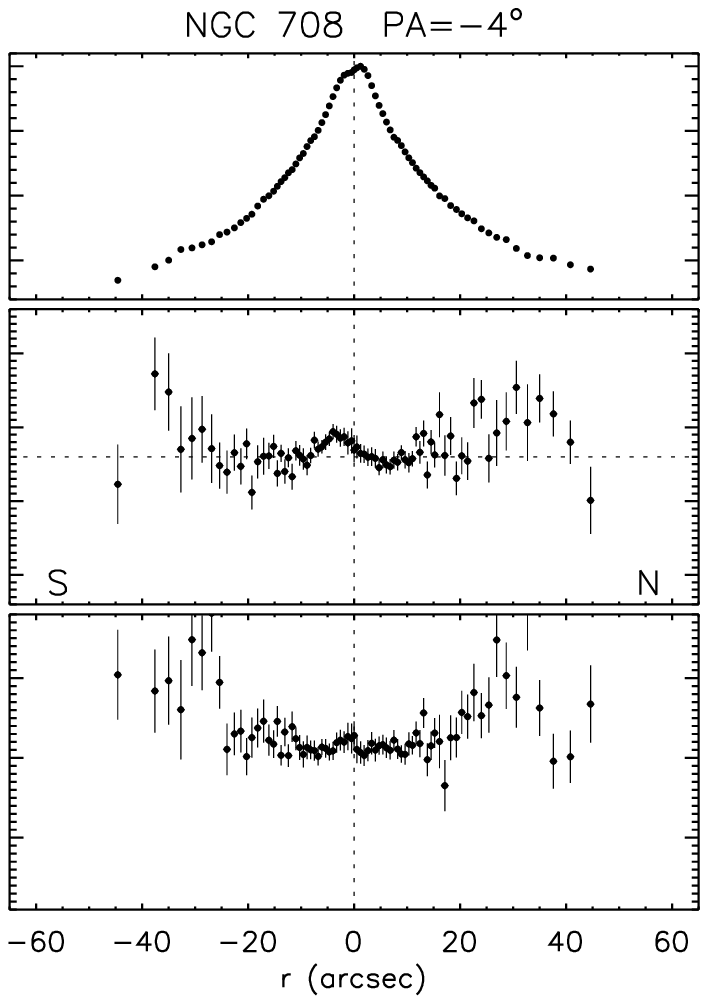}
\plotthree{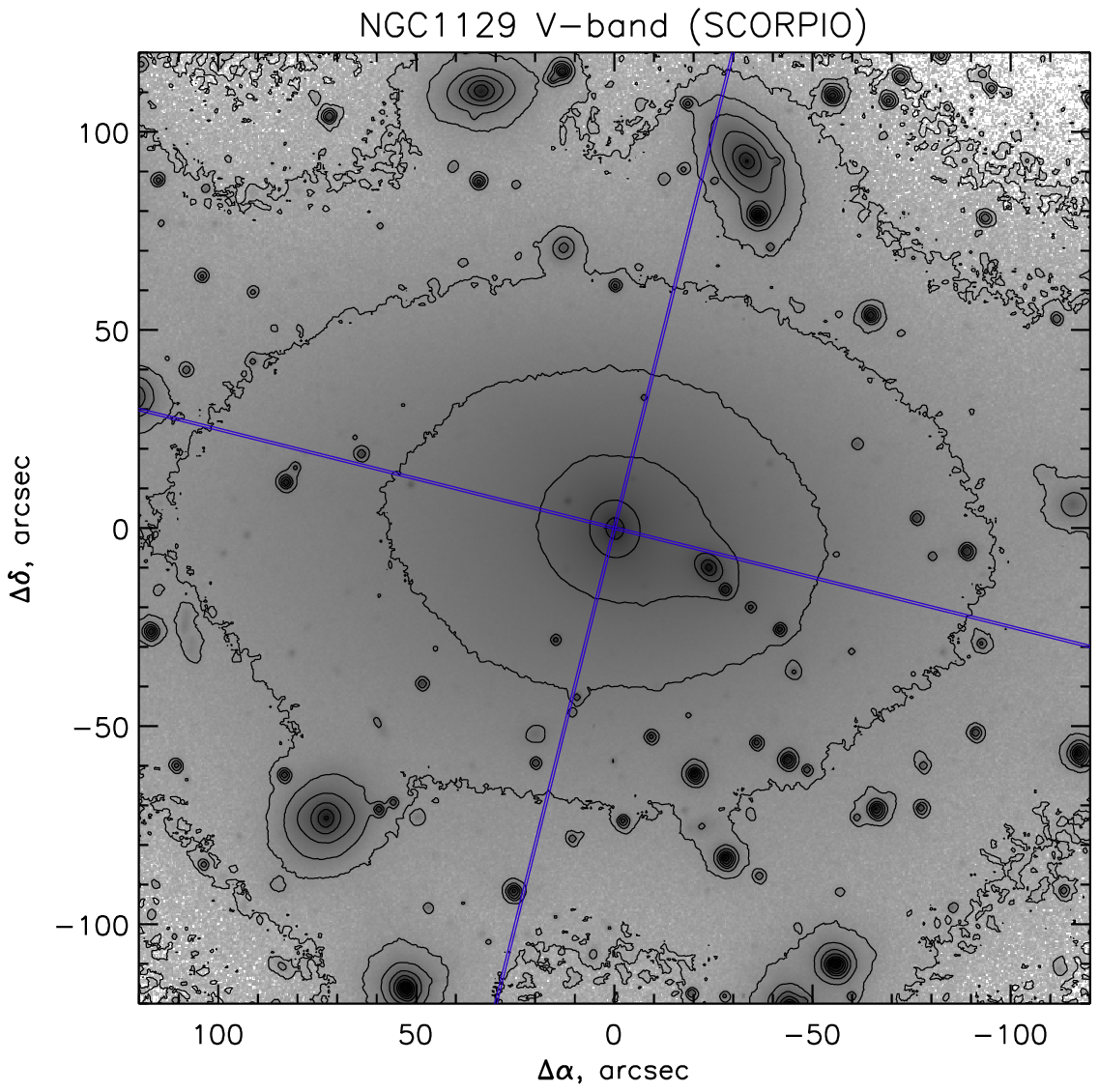}{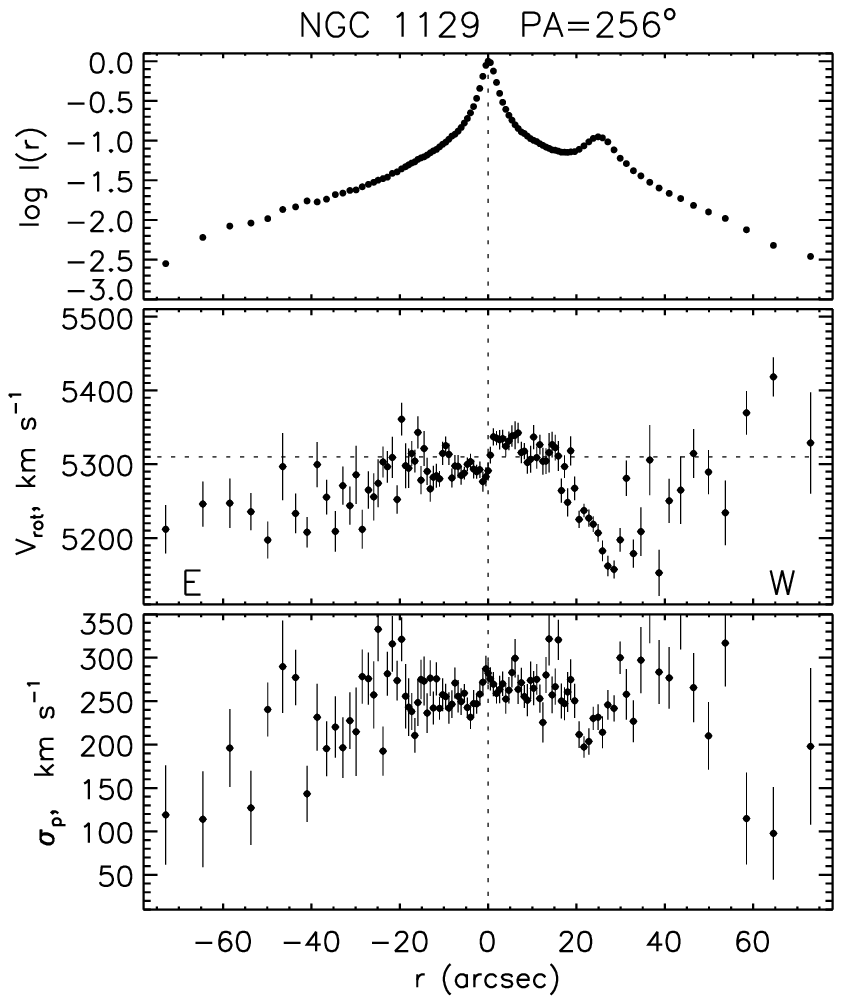}{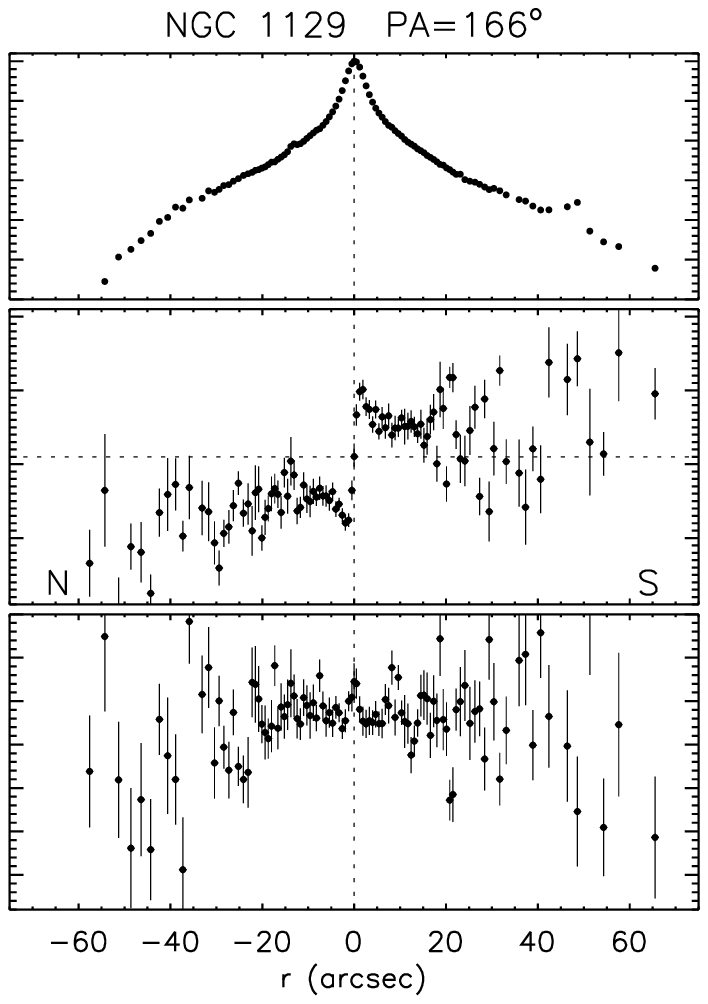}
\plotthree{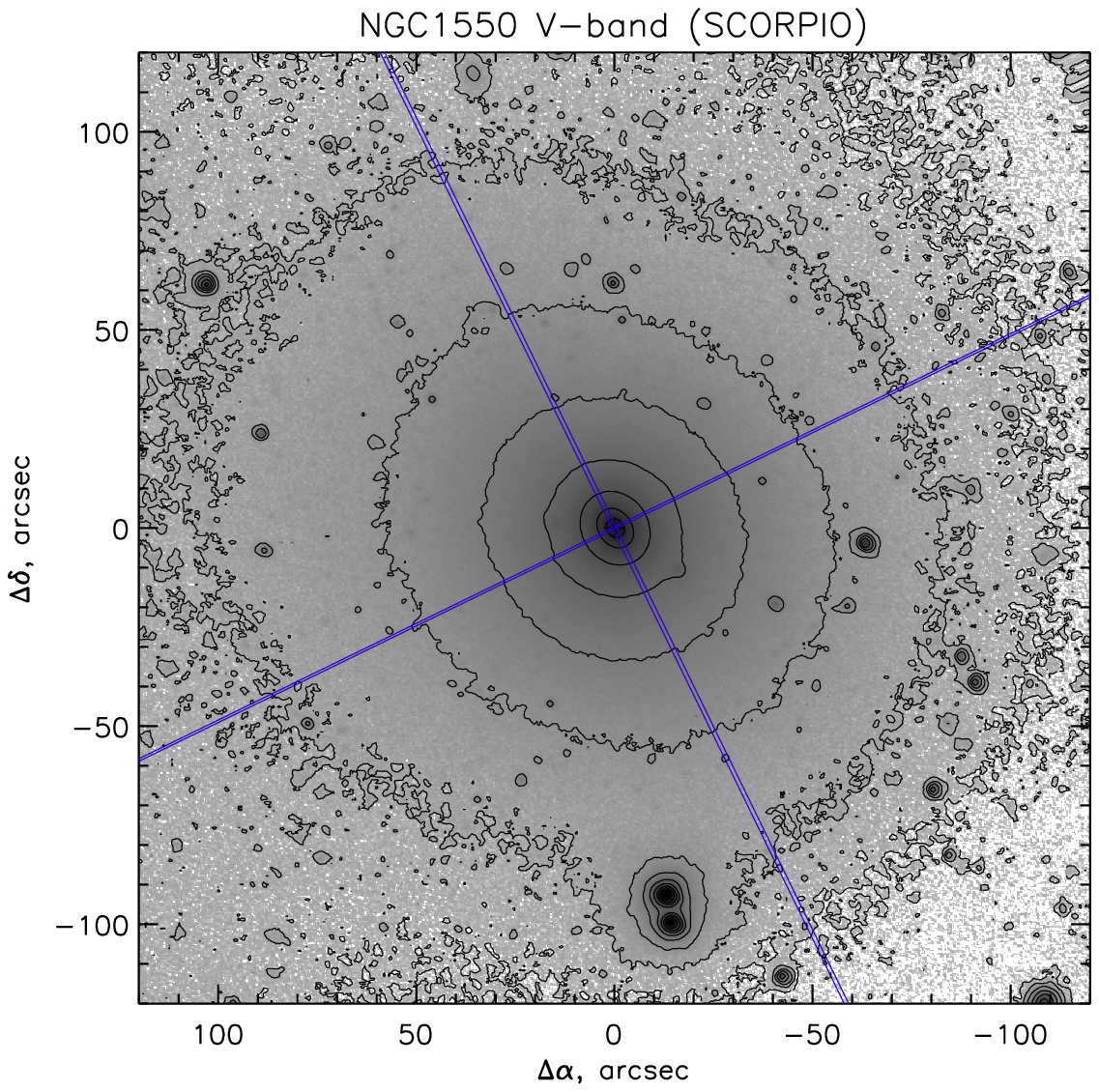}{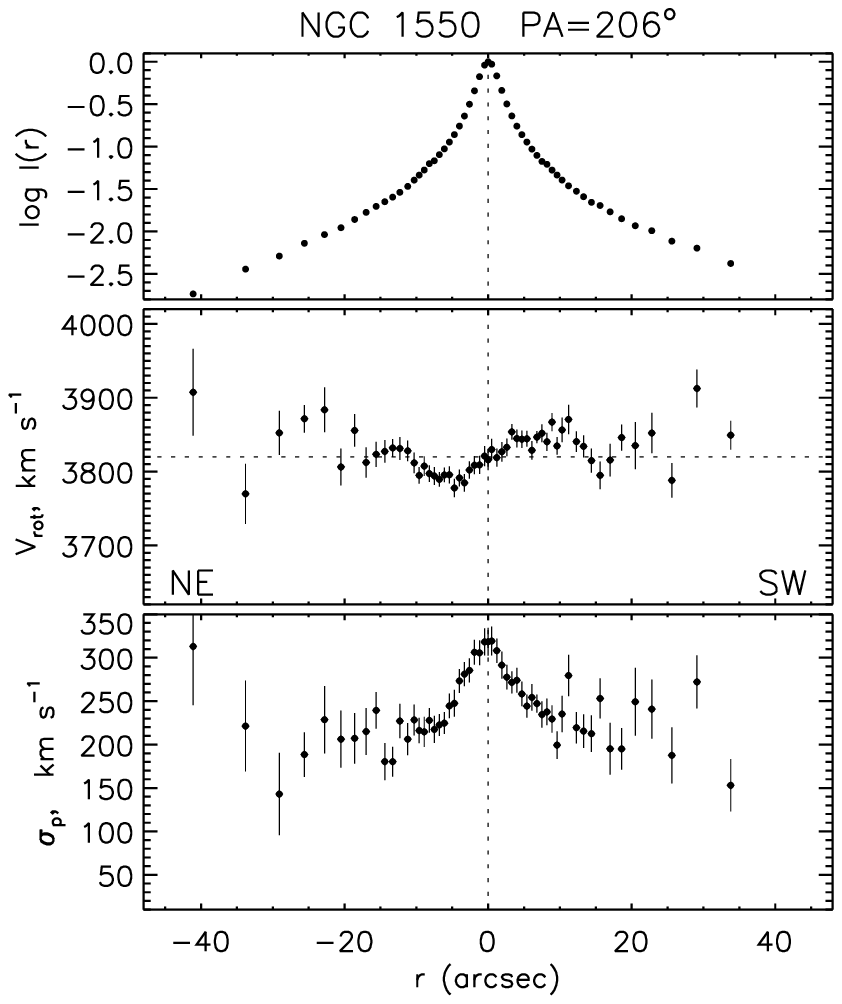}{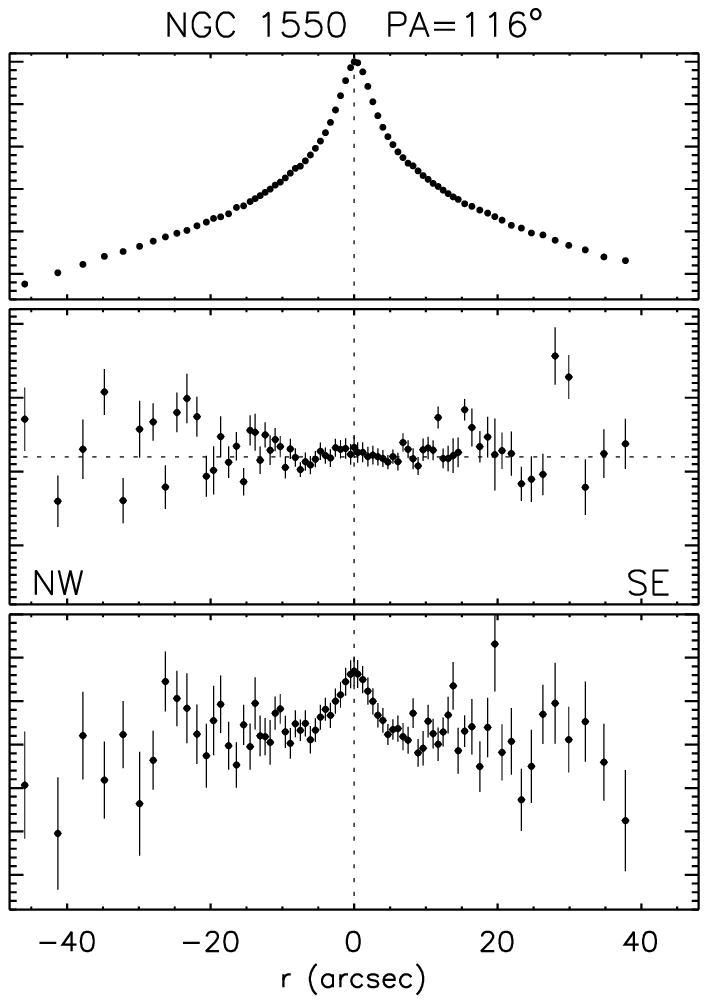}
\caption{The results of the SAO RAS 6-m telescope observations. Left: $V$-band image in logarithmic grey-scale and positions of the spectrograph slits. Middle: the distributions of stellar continuum surface brightness, line-of-sight velocities and velocity dispersion of stars along major axis. The dotted lines mark the position of nucleus and accepted systemic velocity. In the case of NGC~1129 at $r\approx20-30$ arcsec the slit crosses  the companion galaxy. Right: the same for the second slit position. At large radii velocity dispersion uncertainties are likely underestimated  as they do not account for systematic errors coming from the sky subtraction.} \label{fig1_1}
\end{figure*}

The data reduction were made  in a standard way using the IDL-based software package developed at the SAO RAS   \citep{scorpio}. The measurements of the distribution of line-of-sight velocities $V_{\rm rot}$ and stellar
velocity dispersion $\sigma_p$ were carried out by cross-correlating the spectra of galaxies with the spectra of the template star observed on
the same nights. The measurement technique has already been described in our previous papers \citep*{Moiseev2001,Silchenko2010}.
We observed several template stars belonging to the spectral types III G8 - III K5 and also  twilling light sky (i.e. solar spectrum). For the final measurements, we selected a template giving a maximum correlation coefficient. 
During the stellar kinematics parameters estimation we applied logarithmic binning along the slit to provide a sufficient signal-to-noise ratio ($S/N>
15-20$  per bin in each pixel). Also in each beam we calculated  the surface brightness $I(R)$ as an integral intensity of stellar continuum at the range $5040-5140$\,\AA\AA. The  Figure~\ref{fig1_1} shows the results of our spectral observations together with $V$-band images of the galaxies taken at the same nights in the direct image mode of SCORPIO and SCORPIO-2.

\setcounter{figure}{3}
\begin{figure*}
\plotthree{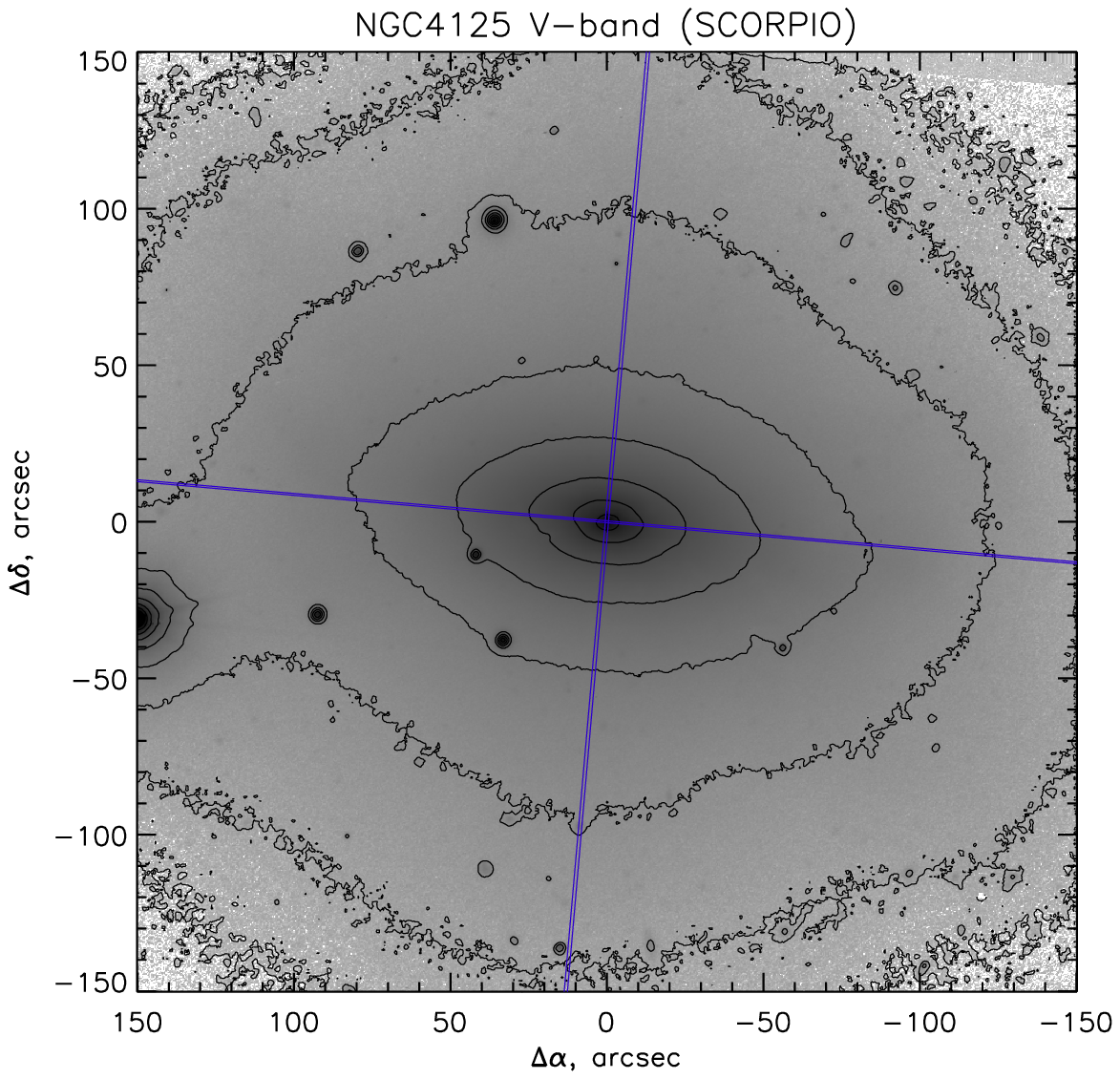}{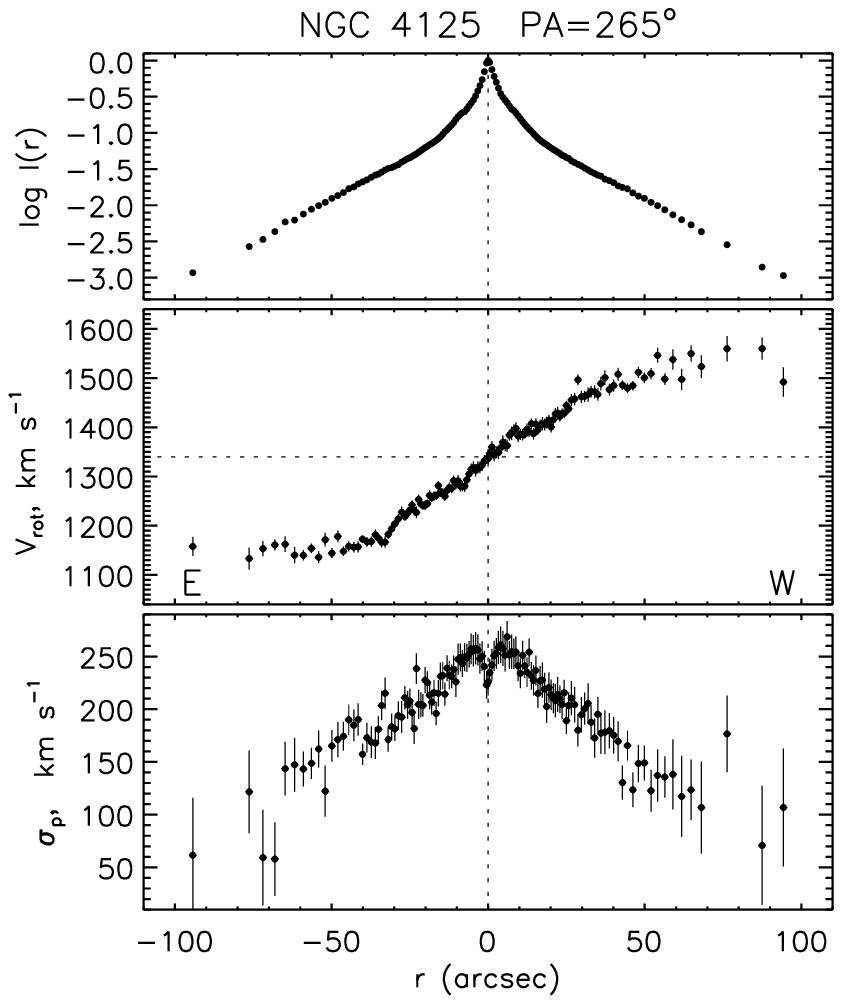}{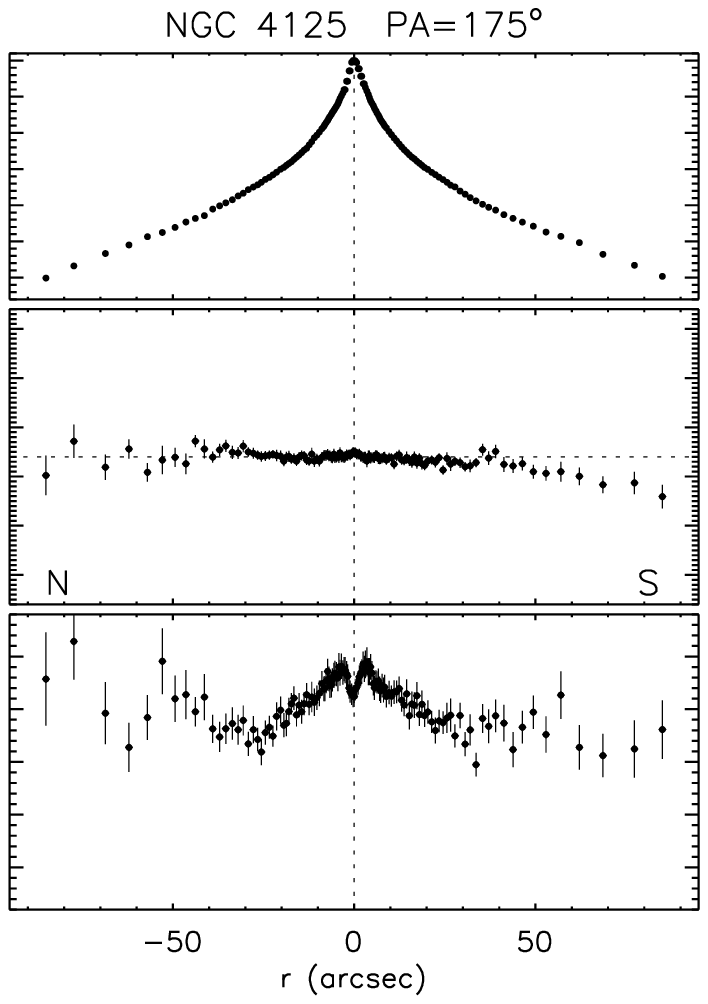}
\plottwoq{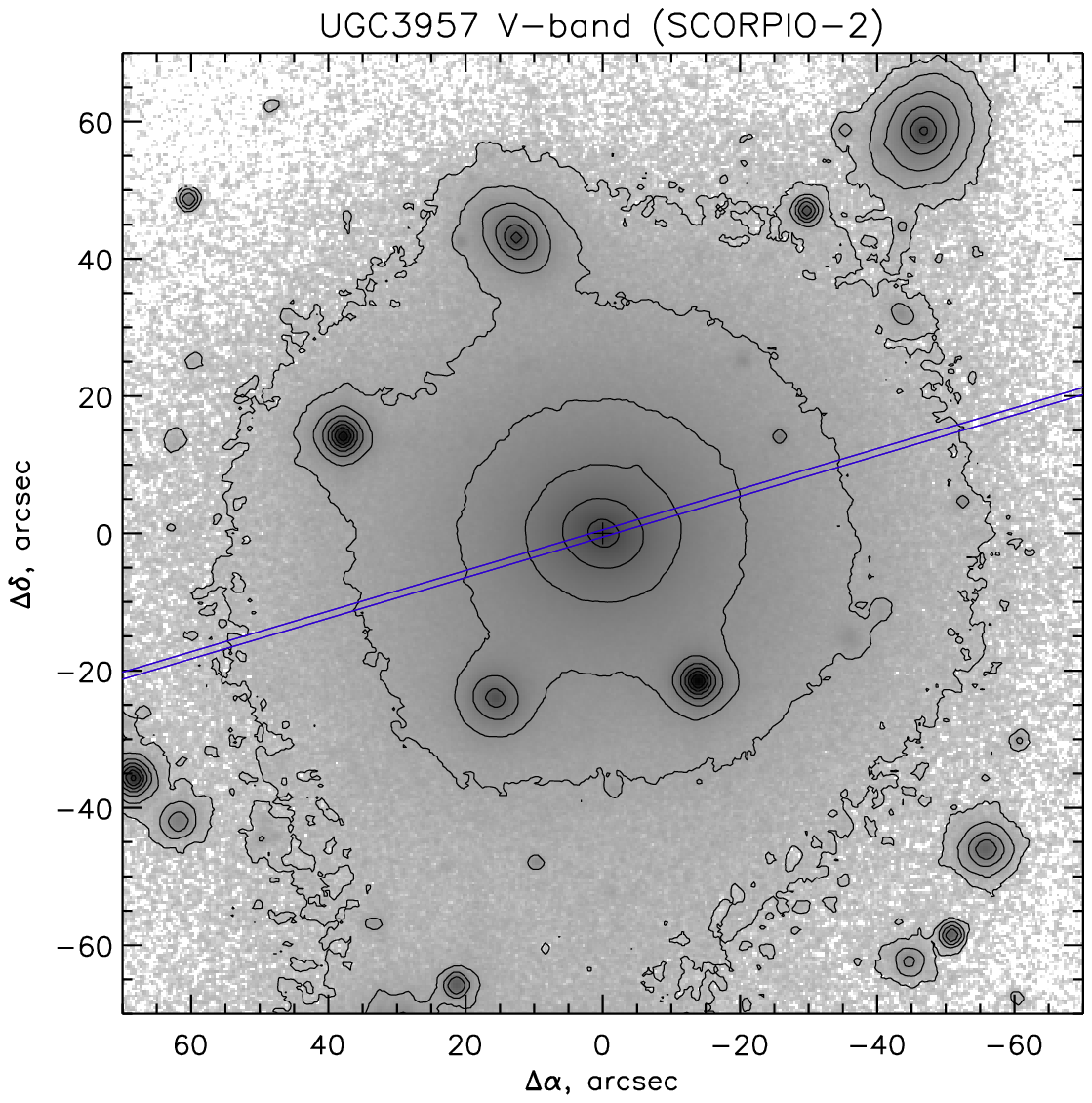}{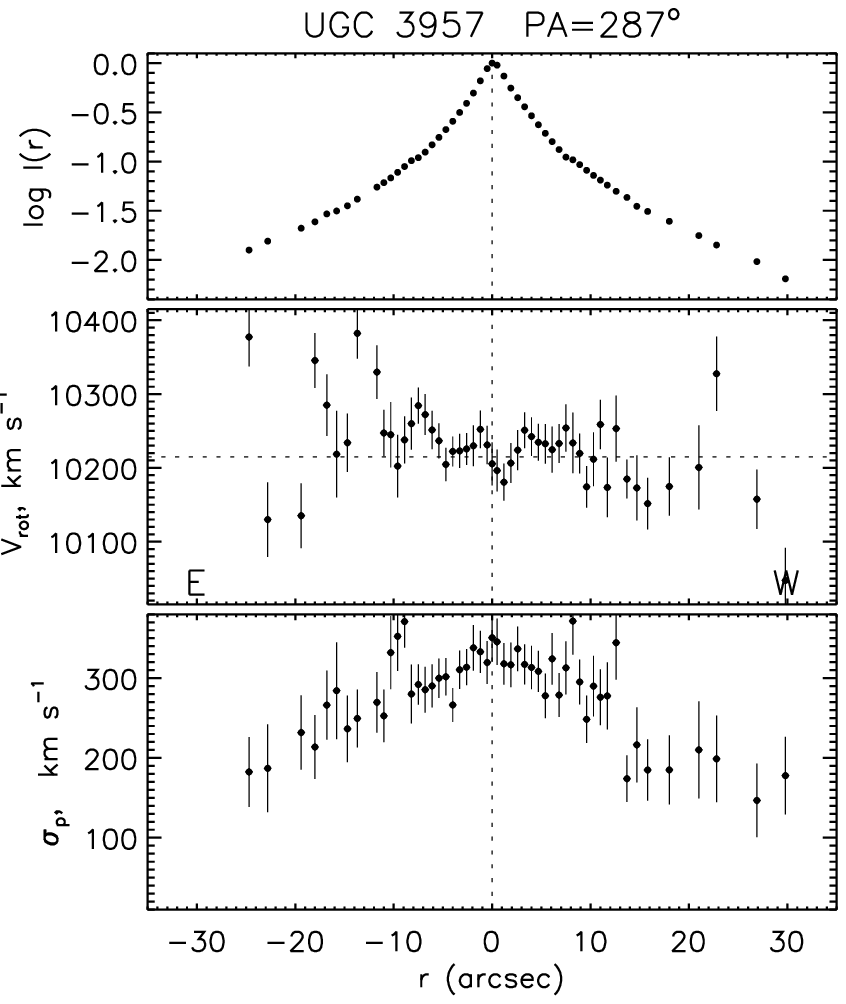}
\caption{(continue)} 
\label{fig1_2}
\end{figure*}

%%%%%%%%%%%%%%%%%%%%%%%%%%

\subsection{Rotation Curves}
\label{subsec:rotcurve}

\subsubsection{Circular speed from X-ray data.}
\label{subsubsec:xray}

Using publicly available Chandra data we have derived the circular speed profiles for galaxies in our sample under the assumption of the hydrostatic equilibrium.
We follow the procedure of the data analysis described in \cite{2010MNRAS.404.1165C}.  Here we only outline the major steps.

First, in each observation we follow the reduction procedure described in  \cite{2005ApJ...628..655V}, i.e. filter out high background periods and apply the latest calibration corrections to the detected X-ray photons, and determine the background intensity.

As a next step we apply a non-parametric deprojection procedure described in \cite{2003ApJ...590..225C,2008MNRAS.388.1062C}. In brief, the observed X-ray spectra in concentric annuli are modeled as a linear combination of spectra in spherical shells; the two sequences of spectra are related by a matrix describing the projection of the shells into annuli.  To account for the projected contribution of the emission from the gas at large distances from the center (i.e., at distances larger than the radial size $r_{max}$ of the region well covered by actual observations) one has to make an explicit assumption about the gas
density/temperature profile. We assume that at all energies the
gas volume emissivity at $r>r_{max}$ declines as a power law with
radius. The slope of this power law is estimated based on the
observed surface brightness profile at $r \lesssim r_{max}$. Since we assume that the same power law shape
is applicable to all energy bands, 
effectively this assumption
implies constant spectral shape and therefore the
isothermality of the gas outside $r_{max}$. The contribution of these
layers is added to the projection matrix with the normalization as an
additional free parameter.   The final projection matrix
is inverted and the shells' spectra are explicitly calculated by
applying this inverted matrix to the data in narrow energy channels.

The resulting spectra are approximated in XSPEC \citep{1996ASPC..101...17A}
with the Astrophysical Plasma Emission Code (APEC) one-temperature optically thin plasma emission model
\citep{2001ApJ...556L..91S}.  The redshift $z$ (from the NASA/IPAC
Extragalactic Database -- NED) and the line-of-sight column density of neutral
hydrogen $N_H$ \citep{Dickey.Lockman.1990} have been fixed at the values
given in Table \ref{tab:sample}. For each shell we determine the emission
measure (and therefore gas density) and the gas temperature. These quantities
are needed to evaluate the mass profile through the hydrostatic
equilibrium equation.  For cool (sub-keV) temperatures and approximately solar
abundance of heavy elements, line emission provides a substantial fraction of
the 0.5-2 keV flux. With spectral resolution of current X-ray missions the
contributions of continuum and lines are difficult to disentangle. As a result
the emission measure and abundance are anti-correlated, which can lead to
a large scatter in the best-fit emission measures. As an interim (not entirely
satisfactory) solution, we fix the abundance at 0.5 solar for all shells, using
the default XSPEC abundance table of \cite{1989GeCoA..53..197A}. We return to this issue below.

Knowledge of the gas number density $n$ and temperature $T$ in each shell allows us to evaluate the $M(R)$ or $V_c(R)$ profile by using the hydrostatic equilibrium equation:

\be
-\frac{1}{\rho}\frac{dP}{dr}=\frac{d\Phi}{dr}=\frac{V_c^2}{r}=\frac{GM}{r^2},
\label{eq:he}
\ee

where $P=nkT$ is the gas pressure, $\rho=\mu m_p n$ is the gas
density ($m_p$ is the proton mass).  The mean atomic weight $\mu$ is assumed to be equal to 0.61. 

The resulting circular speed profiles $V_c^X(r)$ for all galaxies in our sample are shown as black thick lines with errobars represented as black shaded regions in lower panels of Figure \ref{fig:all}. One should keep in  mind that in assuming hydrostatic equilibrium one neglects possible non-thermal contribution to the pressure, arising from turbulence in the thermal gas, cosmic rays, magnetic fields and non-radiating relativic protons \citep[e.g.,][]{2008MNRAS.388.1062C}. So comparing optical and X-ray estimates of the circular speed may provide constraints on the contribution of the non-thermal particles to the gas pressure. High-resolution cosmological simulations of galaxy clusters suggest that the gas motions contribute $\sim 5\%$ of the total pressure support at the center and up to $\sim 15-20 \%$ at $r_{500}$ in relaxed systems \citep[e.g.][]{Lau.et.al.2009, Zhuravleva.et.al.2013}. Recent studies on combining X-ray mass measurements and sophisticated stellar dynamical methods imply up to $\sim 50 \%$ non-thermal support (e.g. \citealt{Shen.Gebhardt.2010}, \citealt{Rusli.et.al.2011}, GT09), although the uncertaintes in model assumptions may be significant \citep{Boute.Humphrey.2012}. As our simple method provides a mass-estimate that is robust and largely insensitive to the orbital anisotropy at the sweet point, we interpret the offset at $R_{\rm sweet}$ between the X-ray and optical measurements as a signature of deviations from hydrostatic equilibrium. In particular, the ratio $\disp f_{nt}=(M_{opt}-M^X)/M_{opt}$ provides an estimate of the fractional contribution of the nonthermal pressure to the total pressure, provided that this fraction does not vary with radius. In this approximation $\disp M^{X,c}(r)=M^X(r)/(1-f_{nt})$ is an estimate of the mass at other radii.

As the gas pressure is assumed to be isotropic, mismatch between the corrected X-ray circular speed $V_c^{X,c}$ and the isotropic one $V_c^{\rm iso}$ derived from the Jeans equation may give a clue regarding the orbital structure of the galaxy. E.g.,  at small radii $\disp V_c^{X,c}>V_c^{\rm iso}$ would suggest more circular orbits, while at larger radii this would correspond to more radial orbits. Of course, the reliability of such analogy strongly depends on the derived $V_c^X$ and $V_c^{\rm iso}$.

\begin{figure*}
\plotwide{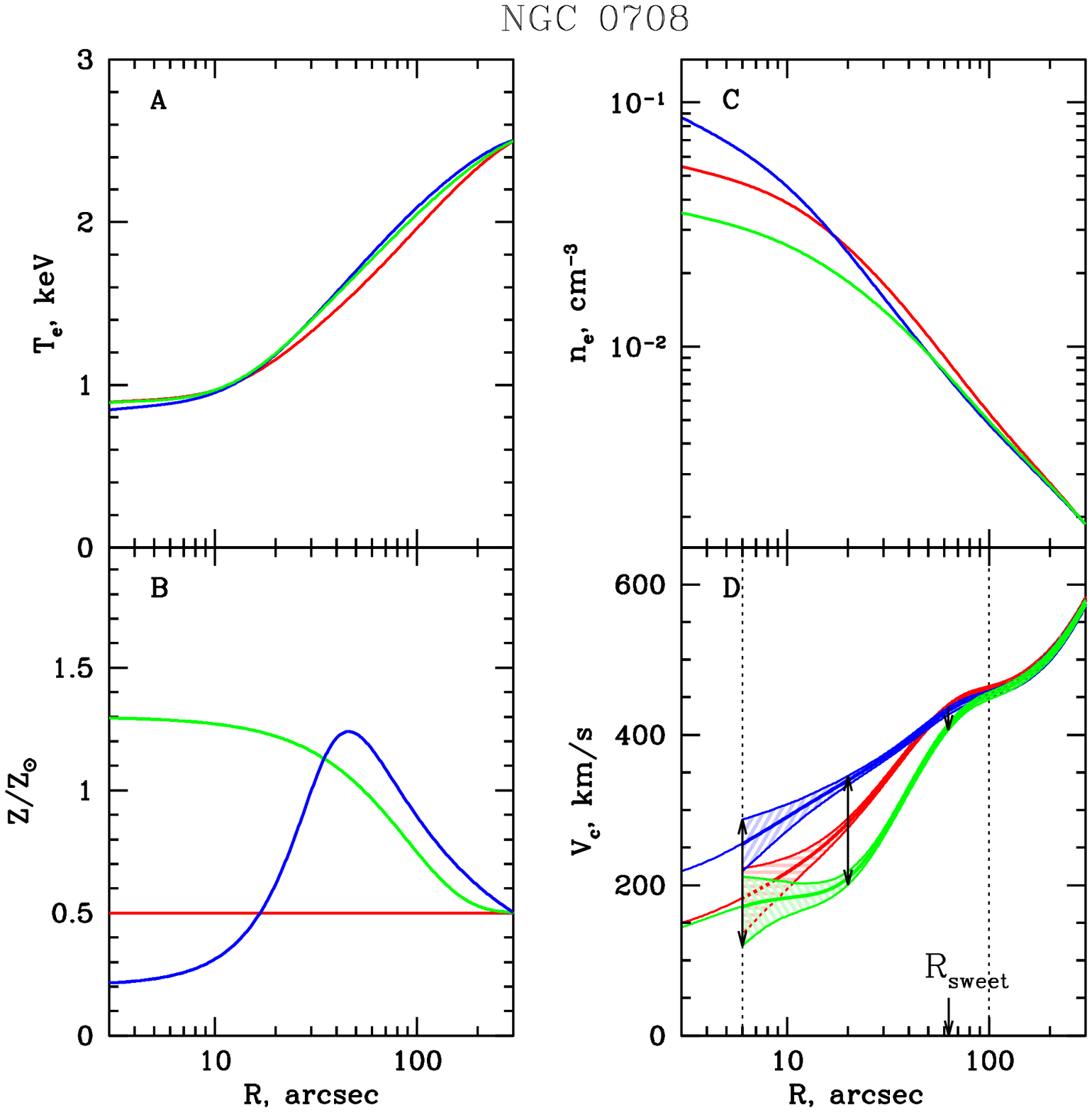}
\caption{The effect of the abundance gradient on the calculated circular speed profile. Panel (A) shows the deprojected temperature for NGC 0708, panel (B) - assumed models for the metallicity, panel (C) - gas density profiles and panel (D) - resulting circular velocity curves with 1$\sigma$ errors from 1000 Monte Carlo realizations. The vertical dotted lines indicate a region of interest where both optical and X-ray data are available. The arrows show the spread in circular speed estimates coming from different abundance profiles.   
\label{fig:ab}
}
\end{figure*}

We now illustrate the impact of our assumption of a flat abundance profile $Z = 0.5 Z_{\odot}$ and estimate arising errors on the inferred circular speed. 
 
At low temperatures ($\lesssim 2 $ keV) metal abundances derived from X-ray spectra with the limited energy resolution of current X-ray missions suffer from the ambiguity of disentangling line emission and continuum. While derived circular velocity is weakly sensitive to the particular value of metallicity in the spectral models, it can be significantly affected by the radial variations of the heavy-element abundance \citep[e.g.][]{Johnson.et.al.2009,2010MNRAS.404.1165C}. 

Since abundance measurements can be biased \citep[e.g.][]{Buote.2000} we tried to make a conservative estimate of the varying abundance profile impact by setting by hand several model metallicity profiles and fitting the deprojected data again, leaving the normalization and temperature as free parameters\footnote{One may consider the stellar metallicity profiles from the absorption indices (Section \ref{subsubsec:stars}) as input $Z(r)$. For our sample galaxies this assumption does not affect the final estimates in a significant way.}.  

%As an example, we show in the lower panel Figure \ref{fig:ab} several profiles for the abundance of heavy metals as a function of radius for NGC 0708 and in the upper panel the resulting circular speed curves with statistical errorbars coming from 1000 Monte Carlo simulations. Here we consider 3 models: (i) flat abundance profile $Z = 0.5 Z_{odot}$ (shown in red), (ii) fit to the deprojected abundance (in blue) and (iii) physically motivated model (in green), where the metal abundance rises to the galaxy center as it is generally expected for elliptical galaxies \citep[e.g.][]{Humphrey.Buote.2006}.    

As an example, we show in  Figure \ref{fig:ab} the derived density, temperature and circular speed profiles\footnote{As eq. (\ref{eq:he}) requares differentiation, to calculate derivatives we smooth density, temperature and pressure profiles following the procedure described in \cite{2010MNRAS.404.1165C}. The typical value of the smoothing width is $\sim 0.55$. } for NGC 0708. The estimated  statistical errorbars come from 1000 Monte Carlo simulations. Here we consider 3 models: (i) flat abundance profile $Z = 0.5 Z_{\odot}$ (shown in red), (ii) fit to the deprojected abundance with a `dip' at the center (in blue) and (iii) physically motivated model (in green), where the metal abundance rises to the galaxy center as is generally expected for elliptical galaxies \citep[e.g.][]{Humphrey.Buote.2006}.  Compared to the flat abundance profile, the metallicity monotonically increasing towards the center  leads to the flattening of the gas density profile and lowering the final circular speed estimate. In contrast, the decreasing to the center $Z(r)$ `boosts' inferred $V_c^X$, as is clearly seen from  Figure \ref{fig:ab}. 

Other galaxies in our sample show only monotonical increase of the deprojected metal abundance to the center, so the spread in final $V_c^X$-esimates is smaller. The circular velocity profiles corresponding to the flat abundance (thick solid black lines) and 1$\sigma$-errors from 1000 Monte Carlo simulations (black shaded area enclosed by two thin black lines) are shown in Figure \ref{fig:all}.

\subsubsection{Optical Rotation Curves}
\label{subsubsec:opt}

All observed galaxies in the sample are quite massive, close to spherical and slowly rotating (except maybe NGC 4125). This makes them suitable for our analysis. According to the algorithm for estimating the circular speed described in Section \ref{subsec:algorithm} we perfom the following steps:

\begin{enumerate}

\item First, given the surface brightness $I_i$, projected velocity dispersion $\sigma_{pi}$ with its errors $\Sigma_{\sigma i}$ and rotational velocity $V_{\rm rot \it i}$ with its errors $\Sigma_{\rm rot \it i}$ along two slits ($i=1,2$) we construct the average profiles 

\be
\label{eq:SB}
I=\frac{I_1+I_2}{2};
\ee

\be
\label{eq:VD}
\disp \sigma^2_p=\frac{I_1\sigma^2_{p1}/\Sigma^2_{\sigma 1}+I_2\sigma^2_{p 2}/\Sigma^2_{\sigma 2}}{I_1/\Sigma^2_{\sigma 2}+I_2/\Sigma^2_{\sigma 2}}.
\ee

If rotation velocity along the slit $V_{\rm rot \it i}$ is not negligible, then instead of $\sigma_p$ we use $V_{\rm rms}$ defined as

\be
\label{eq:Vrms}
V^2_{\rm rms}=\frac{I_1 V^2_{\rm rms \it 1}/\Sigma_{\rm rms \it 1}^2+I_2 V^2_{\rm rms \it 2}/\Sigma_{\rm rms2}^2}{I_1 / \Sigma_{\rm rms \it 1}^2+I_2/ \Sigma_{\rm rms \it 2}^2},
\ee
where $\disp \Sigma_{\rm rms \it i}^2=\Sigma_{\sigma \it i}^2+\Sigma_{\rm rot \it i}^2$; $\Sigma_{\rm rot \it i}$ are the errorbars assigned to rotation velocity measurements.

\item Then we calculate the logarithmic derivatives $\alpha$, $\gamma$ and $\delta$ from the derived profiles using equations (\ref{eq:agd}).

\item Next, we compute the circular speed $V_c(R)$ for isotropic, radial and circular stellar orbits using equations (\ref{eq:main}) in case of reliable data (full analysis) or equations (\ref{eq:agd_simple}) in case noisy or systematics affected observational line-of-sight velocity dispersion profile (simplified analysis). As discussed above, for rotating galaxies one should use $\disp V_{\rm rms}(R)$ instead of $\disp \sigma_p(R)$ in equations (\ref{eq:main}) or (\ref{eq:agd_simple}).

\item $\disp V_c^{\rm iso}(R_{\rm sweet})$ is taken as a final estimate of $V_c$ at the sweet spot $R_{\rm sweet}$ - the radius at which all three curves $\disp V^{\rm iso}_c(R), V^{\rm circ}_c(R)$ and $\disp V^{\rm rad}_c(R)$ are maximally close to each other. At $R_{\rm sweet}$ the sensitivity of the method to the anisotropy parameter $\beta$ is expected to be minimal so the estimation of the circular speed at this particular point is not affected much by the unknown distribution of stellar orbits. In case of the simplified version of the analysis $\disp R_{\rm sweet} \equiv R_2$, where $\disp \alpha = -\frac{d\ln I(R)}{d\ln R}=2$.

\item Finally, we compare derived $V_c$-estimates with the X-ray circular speed at the same radius. Table \ref{tab:table} summarizes the results, providing both optical and X-ray circular speed estimates as well as the temperature of the hot gas at $R_{\rm sweet}$. The scatter in optical $V_c$ arises from differences in observed $I(R)$ and $\sigma_p(R)$ along two slits or from measurement errors of $\sigma_p(R)$ when information is available along one slit only. Errors for X-ray derived $V_c$  comes from 1000 Monte Carlo realizations. In parentheses we present the conservative estimate of errors for the case of radially varying metallicity.    

\end{enumerate}

In Figure \ref{fig:all} we present results of the analysis. The logarithmic slope of the surface brightness profile for each galaxy in our sample is shown in panel A. Thin red lines correspond to slopes measured along the individual slits, while the thick lines show the average profiles (eq. \ref{eq:SB}). The shaded area indicates the scatter in profiles arising from different slits (when available). The derived profiles for $V_c^{\rm iso}$, $V_c^{\rm circ}$ and $V_c^{\rm rad}$ are shown in panel B in blue, magenta and green correspondingly. Again, the thin lines represent the $V_c$-curves resulting from measurements along each individual slit, while the thick lines demonstrate the average profiles coming from equations (\ref{eq:SB})-(\ref{eq:Vrms}). If information is available along two slits, then the shaded regions (paleblue for $V_c^{\rm iso}$, plum for $V_c^{\rm circ}$ and palegreen for $V_c^{\rm rad}$) show the scatter between these slits, in a case when profiles are measured along one slit only, the shaded regions indicate the measurements errors of $\sigma_p(R)$. The circular speed profiles derived from Chandra data are overplotted with errorbars in black. Stellar contribution to the circular velocity derived for the Salpeter and Kroupa IMF is shown in yellow (see Section \ref{subsubsec:stars}).

%Shown in yellow is the contribution from the stellar component only. (\textbf{!!!} )       

%to place constraints on mass-to-light ratio, to estimate the dark matter contribution to the total mass and to place some constraints on the stellar orbit distribution.  

%\begin{table*}
%\centering
%\caption{$V_c$-estimates for our sample of elliptical galaxies derived from optical and X-ray analyses. \label{tab:table}  The columns are: (1) - common name of the galaxy; (2) - sweet radius; (3) - optical $V_c$-estimate at $R_{\rm sweet}$; (4) - optical mass within $R_{\rm sweet}$; (5) - $V_c$-estimate at $R_{\rm sweet}$ from X-ray; (6) - X-ray mass within $R_{\rm sweet}$.}
%\begin{tabular}{lccccc}
%\hline
%Name   & $R_{\rm sweet}$,arcsec  & $V_c^{\rm iso}$, $\kms$& $M_{\rm opt}$, $10^{11} M_{\odot}$  & $V_c^X$, $\kms$ &  $M_{\rm X}$, $10^{11} M_{\odot}$ \\
%(1)   & (2) & (3) & (4) & (5) & (6) \\
%\hline
%
%NGC1550        &  30.9 & 382 & 2.6 & 383 & 2.6 \\
%NGC1129        &  45.7 & 444 & 7.2 & 464 & 7.8 \\
%NGC4125        &  44.7 & 375 & 1.7 & 322 & 1.3 \\
%NGC0708        &  63.1 & 371 & 6.5 & 438 & 9.0 \\
%UGC3957        &  14.8 & 476 & 5.2 & 504 & 5.8  \\
%
%\hline
%\end{tabular}
%\end{table*}

\begin{table*}
\centering
\caption{$V_c$-estimates for our sample of elliptical galaxies derived from optical and X-ray analyses. \label{tab:table} The columns are: (1) - common name of the galaxy; (2) - sweet radius; (3) - optical $V_c$-estimate at $R_{\rm sweet}$; (4) - $V_c$-estimate at $R_{\rm sweet}$ from X-ray, in parentheses are presented the conservative error estimates ; (5) - gas temperature at $R_{\rm sweet}$.}
\begin{tabular}{lcccc}
\hline
Name   & $R_{\rm sweet}$,arcsec  & $V_c^{\rm iso}$, $\kms$& $V_c^X$, $\kms$ &  kT, keV \\
(1)   & (2) & (3) & (4) & (5) \\
\hline

NGC 708        &  63.1 & $371^{+53}_{-53}$ &  $437^{+4}_{-4}$ $(_{-31})$ & 1.7 \\[1ex]
NGC 1129        &  45.7 & $444^{+31}_{-44}$ & $464^{+20}_{-25}$  &  3.0 \\[1ex]
NGC 1550        &  30.9 & $382^{+12}_{-19}$ & $383^{+5}_{-6}$ $_{(-29)}$ & 1.2 \\[1ex]
NGC 4125        &  44.7 & $375^{+45}_{-36}$ & $322^{+7}_{-9}$ $(_{-18})$ &  0.5 \\[1ex]
UGC 3957        &  14.8 & $476^{+43}_{-43}$ &  $518^{+45}_{-66}$ $^{(+47)}$ & 2.2 \\[1ex]

\hline
\end{tabular}
\end{table*}

\subsubsection{Comments on individual galaxies.}
\label{subsubsec:comments}

% figures...

%\begin{subfigures}
%\begin{figure*}
%\plottwo{N1550_V_slitpos.eps}{ngc1550_xo_pap_v2.ps}
%\caption{NGC 1550. {Left:} V-band image and position of slits.
%{Right:} Panel A - the surface brightness slope $\alpha = \disp d\ln I(R)/d\ln R$. Panel B - circular velocity profiles for isotropic (blue lines and paleblue shaded area), pure radial (green lines and palegreen shaded area) and pure circular (magenta lines and plum shaded area) orbits. The $V_c(R)$ derived from Chandra data under the assumption of hydrostatic equilibrium is shown as the black thick line. The shaded area shows statistical errorbars for the flat and varying with radius metallicity.  
%%The stellar contribution to the circular speed profile is presented in yellow.
%\label{fig:1550}
%}
%\end{figure*}

%\begin{figure*}
%\plottwo{N1129_V_slitpos.eps}{ngc1129_xo_pap_v2.ps}
%\caption{NGC 1129.
%\label{fig:1129}
%}
%\end{figure*}

%\begin{figure*}
%\plottwo{N4125_V_slitpos.eps}{ngc4125_xo_pap_v2.ps}
%\caption{NGC 4125.
%\label{fig:4125}
%}
%\end{figure*}

%\begin{figure*}
%\plottwo{N708_SCORPIO.ps}{ngc0708_xo_pap_v2.ps}
%\caption{NGC 708. Arrows are the same as in Figure \ref{fig:ab} and indicate the conservative lower and upper limits on X-ray circular speed coming from radial variations of metal abundance. The right arrow (the shortest one) is located at the optical sweet radius. 
%\label{fig:708}
%}
%\end{figure*}

%\begin{figure*}
%\plottwo{U3957_slitpos.eps}{ugc3957_xo_pap_v2.ps}
%\caption{UGC 3957.
%\label{fig:3957}
%}
%\end{figure*}
%\end{subfigures}

\begin{figure*}
\plotfive{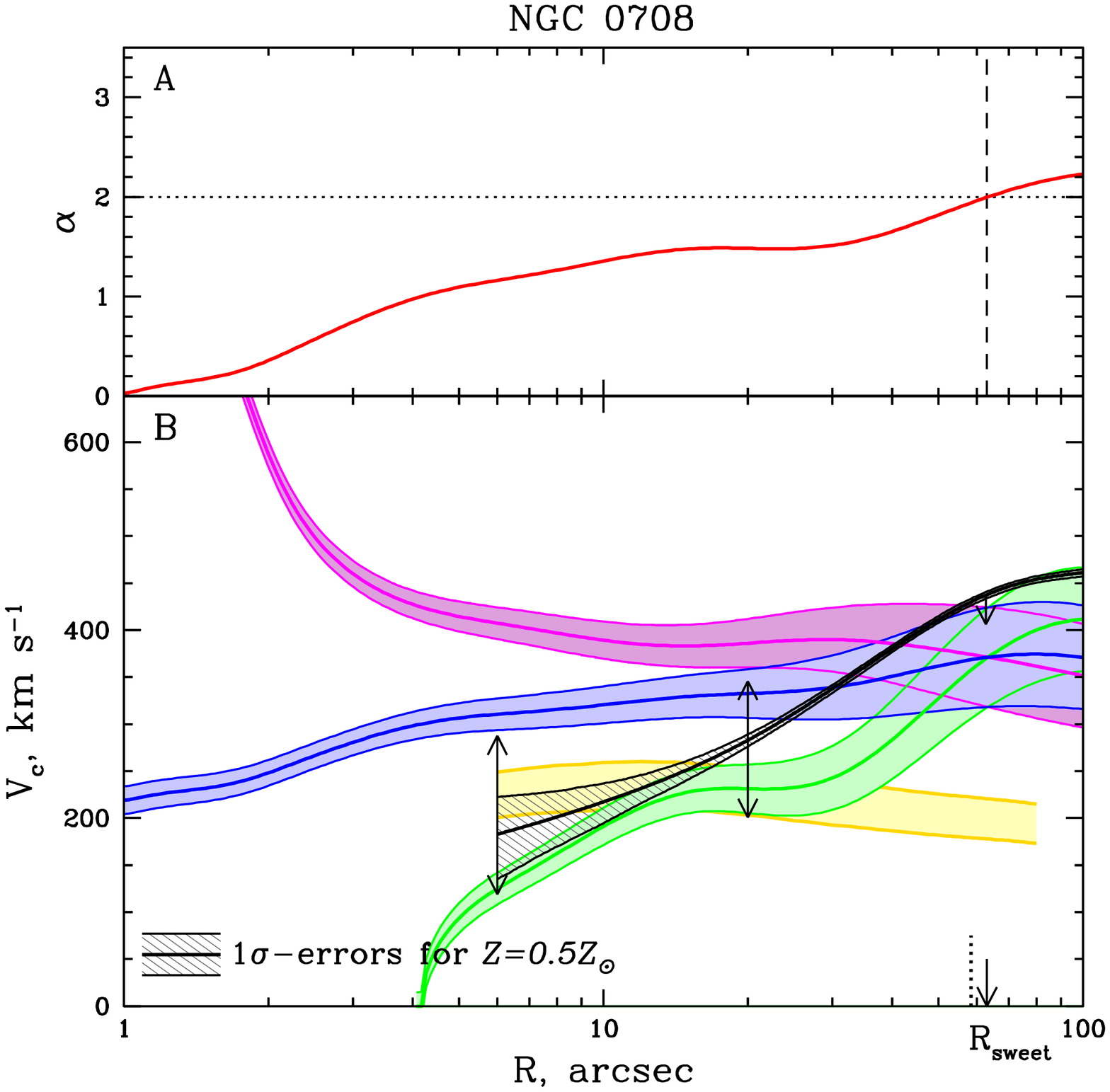}{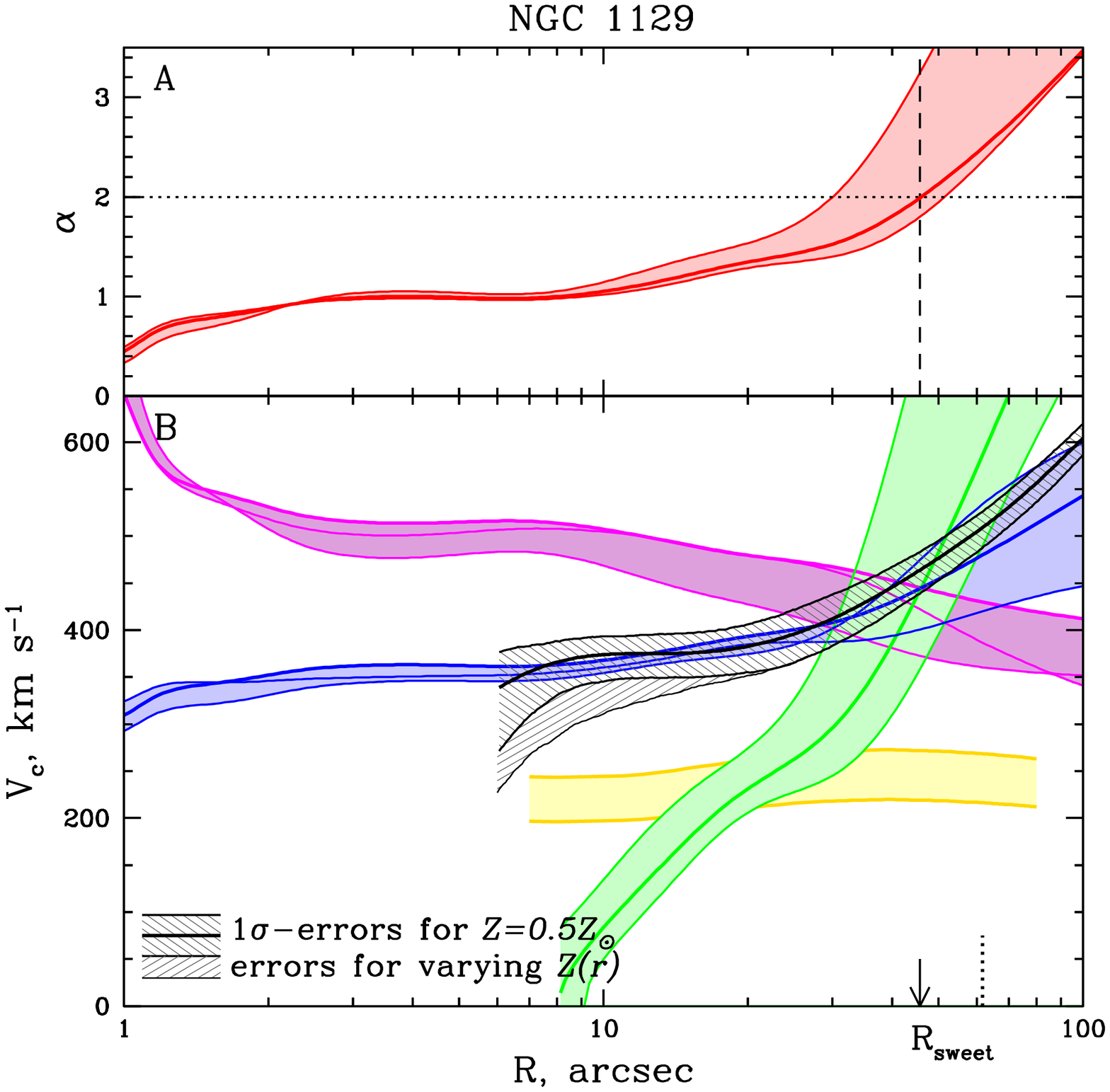}{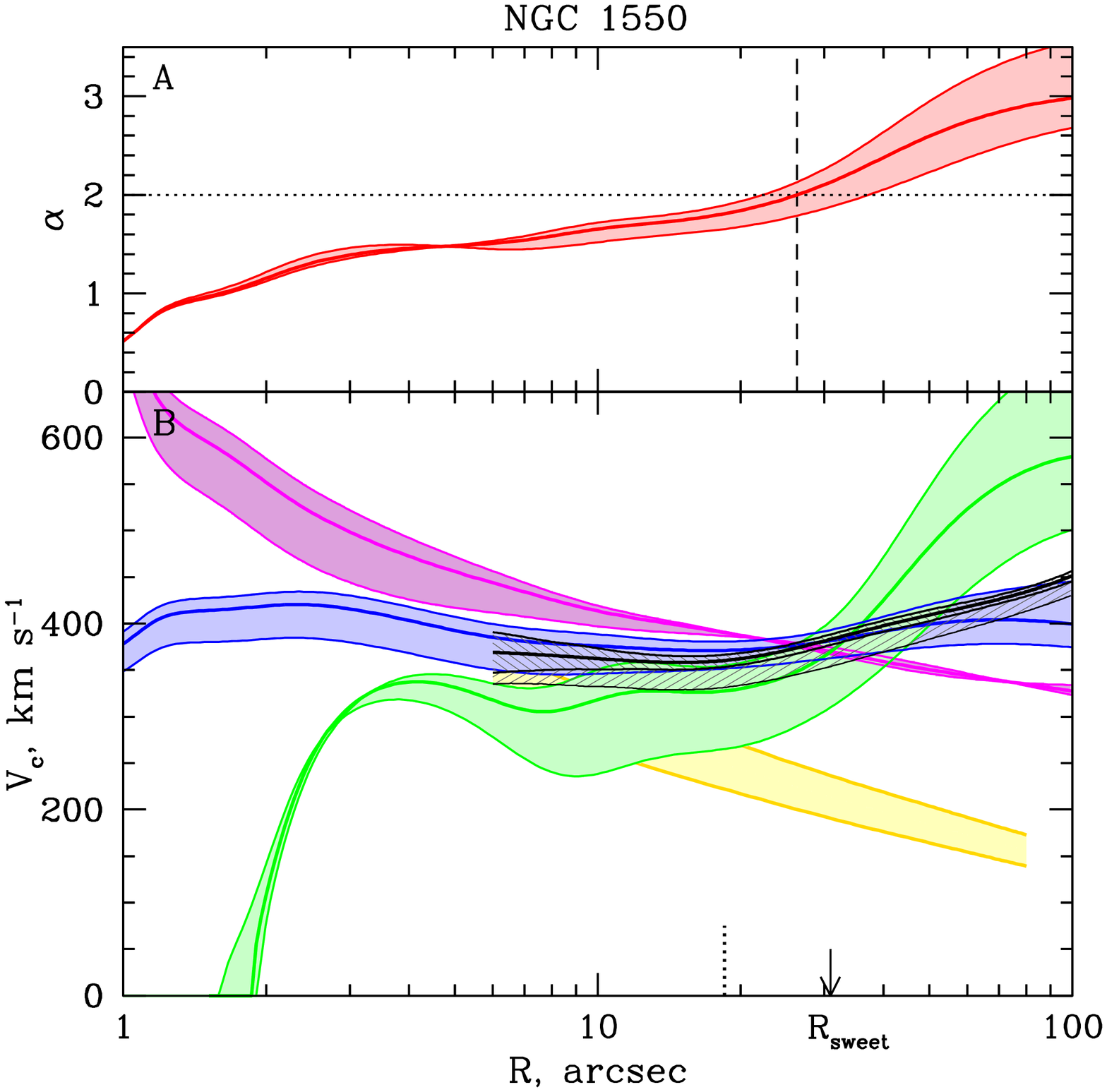}{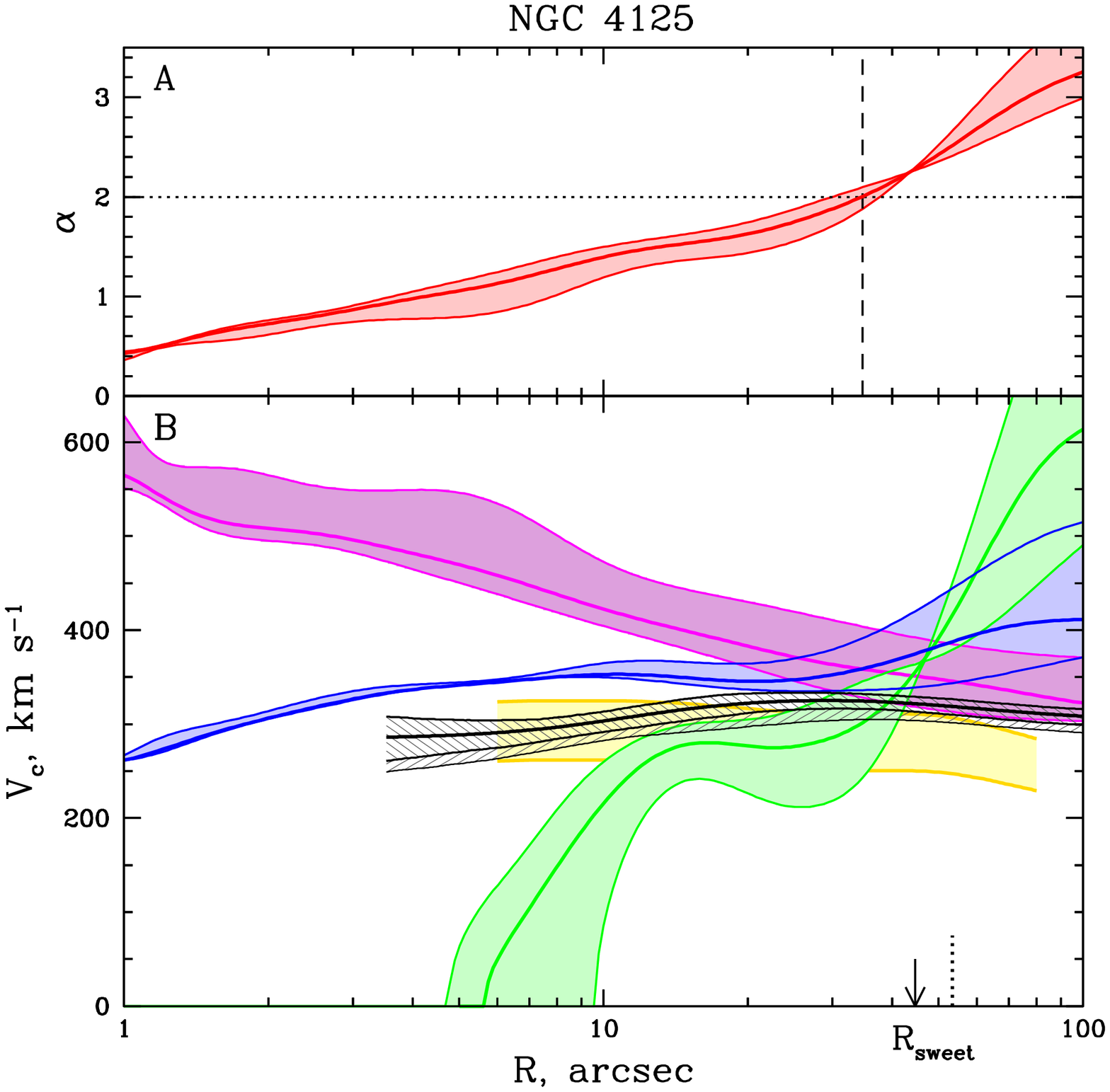}{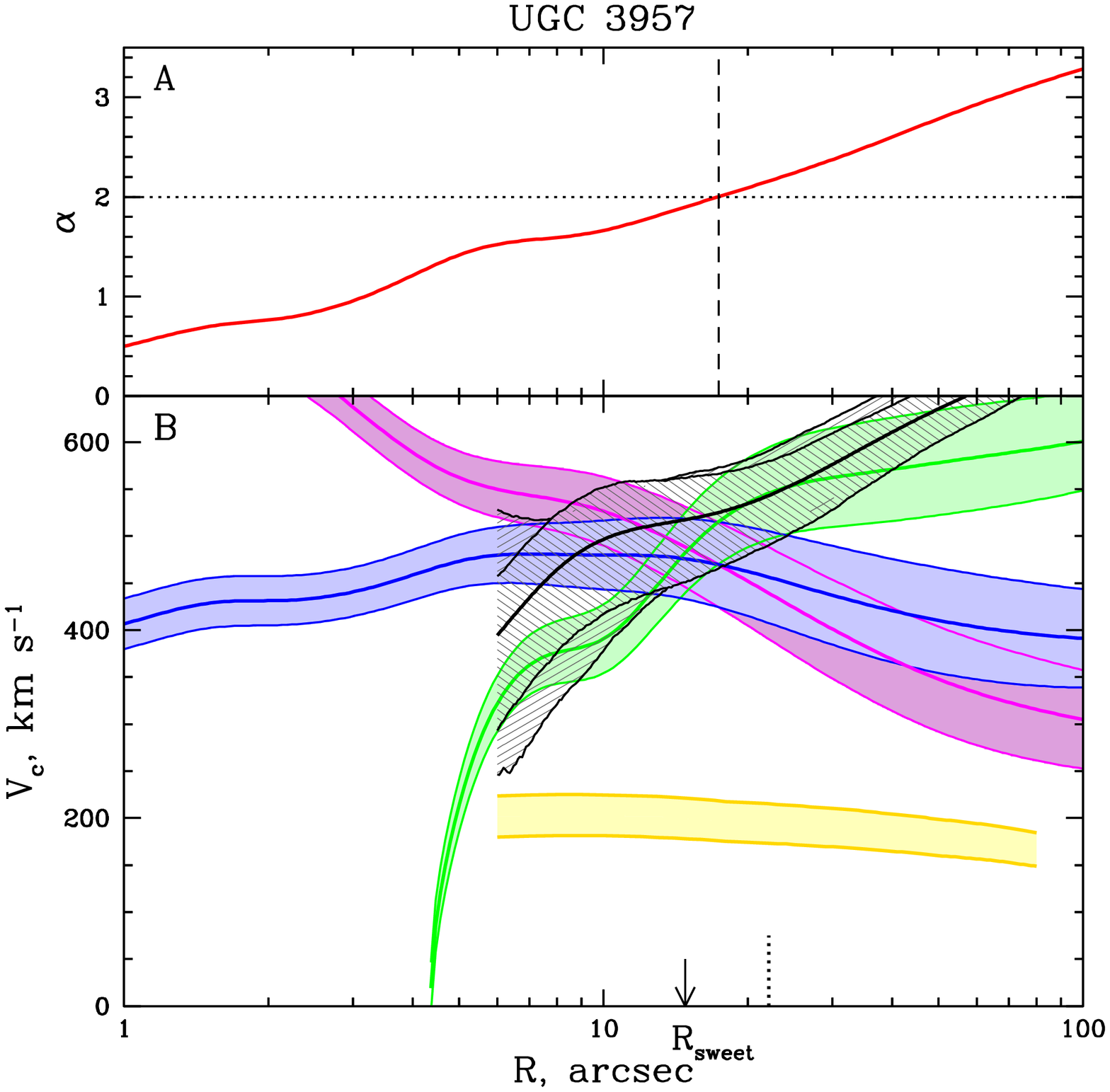}
\caption{Panel A - the surface brightness logarithmic slope $\alpha$ (red curve) and location of $R_2$ where $\disp \alpha = 2$ (vertical dashed line). Panel B - $V_c$-profiles for isotropic (blue lines and paleblue shaded area), pure radial (green lines and palegreen shaded area) and pure circular (magenta lines and plum shaded area) orbits. The $V_c(r)$ derived from Chandra data under the assumption of hydrostatic equilibrium is shown as the black thick line. The shaded area shows statistical errorbars for the flat and varying with radius metallicity. For NGC 708 arrows are the same as in Figure \ref{fig:ab} and indicate the conservative lower and upper limits on X-ray based $V_c(r)$ coming from radial variations of metal abundance. The stellar contribution to the circular speed profile is presented in yellow (see Section \ref{subsubsec:stars}). Location of $\disp R^{slit}_{\rm eff}$ defined from the de Vaucouleurs fit to the long-slit surface brightness profile is marked with the dotted line.}
\label{fig:all}
\end{figure*}

\begin{itemize}

\item NGC0708

NGC 0708 (Figure \ref{fig:all}, upper left corner) is a cD galaxy located at the center of Abell 262 galaxy cluster. The surface brightness and the projected velocity dispersion profiles are available for two slit positions oriented at P.A. = $-4^{\circ}$ and at P.A. = 215$^{\circ}$. The surface brightness profile  along the slit at P.A. = $-4^{\circ}$  declines very slowly, the logarithmic slope $\disp \alpha = - d \ln I(R) / d \ln R$ does not exceed 1.5 in observed range of radii leading to a diverging total stellar mass. Such a behaviour may be a result of influence of cluster gravitational potential. So for our analysis we use information along the slit at P.A. = $215^{\circ}$  only. The projected velocity dispersion is close to being flat at $R \lesssim 30''$ and gets systematics affected at larger radii. So we use the simplified version of the analysis. Results of our analysis are presented in the upper left corner of Figure \ref{fig:all}. The surface brightness slope (for the slit at P.A. = 215$^{\circ}$) is shown in panel A, derived circular velocity profiles different types of orbits are plotted in panel B. The shaded areas indicate uncertaintes in derived $V_c^{\rm iso}$ (paleblue), $V_c^{\rm circ}$ (plum) and $V_c^{\rm rad}$ (palegreen), coming from measurement errors of $\sigma_p(R)$. The sweet radius where the sensitivity of the method to the anisotropy is minimal is located at $63''$, i.e slightly beyond the range of radii where optical data are available. Although the reliability of such the estimate is unclear the extrapolated $V_c^{\rm iso}$ lies quiet close to the circular speed curve derived from X-ray analysis.

\item NGC1129

NGC 1129  is a giant elliptical galaxy located in the center of a poor cluster AWM 7. In Figure \ref{fig:all} (upper right) are shown results of optical and X-ray analyses. Before estimating the circular speed from optical data we have excluded regions where the surface brightness profile seems to be contaminated by projection of companions. The exclusion is done on the basis of visual inspection. So we consider the surface brightness profile along the slit positioned at P.A. = 166$^{\circ}$ in radial range from $-55''$ till $41''$ and in case of P.A. = 256$^{\circ}$  slit - at $R \le 0''$. The projected velocity dispersion profile looks nearly flat at $R \lesssim 20''$ and is getting  noisy at $R \gtrsim 20''$, so we assume $\sigma_p(R)\equiv const=257$ $\kms$ (the surface brigness weighted average value).  Optical $V_c^{\rm iso}$-estimate in the sweet region (which is coincident with a range of radii where $\alpha \approx 2$) is consistent with the circular speed derived from hydrostatic equilibrium of hot gas in the galaxy. Moreover, $V_c^{\rm iso}$ and $V_c^X$ agree within errorbars over the range of radii where both optical and X-ray data are available. It should be noted that NGC1129 shows significant minor axis rotation, indicating a triaxial intrinsic shape of the galaxy.

\item NGC1550

NGC 1550 is a S0 galaxy lying at the center of a luminous galaxy group. The surface brightness and the projected velocity dispersion profiles are available for two slit positions oriented at P.A. = 116$^{\circ}$ and at P.A. = 206$^{\circ}$. Rotation velocity is consistent with zero. The profiles do not have any peculiar features so we use all available information to estimate the circular speed. The results of our analysis are shown in Figure \ref{fig:all}, left side of a middle panel.  Note that the circular velocity corresponding to the isotropic distribution of stellar orbits is nearly constant over the whole available range of radii and it coincides within errorbars with the X-ray circular speed profile. This fact could indicate that the gravitional potential of NGC 1550 is close to isothermal and the galaxy is dynamically relaxed with hot gas being in hydristatic equilibrium.

\item NGC4125

NGC4125 (Figure \ref{fig:all}, middle right side) is a E6 galaxy located at the center of NGC 4125 group of galaxies. It is the only galaxy in our sample with significant rotation. To take the rotation into account we use $V_{\rm rms}(R)=\sqrt{\sigma_p(R)^2+V_{\rm rot}(R)^2}$ instead of $\sigma_p(R)$  in equations (\ref{eq:main}). The isotropic circular speed $V_c^{\rm iso}$ slightly exceeds $V_c^X$ over the whole range of radii where the optical observations are available what may indicate the the non-thermal pressure support at the level of $f_{nt} \approx 36 \%$ at the sweet point.

\item UGC3957

UGC3957 (Figure \ref{fig:all}, lower panel) is an elliptical galaxy at the center of UGC 03957 group. It has been observed using only one slit positioned at P.A. = 287$^{\circ}$. As in case with NGC 0708 shaded areas indicate uncertaintes in derived $V_c$-profiles, arising from measurement errors of $\sigma_p(R)$. At the sweet point $V_c$-estimate from optical data agrees with X-ray derived one. The discrepancy between optical $V_c(R)$ and X-ray $V_c^X(R)$ may indicate that at $r \gtrsim 20''$ anisotropy parameter $\beta>0$ if the hydrostatic equilibrium approximation is valid.

\end{itemize}

\subsubsection{Stellar populations: properties, mass-to-light ratios, contributions to the total mass}
\label{subsubsec:stars}

By using the same SCORPIO/BTA long-slit spectral data, we have calculated Lick
indices H$\beta$, Mgb, Fe5270, and Fe5335 along the slit, to derive the ages
and chemical abundances which are in turn used to estimate mass-to-light ratios
of the stellar component varying along the radius and to calculate properly the 
mass contributed by the stellar component within the radius $R_{\rm sweet}$. Our
approach to the Lick index calibrations can be found in \citet{webaes}. 

\begin{figure}
\centering
\plotone{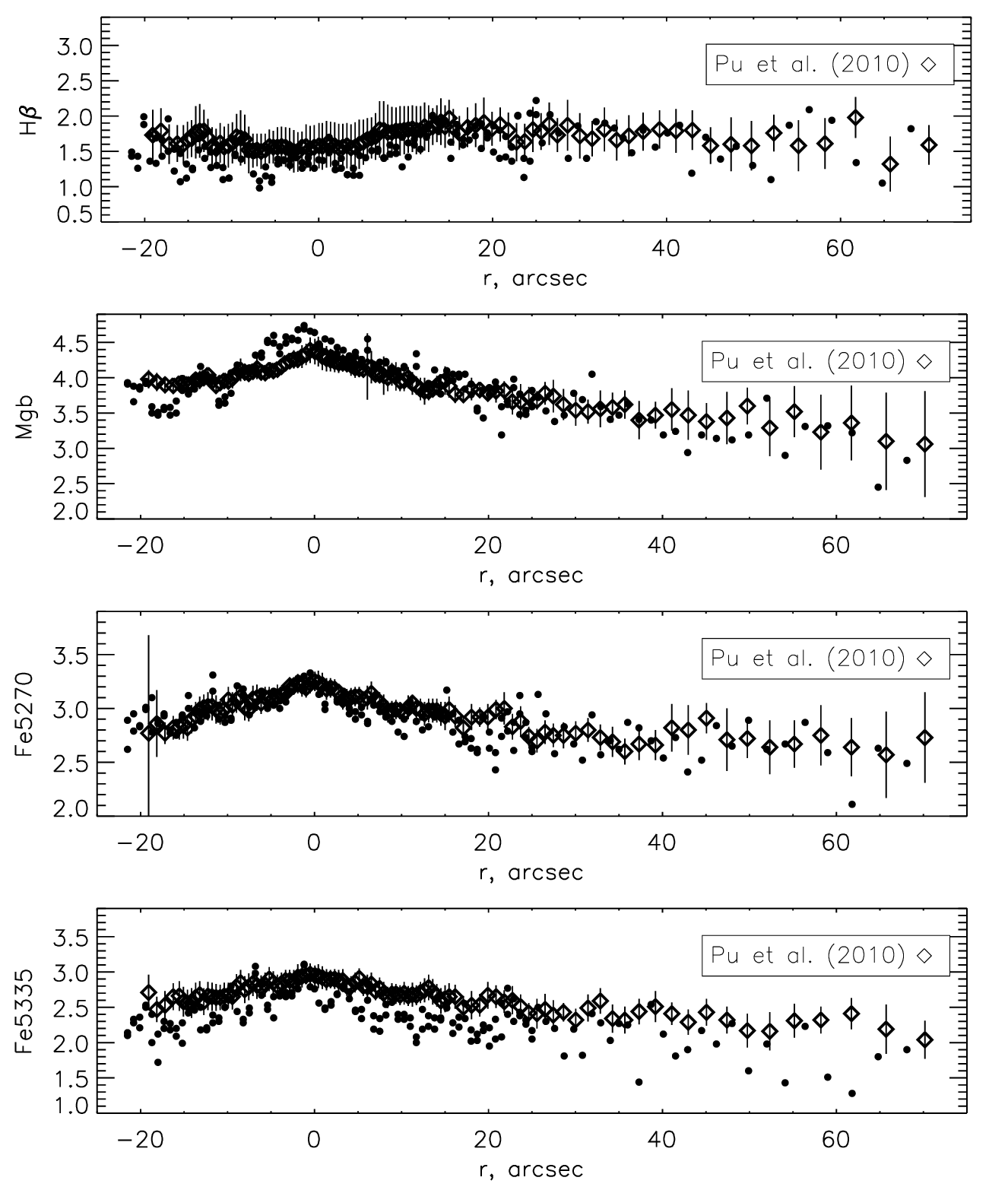}
\caption{The comparison of the Lick index profiles along the major axis in NGC~4125, according to our data and to the data by \citet{Pu2010}.}
\label{n4125indcomp}
\end{figure}

\begin{figure*}
\plottwo{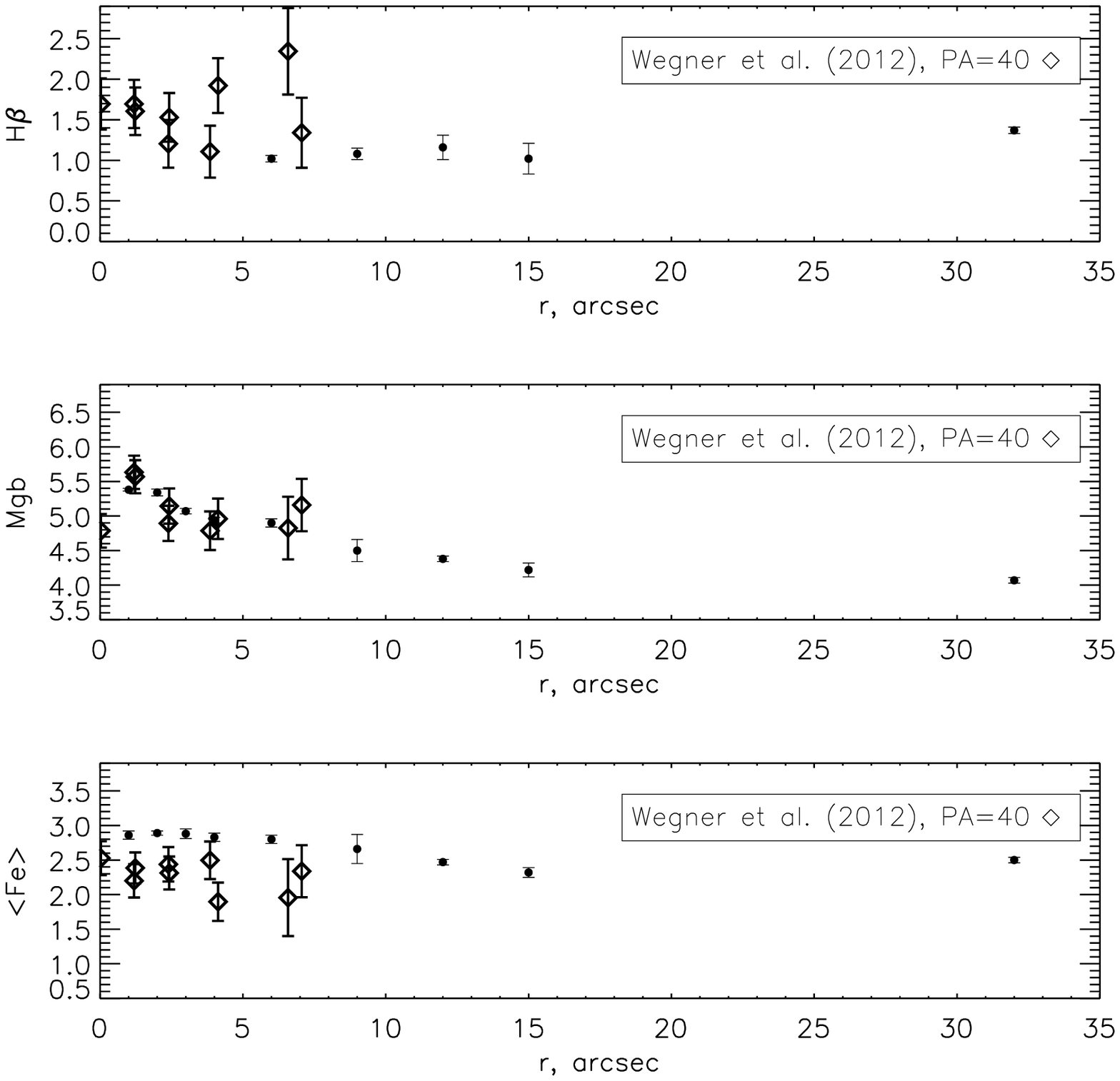}{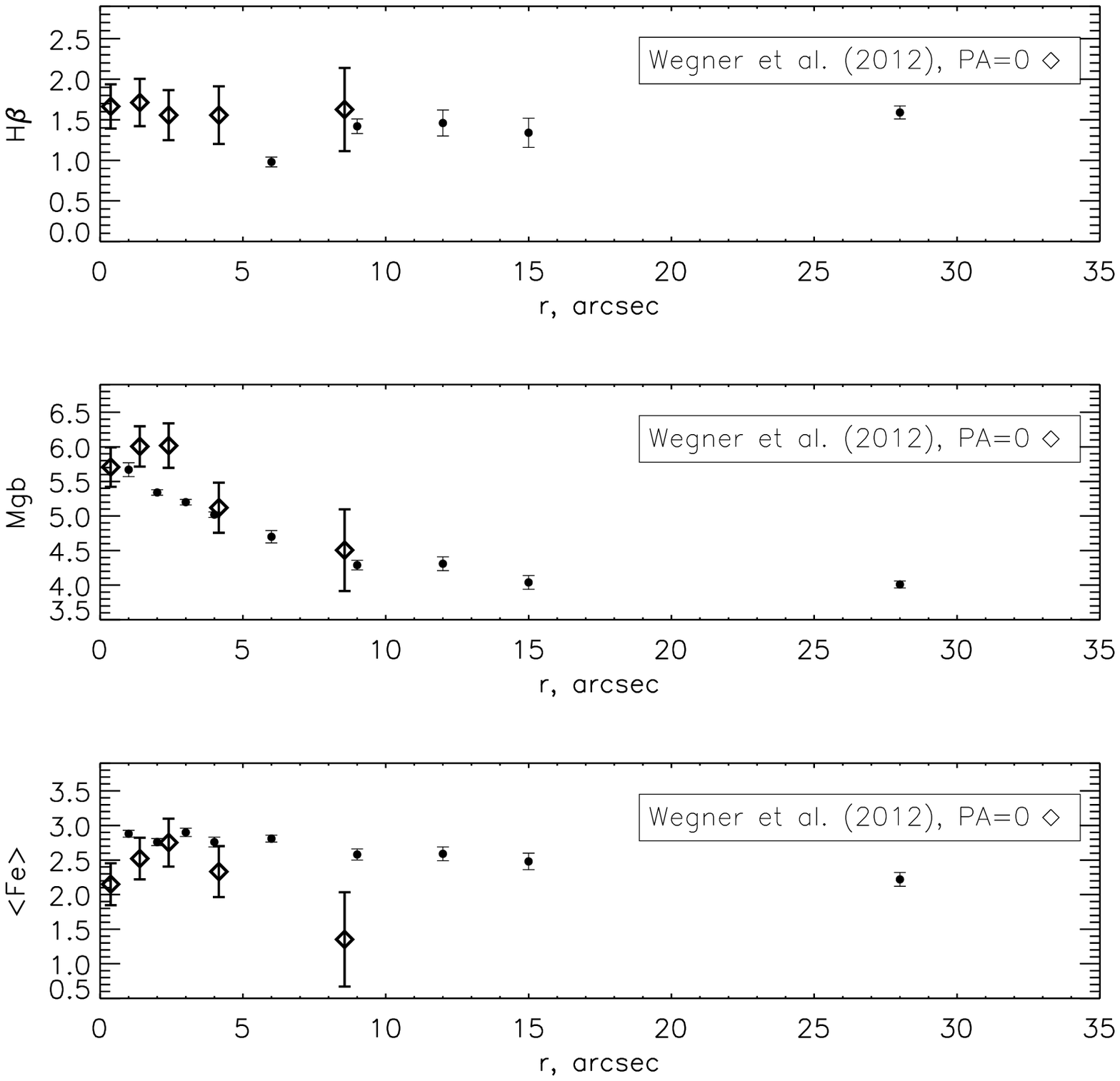}
\caption{The comparison of the Lick index profiles in NGC~708, according to
our data and to the data by \citet{Wegner2012}, in two slit orientation.} \label{n708indcomp}
\end{figure*}

Our spectral data are rather deep and have provided the profiles
of high-precision Lick indices up to $1.1-2.2$ effective radii from the center
in four galaxies of five. Figure~\ref{n4125indcomp} presents the comparison of
the major-axis Lick index profiles in NGC~4125 according to our measurements with
those by \citet{Pu2010} obtained at the 9.2m Hobby-Eberly telescope. The agreement
is rather good, and the index point-to-point scatters are comparable. 
Figure~\ref{n708indcomp} compares our data for NGC~708 with the data from \citet{Wegner2012}
at two slit positions (we don't compare H$\beta$ measurements near the
center because we have not been able to correct them properly for the emission
contamination). This time our data are much more precise, and the Lick index profiles
are much more extended that the data by \citet{Wegner2012} obtained at the 2.4m Hiltner Telescope.

\begin{figure*}
\plotfour{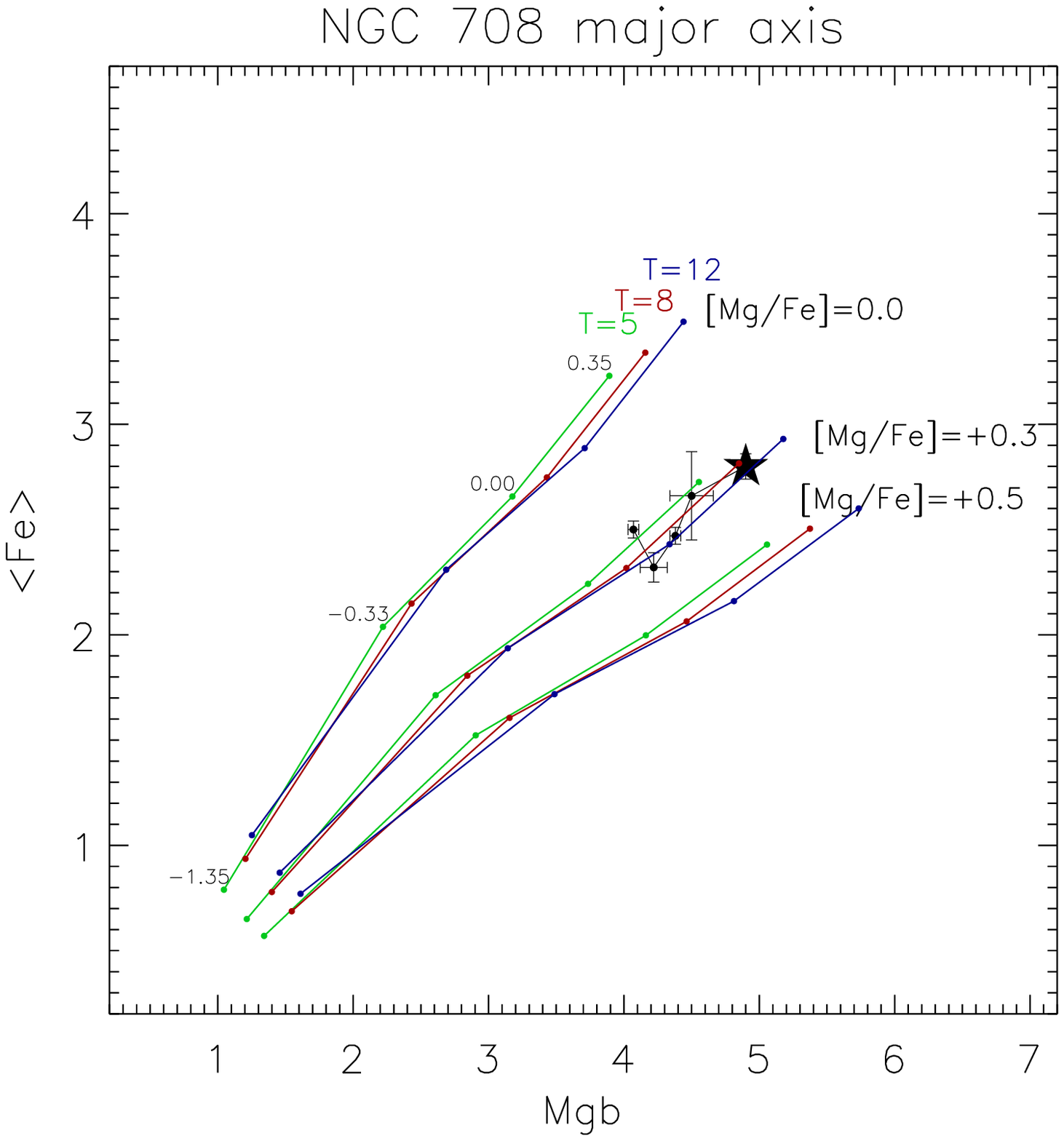}{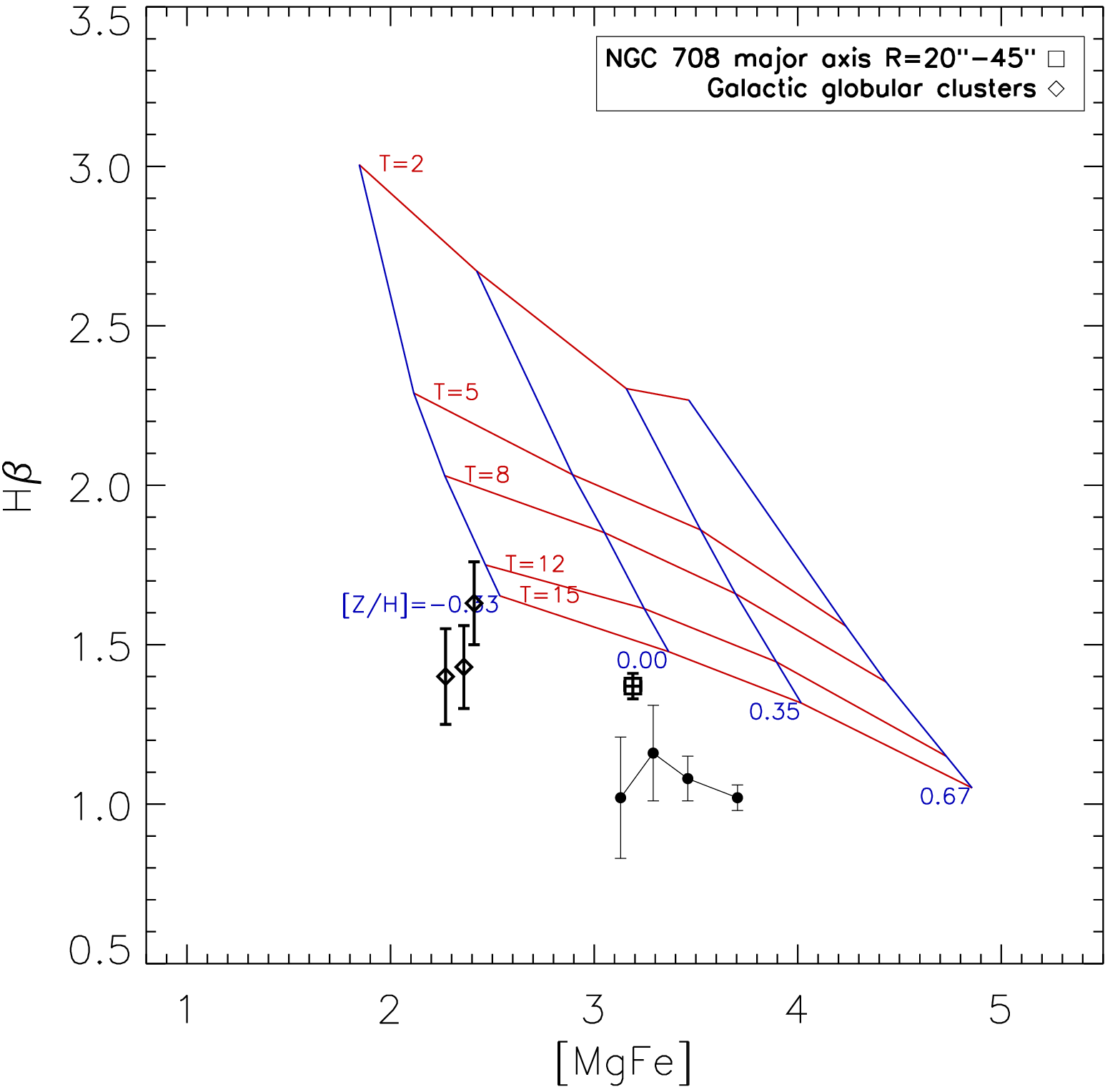}{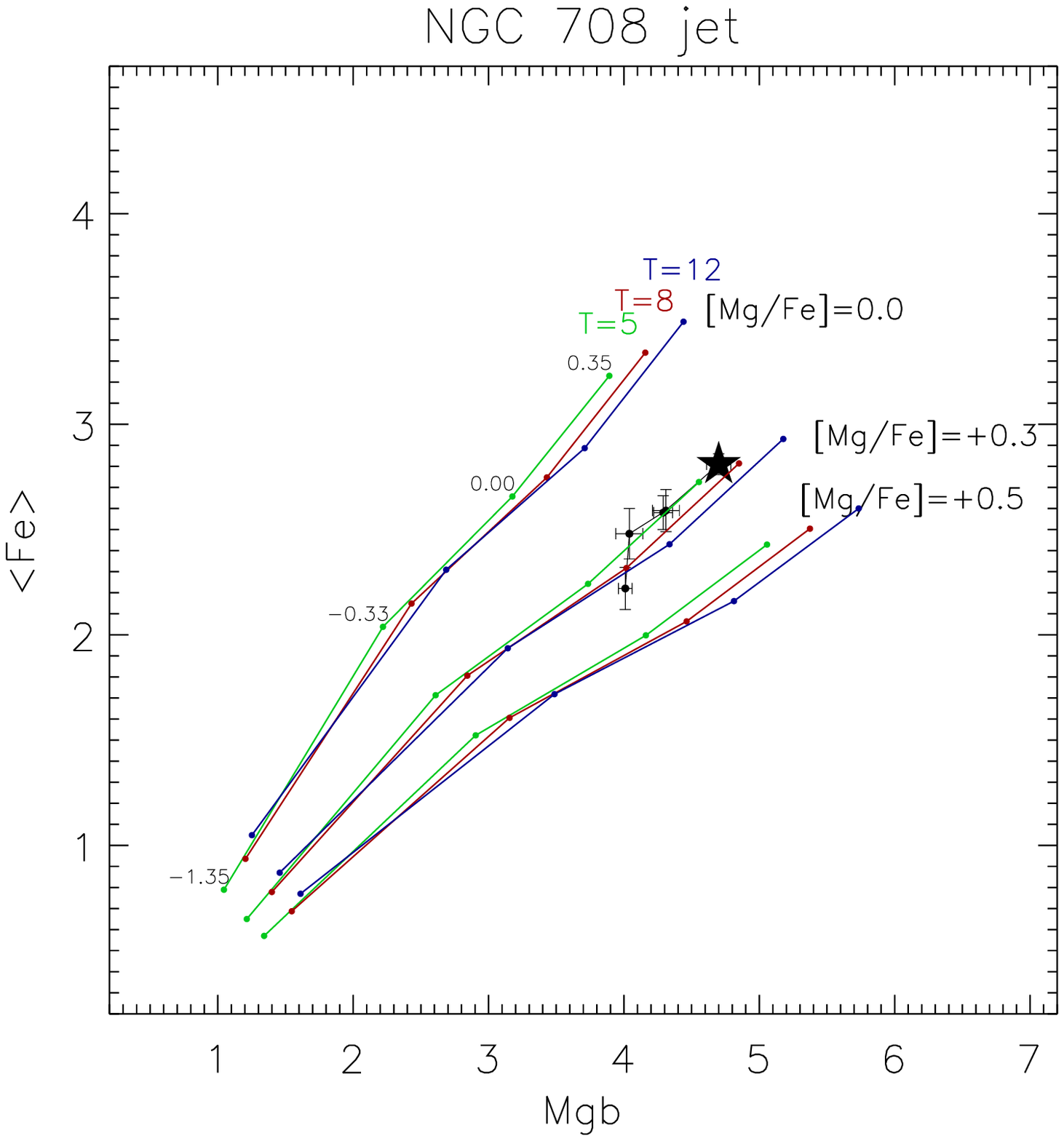}{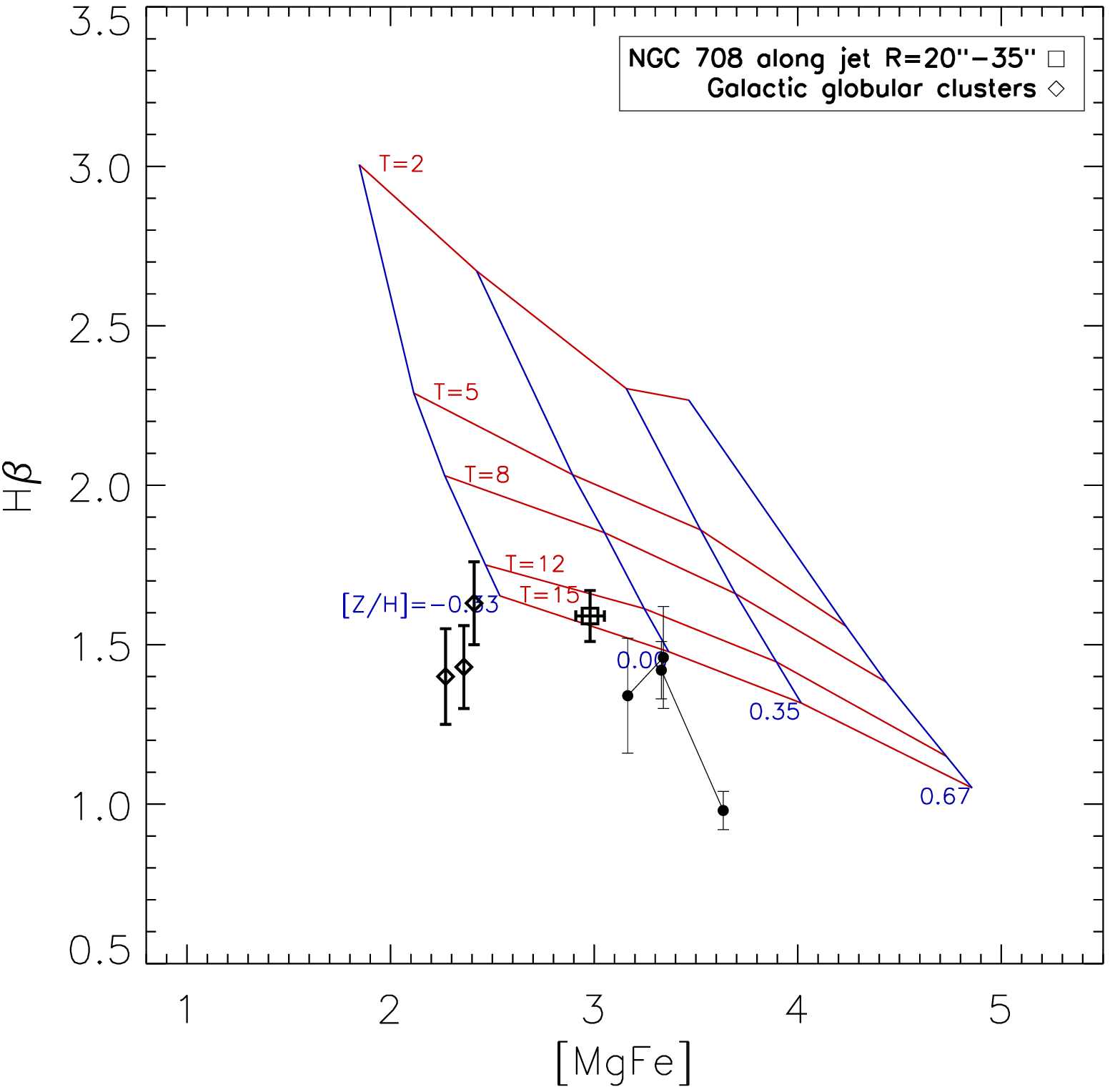}
\caption{Diagnostic Lick index -- index diagrams for all our galaxies in all slit orientations: {\bf (left side)} -- The $\langle \mbox{Fe} \rangle$ vs Mgb diagram. The simple stellar population models by \citet{thomod} for three different magnesium-to-iron ratios ($0.0$, $+0.3$, and $+0.5$) and three different ages (5, 8, and 12 Gyr) are plotted as reference. The small signs
along the model curves mark the metallicities of $+0.35$, $0.00$, $-0.33$, and $-1.35$, if one takes the signs from right to left; {\bf (right side)} -- The age-diagnostic diagram for the stellar populations in the central parts of the galaxies under consideration. The stellar population models by \citet{thomod} for [Mg/Fe]$=+0.3$ (for NGC~4125 -- for [Mg/Fe]$=+0.0$) and five different ages (2, 5, 8, 12 and 15 Gyr, from top to bottom curves) are plotted as reference frame; the blue lines crossing the model metallicity sequences mark the metallicities of $+0.67$, $+0.35$, $0.00$, $-0.33$ from right to left.}
\label{diag}
\end{figure*}

We have confronted our Lick index measurements along the radii in the galaxies under consideration to the Simple Stellar Population (SSP) models by \citet{thomod} which allow to vary magnesium-to-iron ratio. Indeed, giant elliptical galaxies are known to be magnesium-overabundant \citep{trager} so it must be taken into account when age diagnostics 
are applied. By confronting  $\disp \langle \mbox{Fe} \rangle \equiv (\mbox{Fe5270}+\mbox{Fe5335})/2$ vs Mgb, we have found that indeed in four galaxies [Mg/Fe]$=+0.3$ being constant along the radius, 
while only in NGC~4125 [Mg/Fe]$=+0.1$ with slightly different behaviour along the major and the minor axes: in the latter cross-section, at large radii the [Mg/Fe] comes to $+0.3$ (Figure~\ref{param}).  This 
difference, together with the fast rotation along the major axis, gives an evidence for an embedded stellar disc in NGC~4125; so we would prefer to give more weight to the stellar mass-to-light ratio profile along the minor axis (see below). The estimates of the SSP-equivalent (mean, star luminosity-weighted) ages made by confronting the H$\beta$ index to the complex metal-line index [MgFe] (Figure~\ref{diag}) indicate mostly old stellar
population, older than 8~Gyr, beyond the very centers of the galaxies; however the stellar nuclei of UGC~3957, NGC~1129, and NGC~1550 may be as young as 5~Gyr old (in NGC~708 we cannot estimate the age of the nuclear stellar population because of the very strong gaseous emission contaminating the H$\beta$ index).

% Figure 9

\setcounter{figure}{8}
\begin{figure*}
\plottwoo{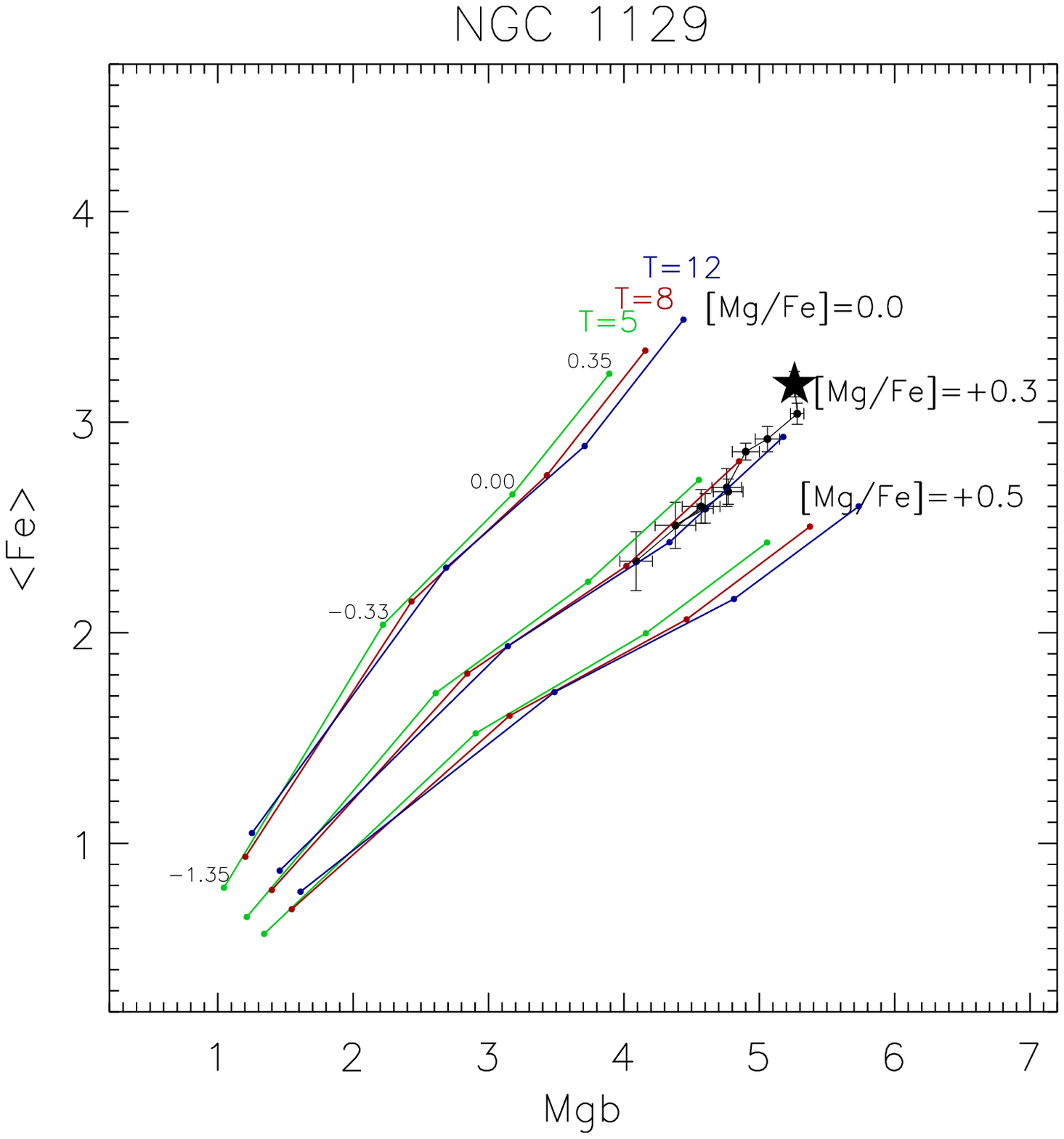}{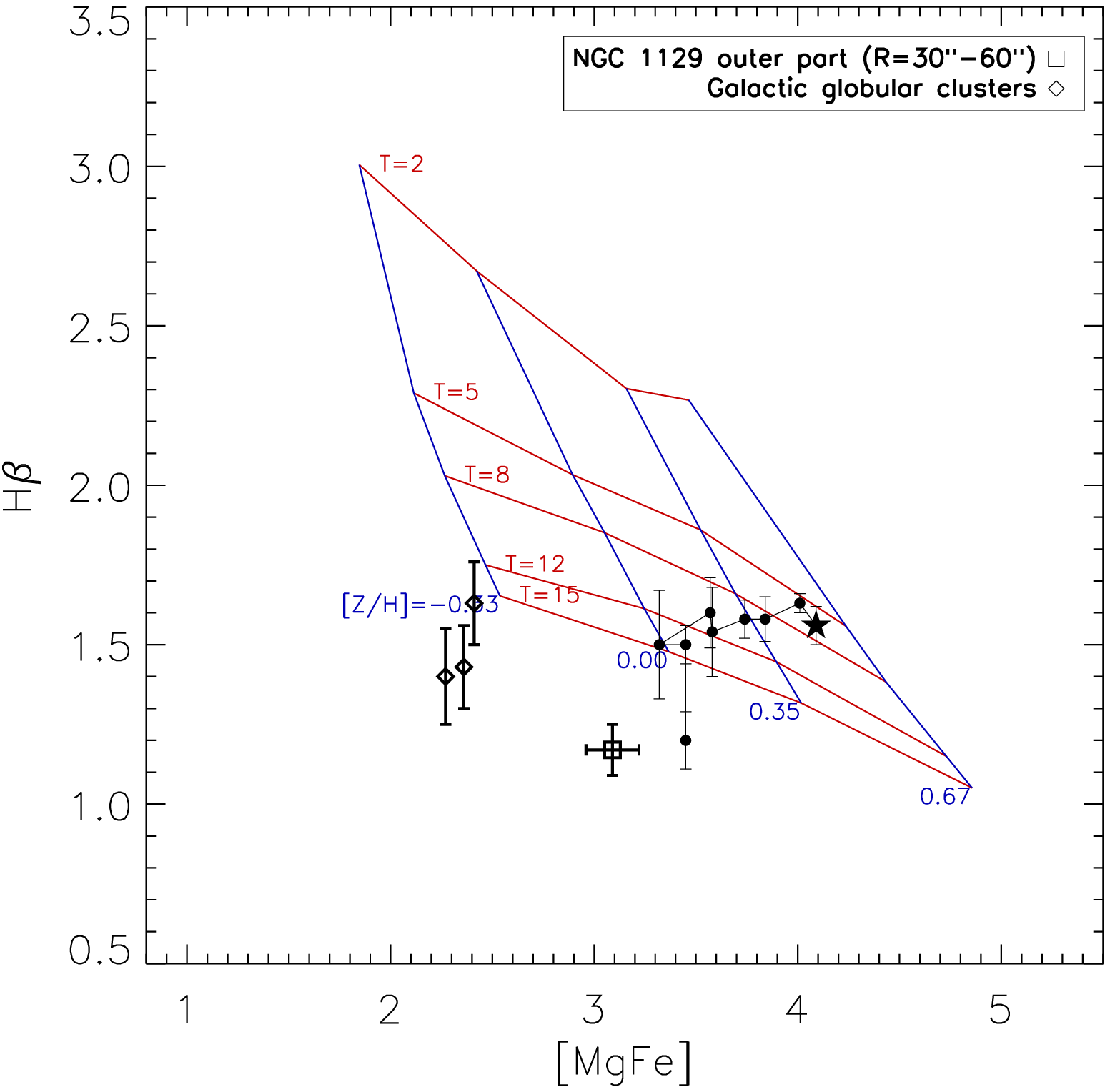}\vspace{2mm}\hfil
\plottwoo{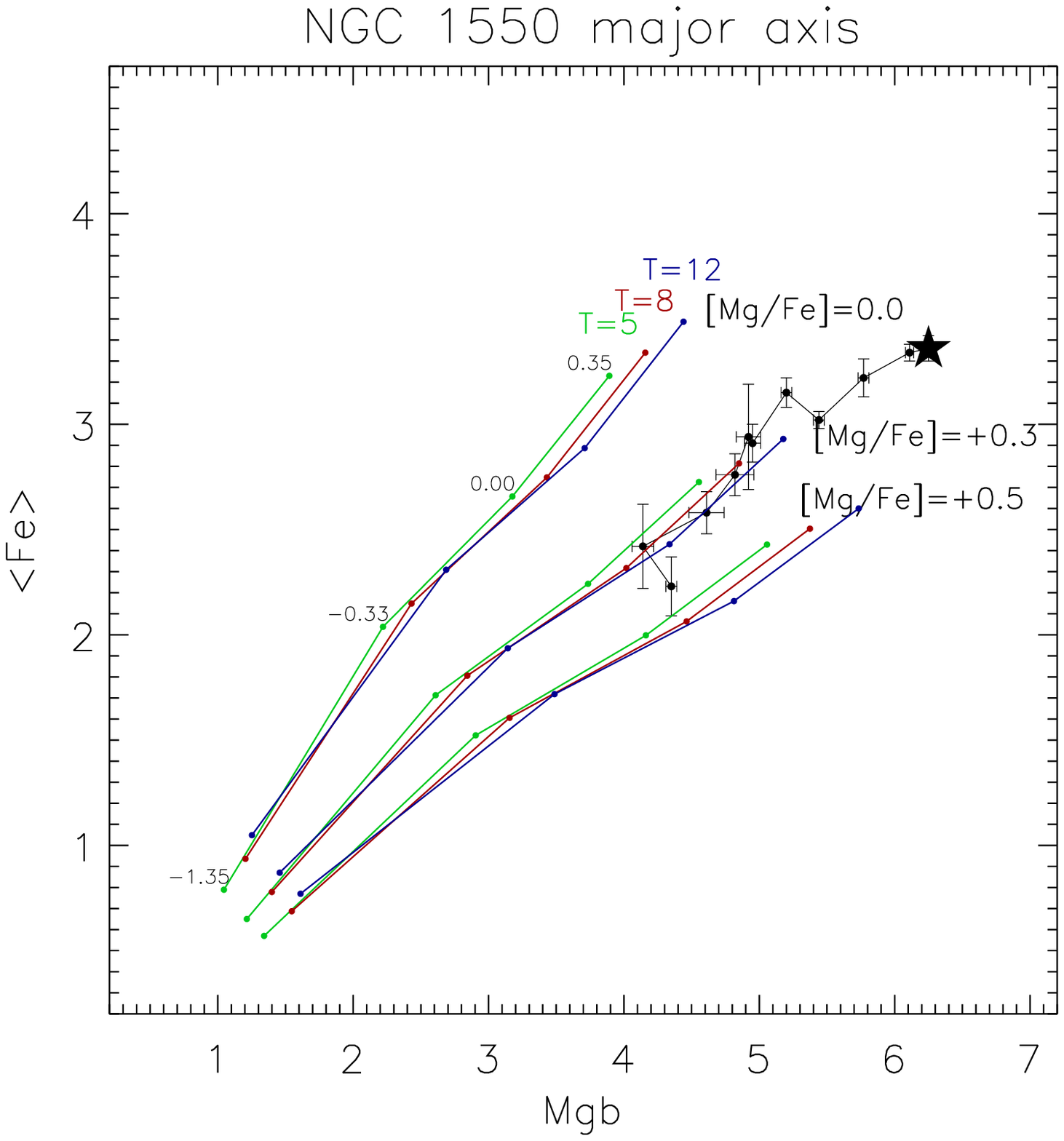}{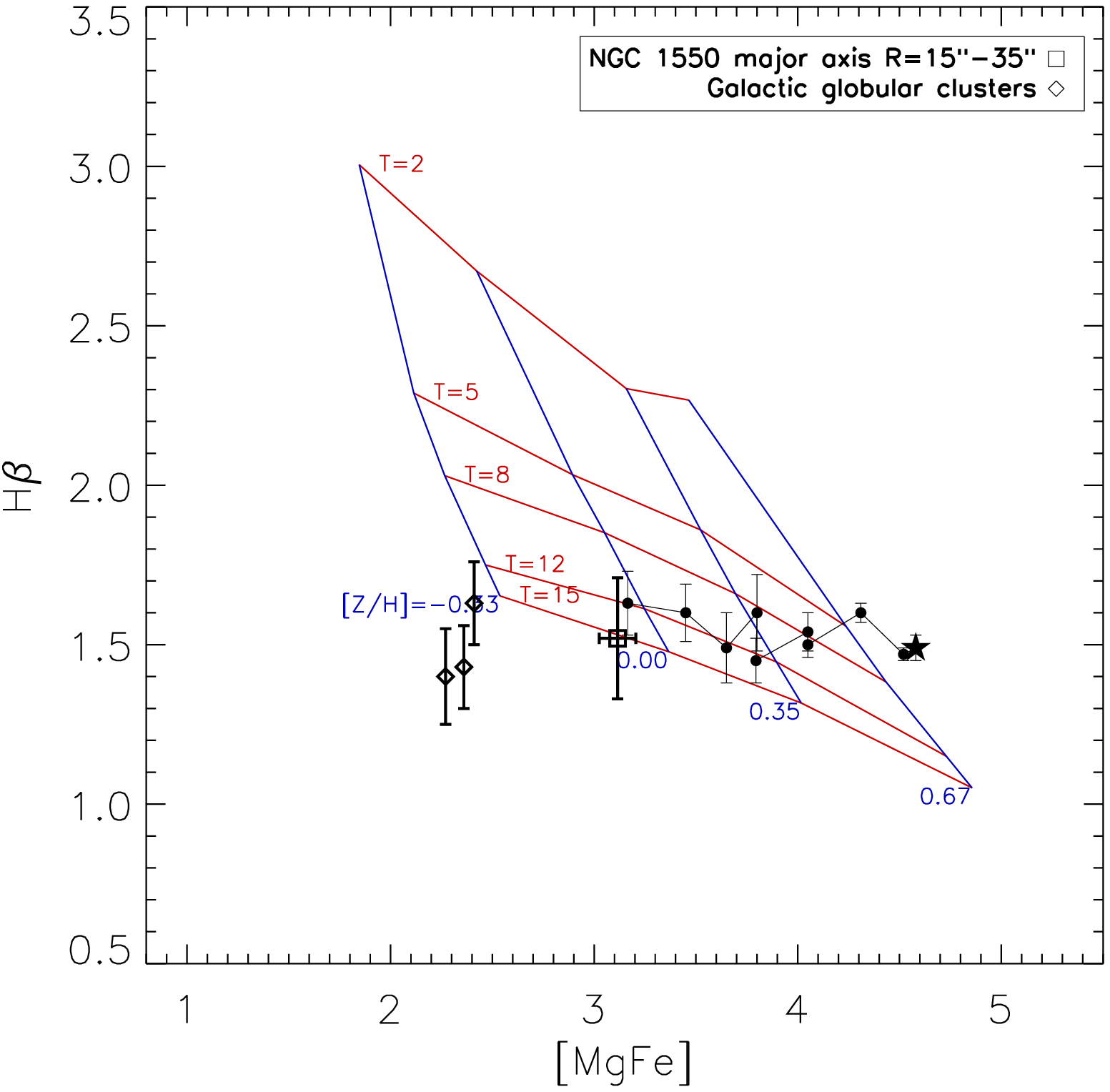}\vspace{2mm}\hfil
\plottwoo{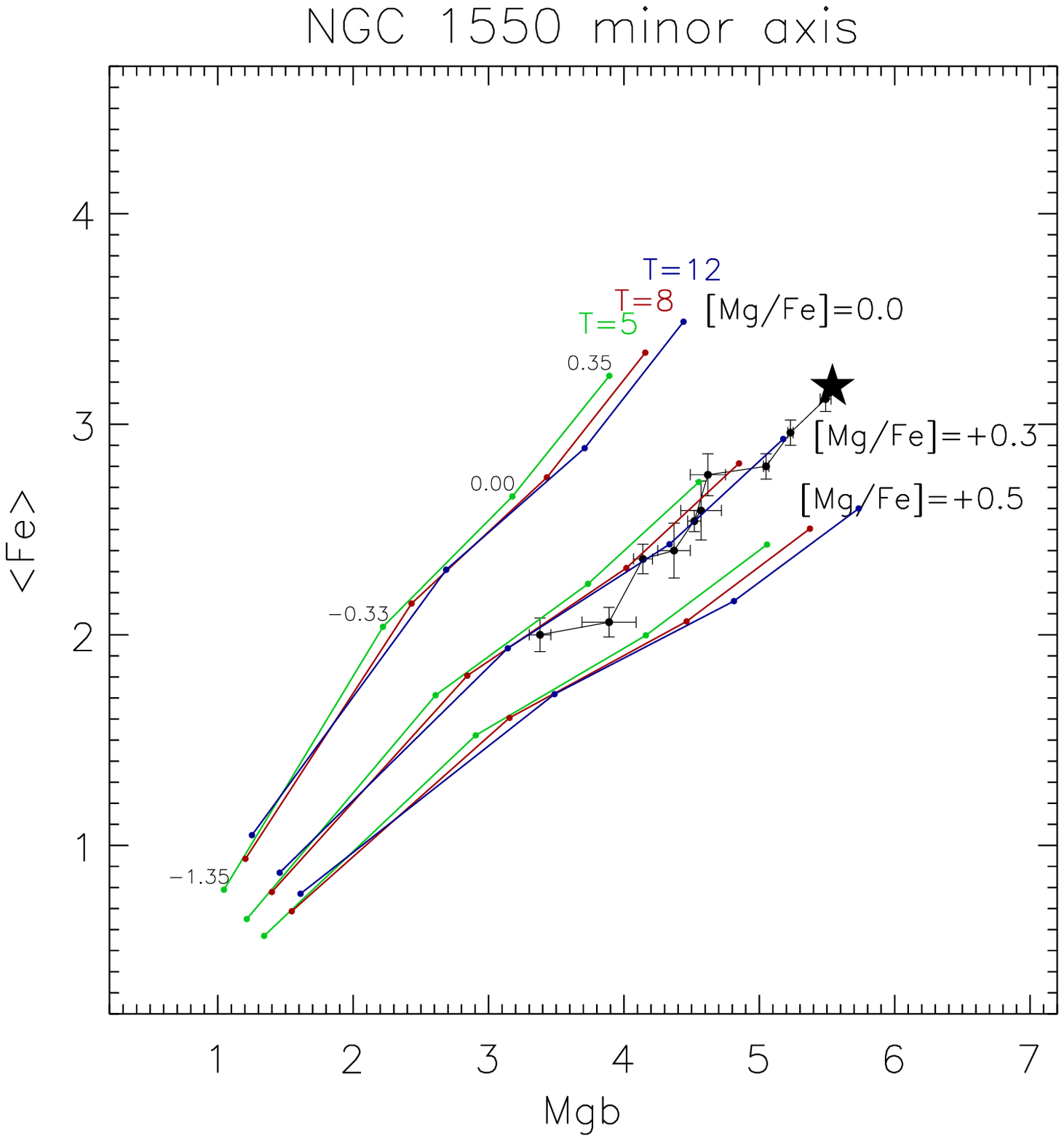}{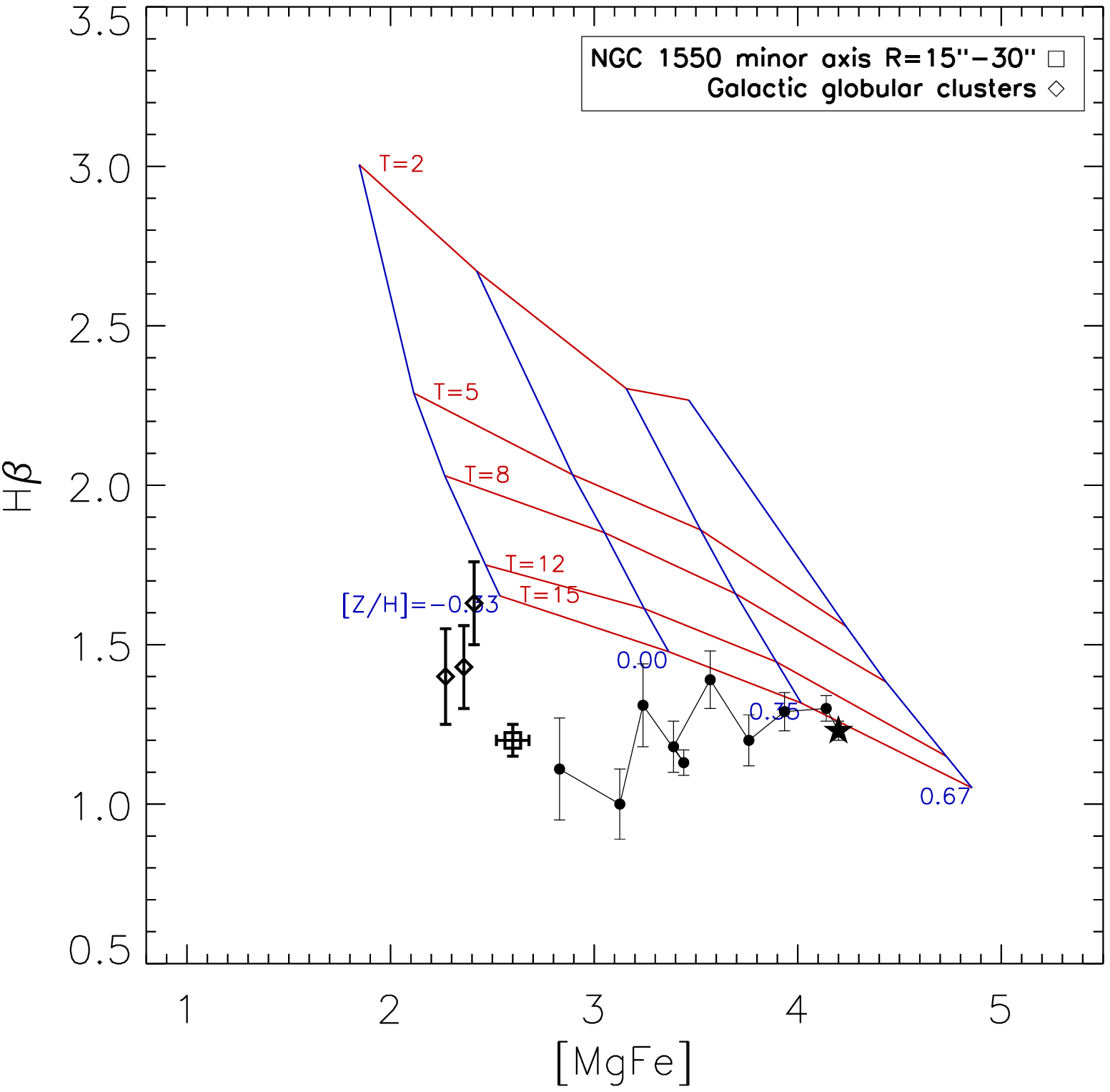}
\caption{(continue)} 
\end{figure*}

\setcounter{figure}{8}
\begin{figure*}
\plottwoo{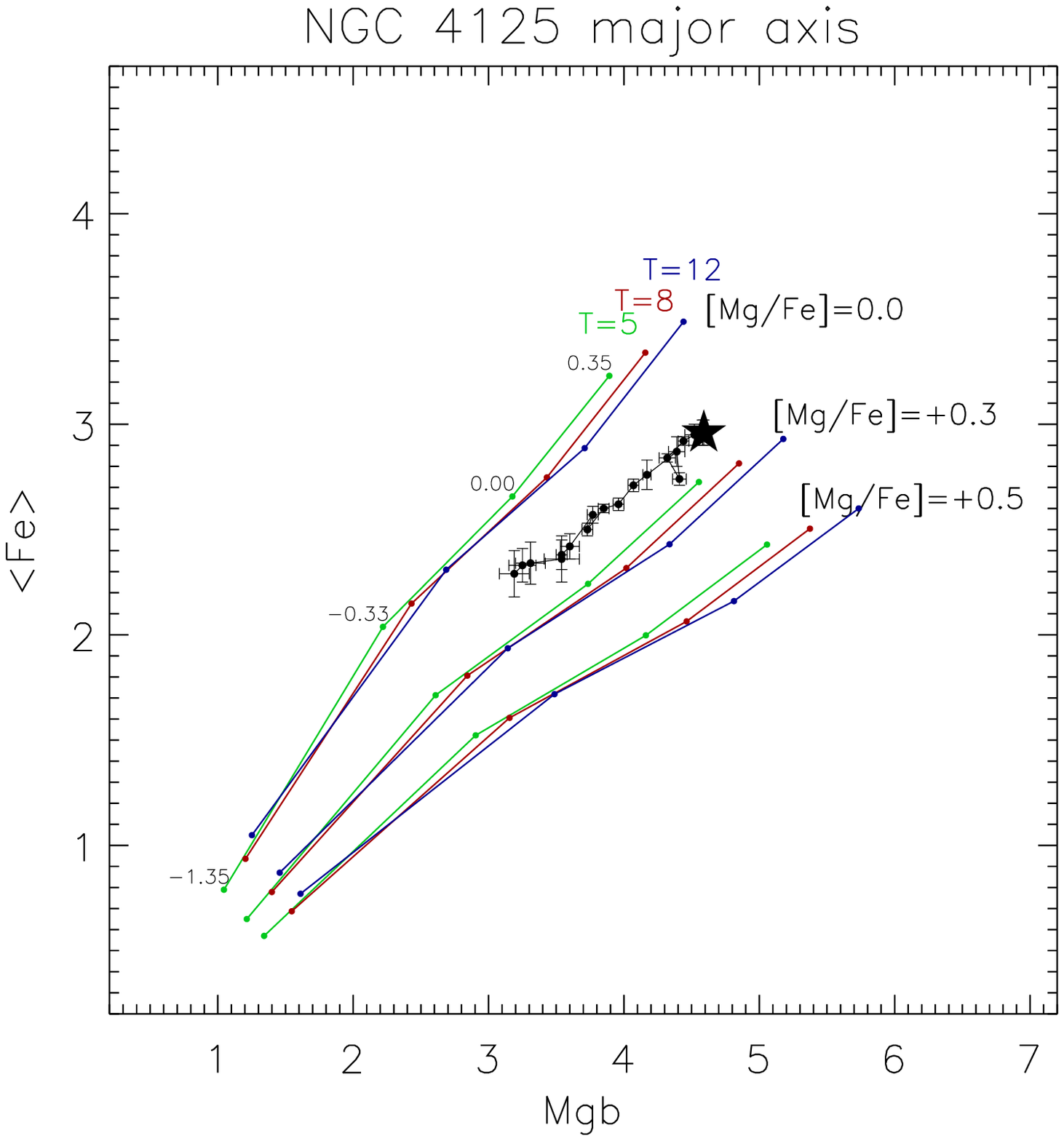}{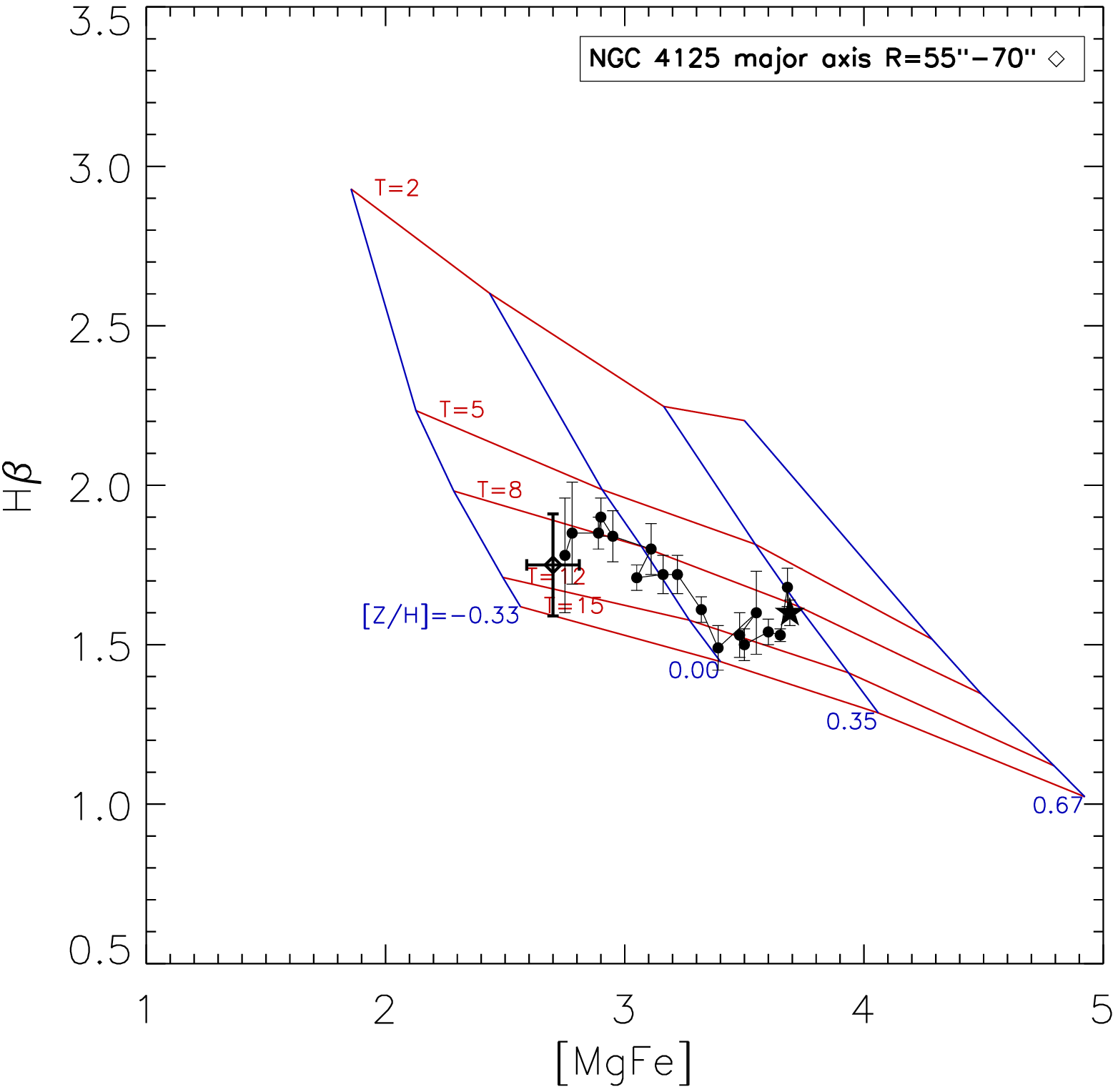}\vspace{2mm}\hfil
\plottwoo{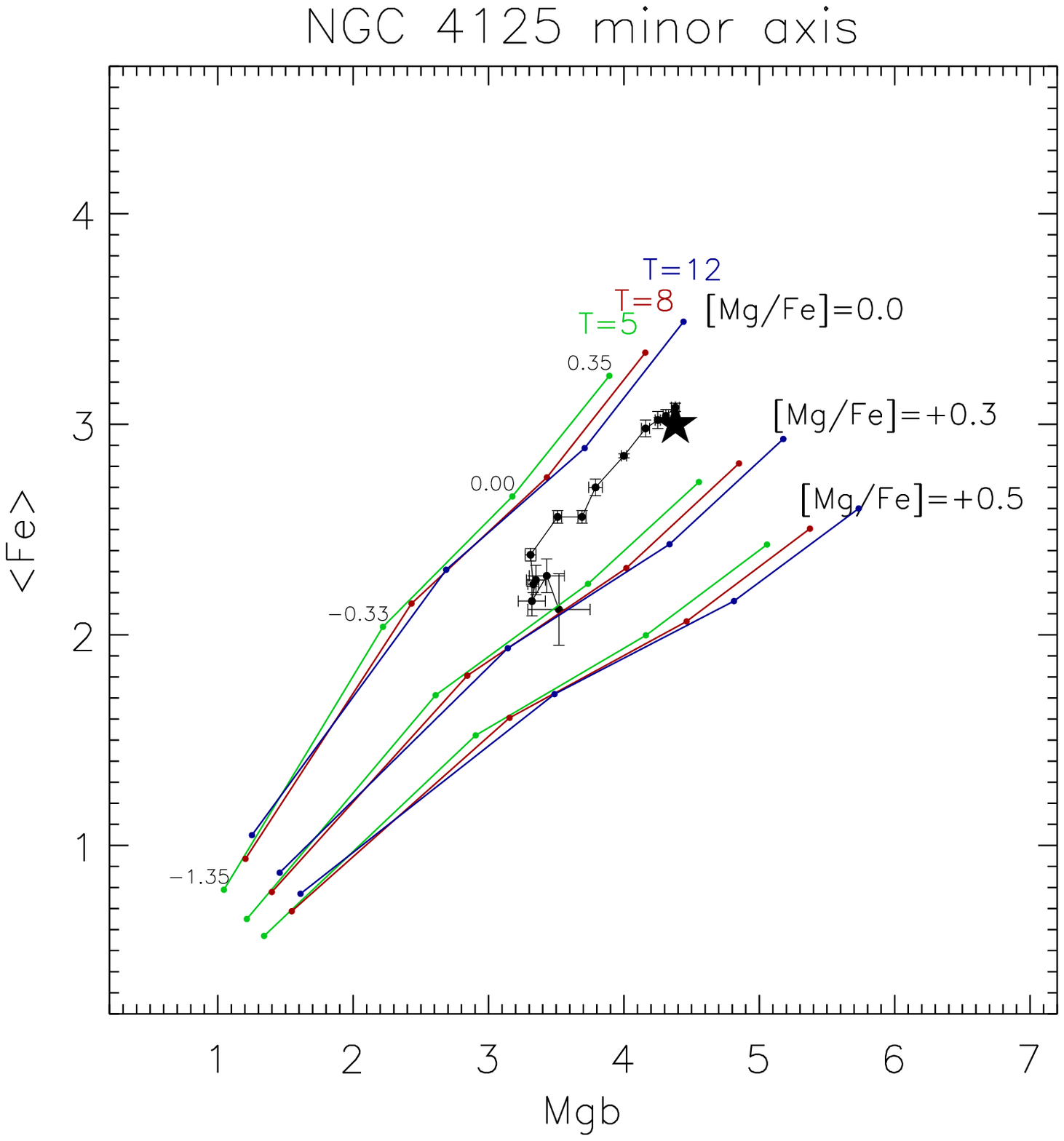}{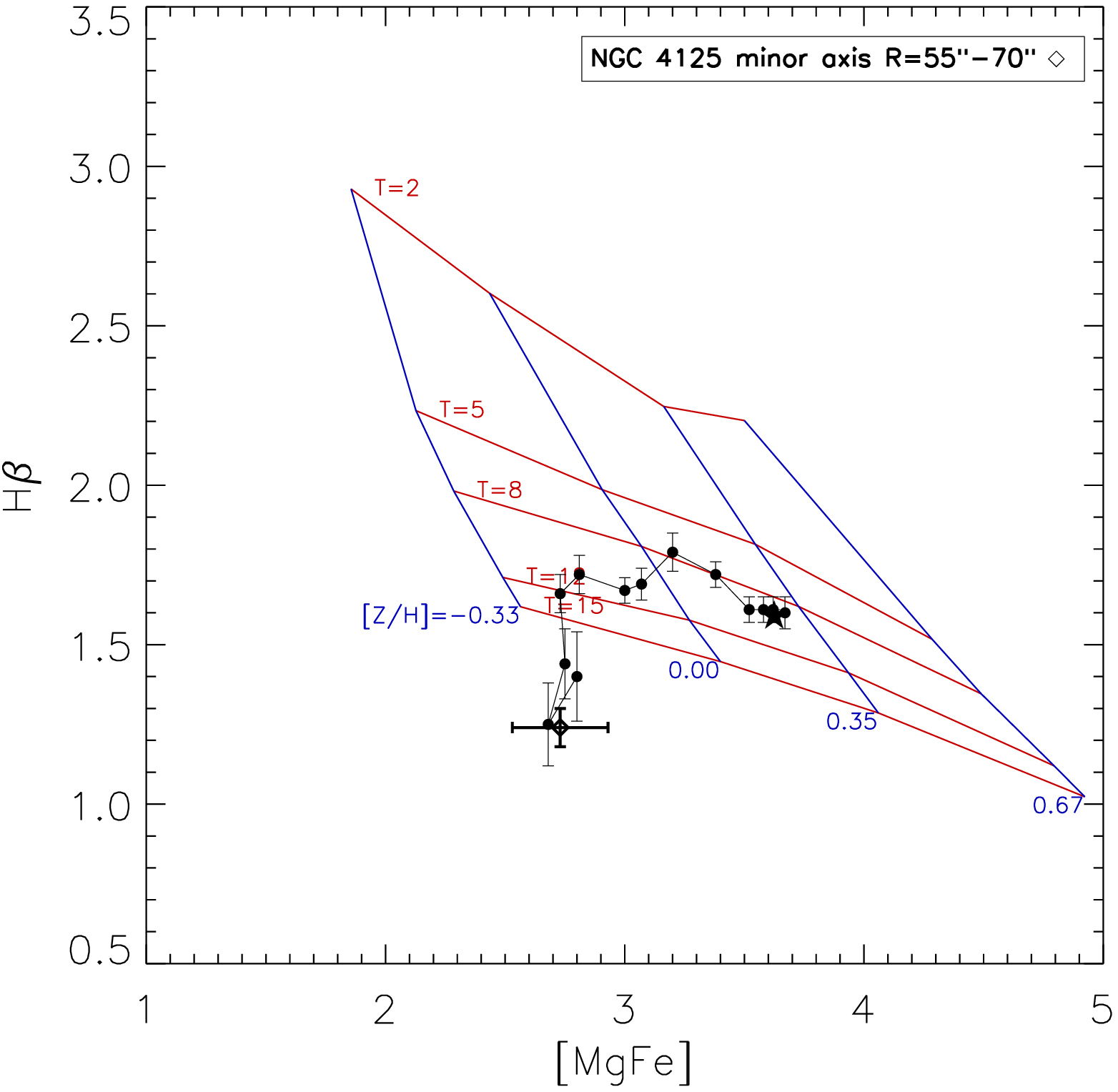}\vspace{2mm}\hfil
\plottwoo{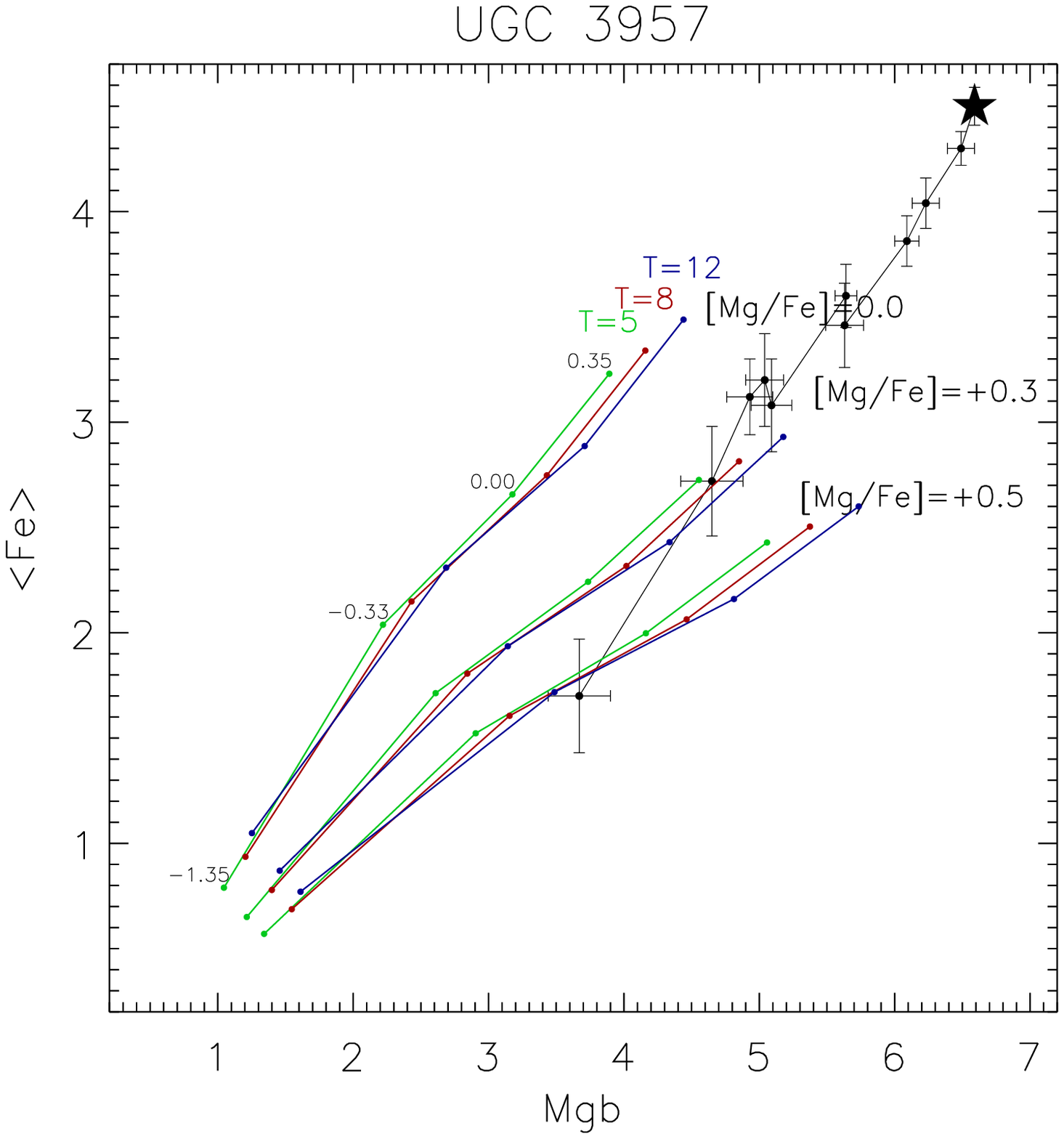}{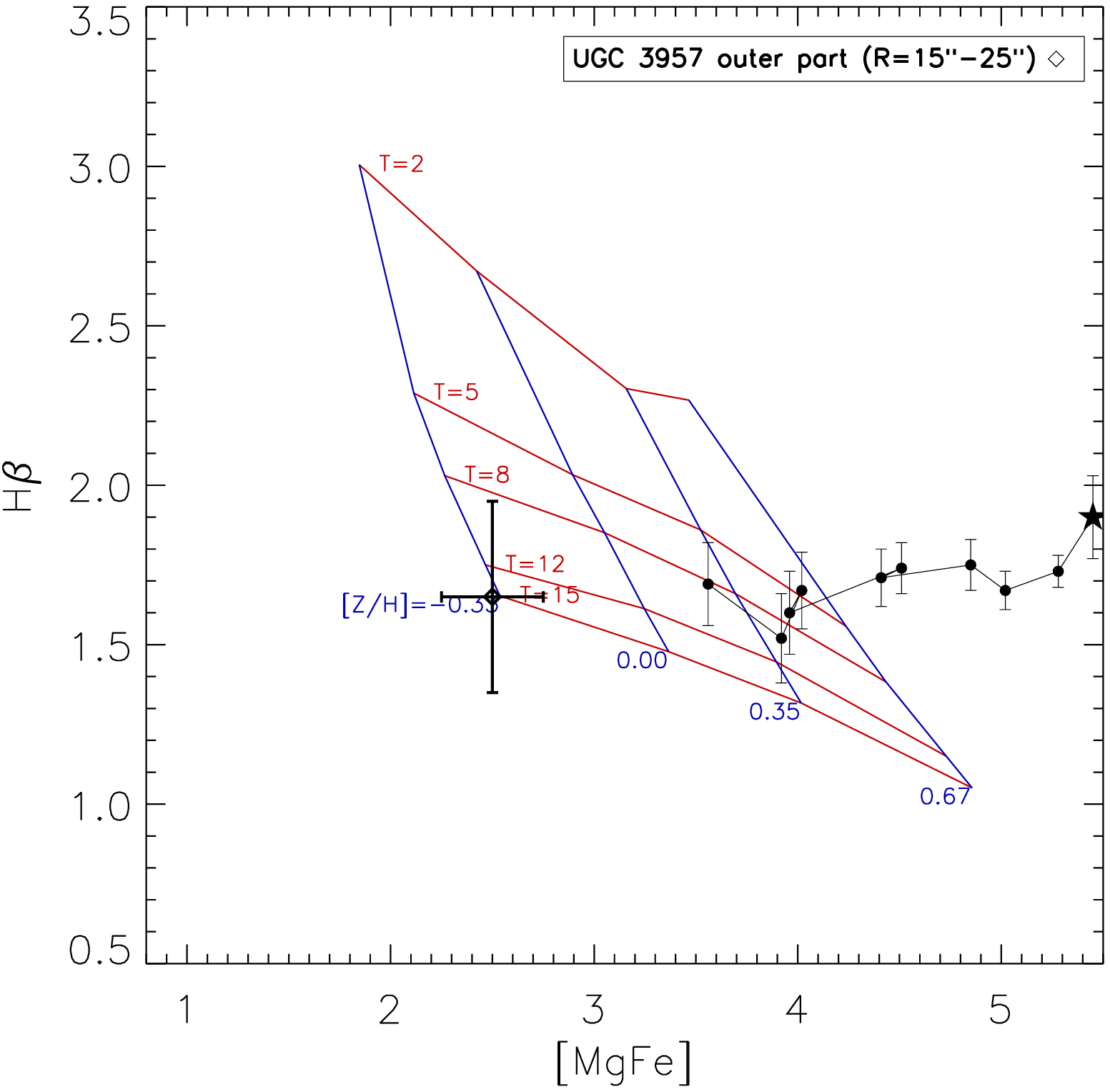}
\caption{(continue)} 
\end{figure*}

% Figure 10
\begin{figure*}
\plotfiveq{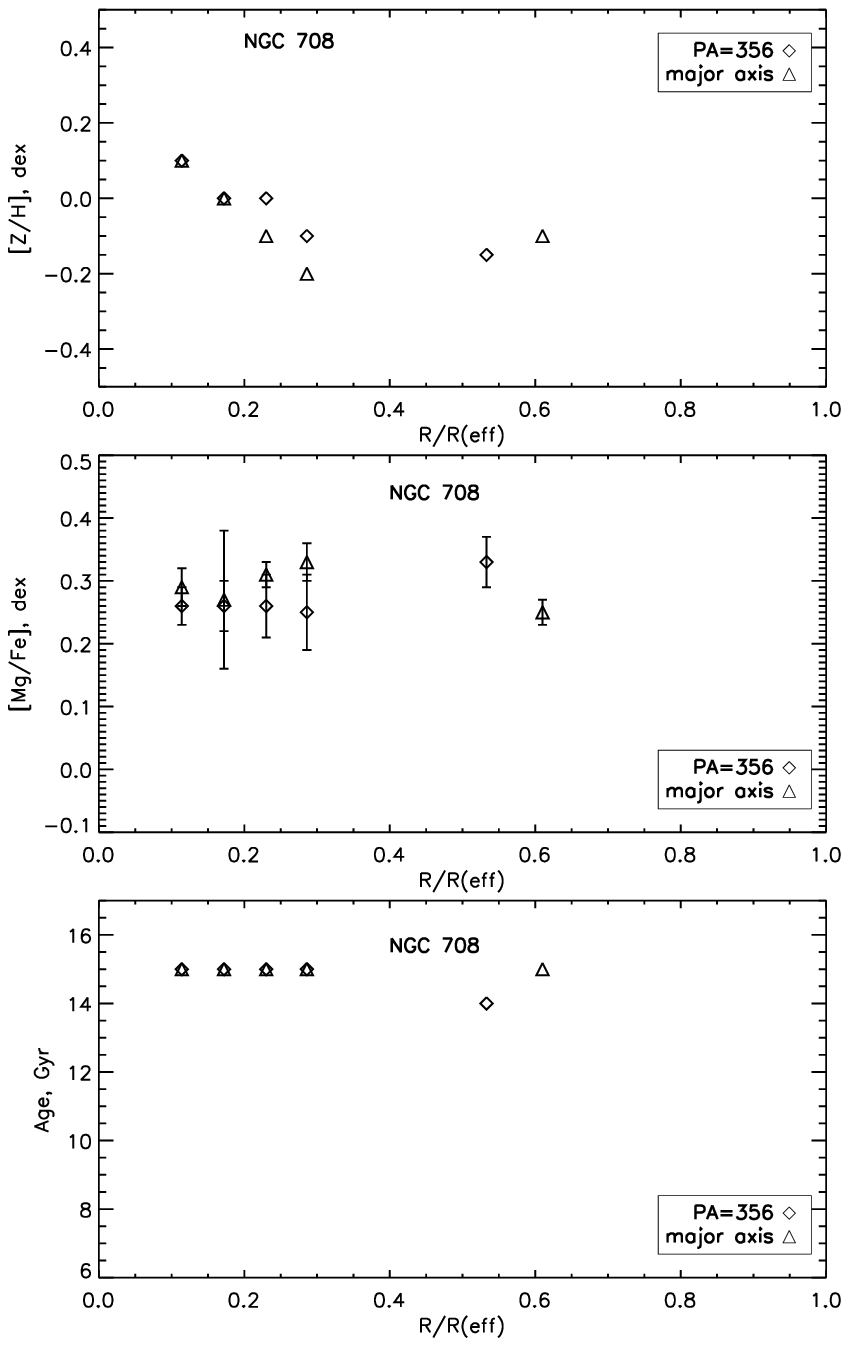}{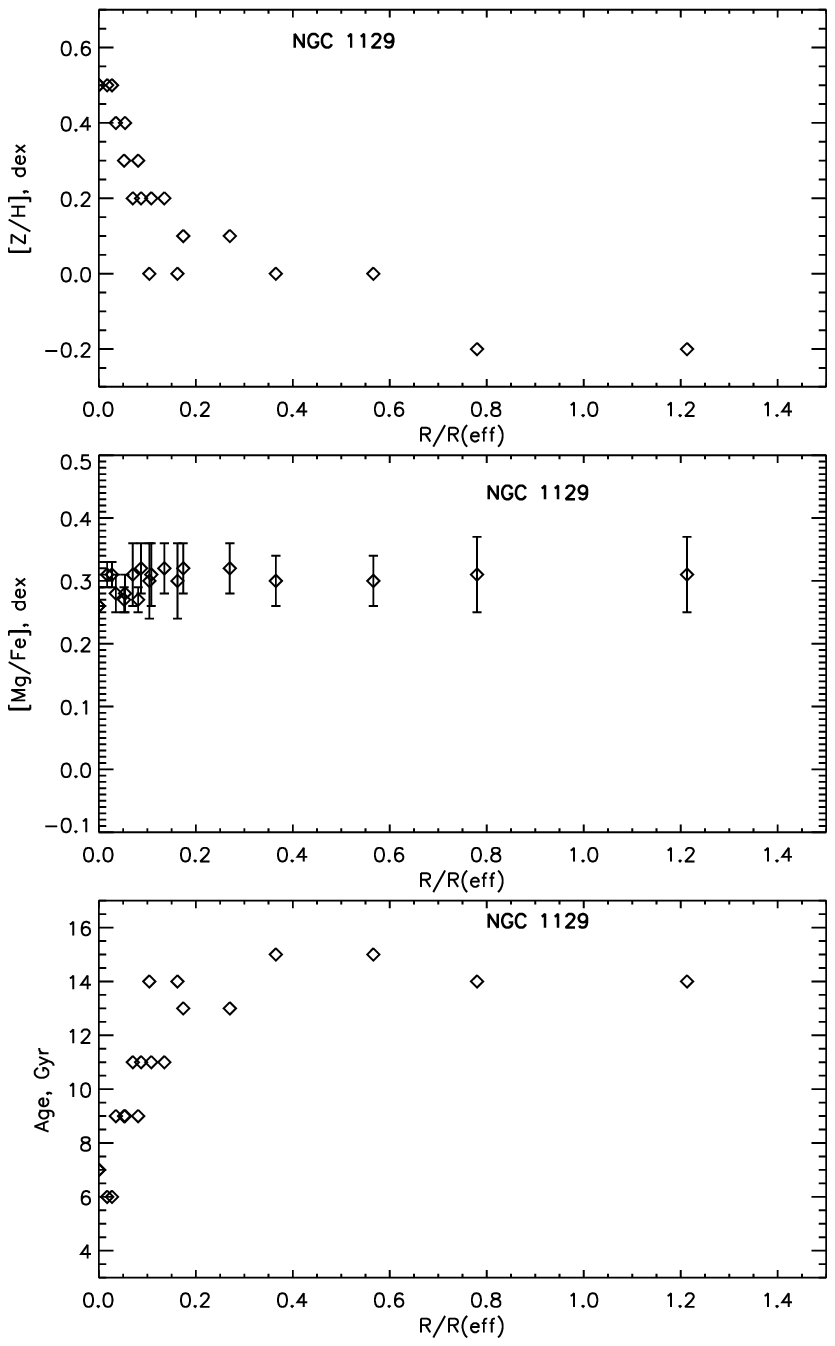}{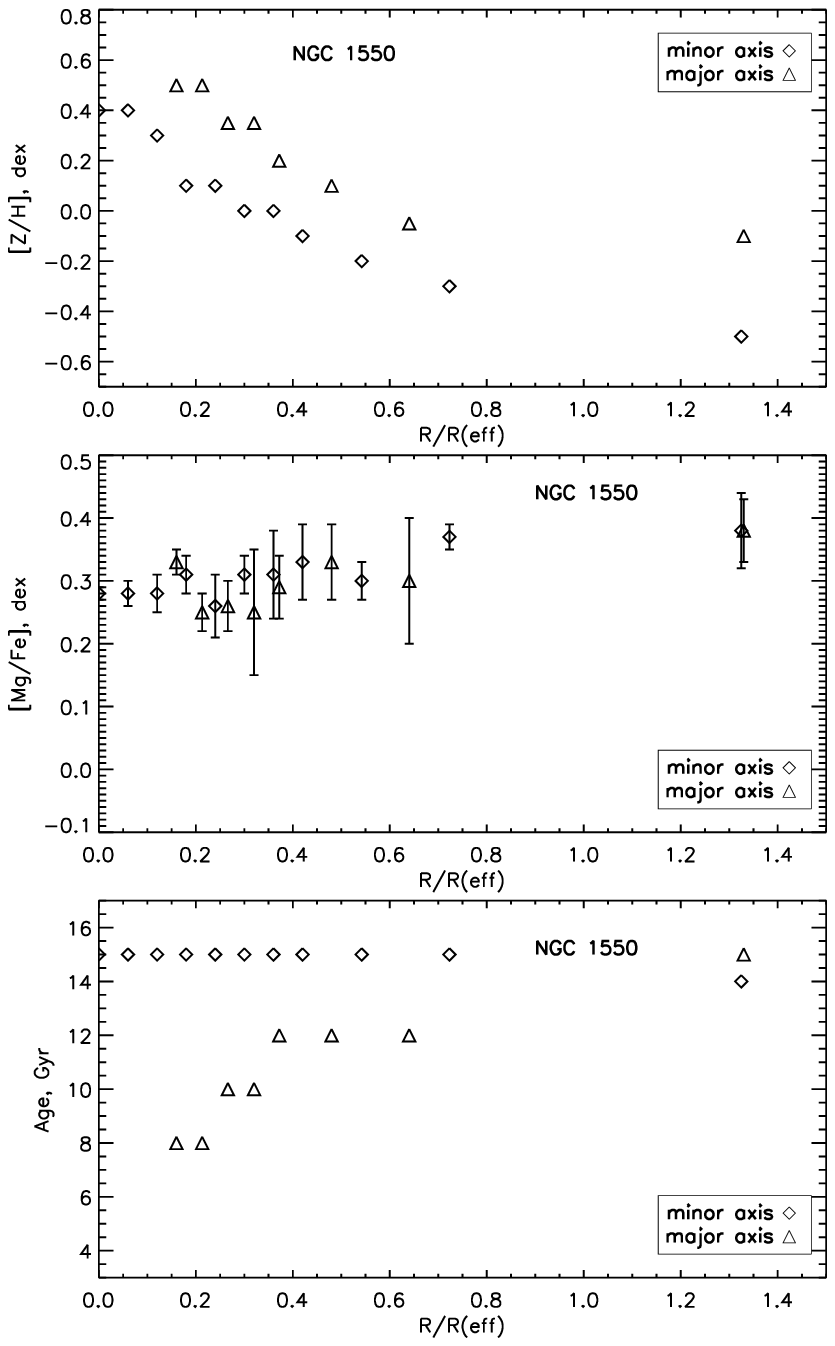}{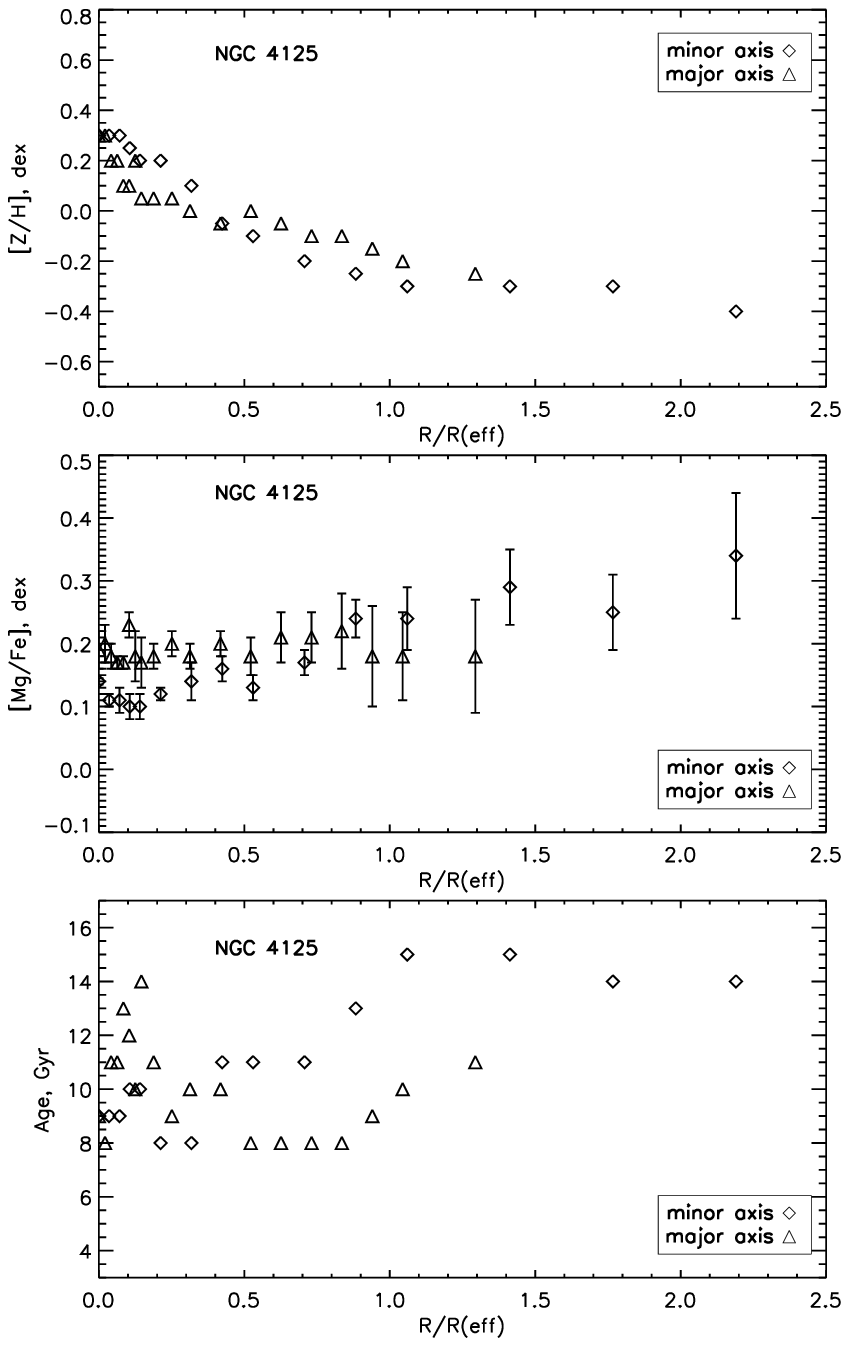}{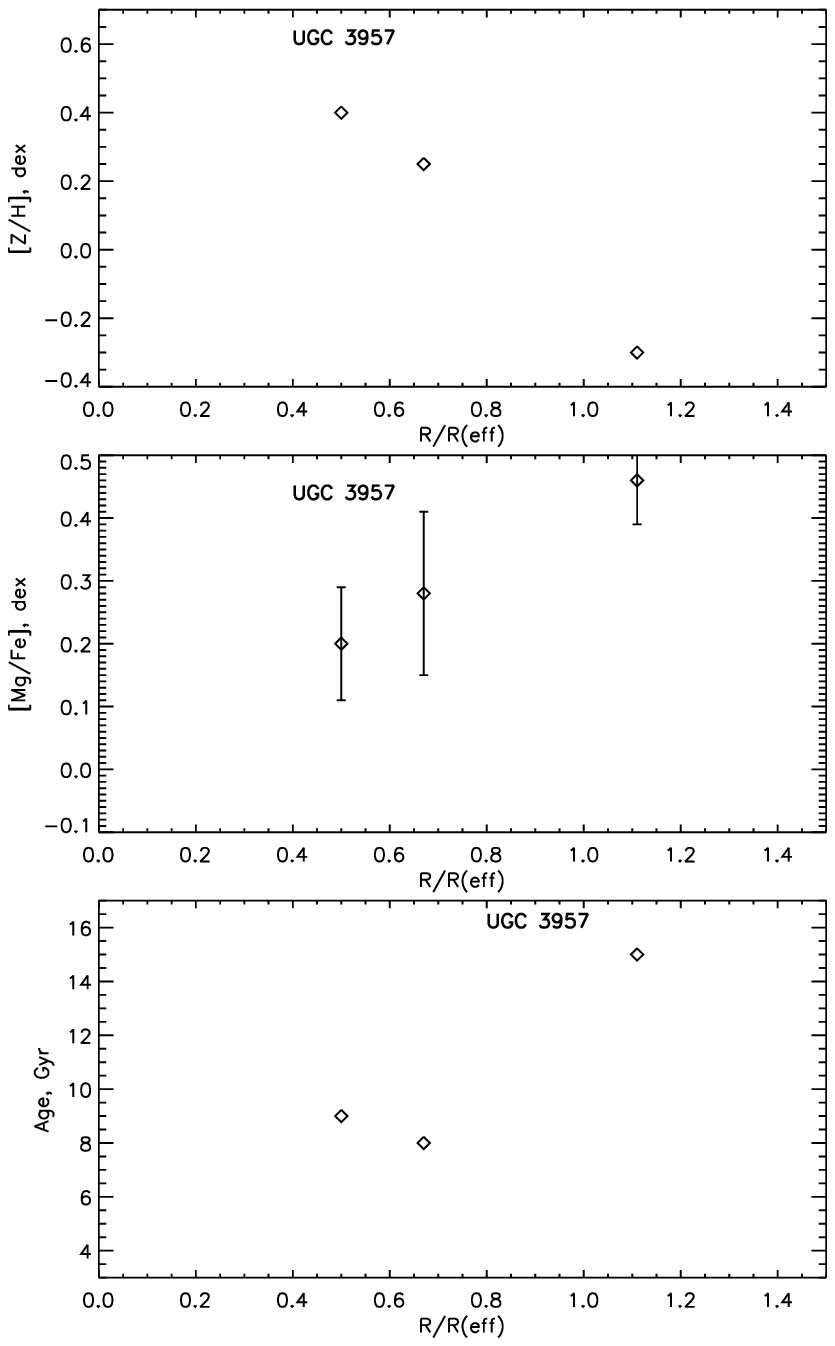}
\caption{The radial variations of the stellar population parameters along the
radius in five elliptical galaxies under consideration; in NGC~1129 both cross-sections are united since they give coincident dependencies.}
\label{param}
\end{figure*}

% Table
\begin{table*}
\caption{Metallicity gradient within and beyond the half effective radius.}\label{metgrad}
\begin{tabular}{lcc}
\hline
Galaxy & $\Delta \mbox{[Z/H]} / \Delta \log R$, dex per dex ($R<0.5 R_{\rm eff})$ &
$\Delta \mbox{[Z/H]} / \Delta \log R$, dex per dex ($R>0.5 R_{\rm eff}$) \\
\hline 
NGC~0708 maj. axis & $-0.74\pm 0.07$ & $0$? \\
NGC~0708 $PA=-4$ & $-0.45 \pm 0.11 $ & $0$? \\
NGC~1129 maj. axis & $-0.41\pm 0.07$ & the same? \\
NGC~1129 min. axis & $-0.50\pm 0.15$ & the same? \\
NGC~1550 min. axis & $-0.69\pm 0.04$ & the same? \\
NGC~4125 min. axis & $-0.52\pm 0.05$ & $-0.27\pm 0.18$ \\
UGC~3957 & -- & $-2.07\pm 0.33$ \\ 
\hline
\end{tabular}
\end{table*}

% Figure 11
%\begin{figure*}
%\plotfiveq{met708.eps}{met1129.eps}{met1550.eps}{met4125.eps}{met3957.eps}
%\caption{The radial variations of the stellar population metallicity along the
%radius in five elliptical galaxies under consideration; in NGC~1129 both cross-sections are united since they give coincident dependencies.}
%\label{metal}
%\end{figure*}

Figure~\ref{param} presents in particular the metallicity radial variations in five ellipticals. The metallicity values [Z/H] are plotted against normalized radius, $R/R_{\rm eff}$, taking into account different values of $R_{\rm eff}$ along the major and the minor axes. The centers of all galaxies demonstrate supersolar metallicity, which is even beyond the model grid of \citet{thomod} in the most massive and luminous galaxy, UGC~3957; however in the outer parts the stellar metallicity drops below the solar value everywhere. The metallicity gradients in our sample ellipticals are negative and
can be estimated mostly as from $-0.4$ to $-0.7$ dex per dex. In NGC~4125 and NGC~1550 the outer metallicity profiles along the major axis go on above the minor-axis profiles that reveals once more the probable presence of the discs aligned with the major axis, formed in some dissipative events including heavy-element enrichment.
We have estimated the metallicity gradients in the spheroids within $0.5R_{\rm eff}$, $R<0.5 R_{\rm eff}$, 
and beyond $0.5R_{\rm eff}$, $R>0.5 R_{\rm eff}$ (Table~\ref{metgrad}), 
because earlier we have found breaks of the metallicity gradients just near this radius in another sample of elliptical galaxies studied with the long-slit spectroscopy of the SCORPIO/BTA \citep{webaes}. Now we have found breaks between steep metallicity gradients in the centers and nearly zero ones  in the outer parts at $0.5R_{\rm eff}$ only in 
two galaxies having the lower mass -- in NGC~0708 and NGC~4125. 
In massive NGC~1129, NGC~1550, and UGC~3957 the outer metallicity
gradients look as steep as the inner ones. Perhaps, for these galaxies we have not reached the radius of break because in the central Coma cluster galaxy NGC~4889 the metallicity gradient break is found at $R=1.2 R_{\rm eff}$ \citep{coccato10}; perhaps the position of break radius correlates with the mass of a galaxy. However, the inner metallicity gradients in our galaxies (and the outer one in UGC~3957) are all steeper than --0.3 dex per dex; it means that these inner parts of the elliptical galaxies under consideration could not be formed by major merger \citep{koba2004}.

\begin{figure*}
\plotfiveq{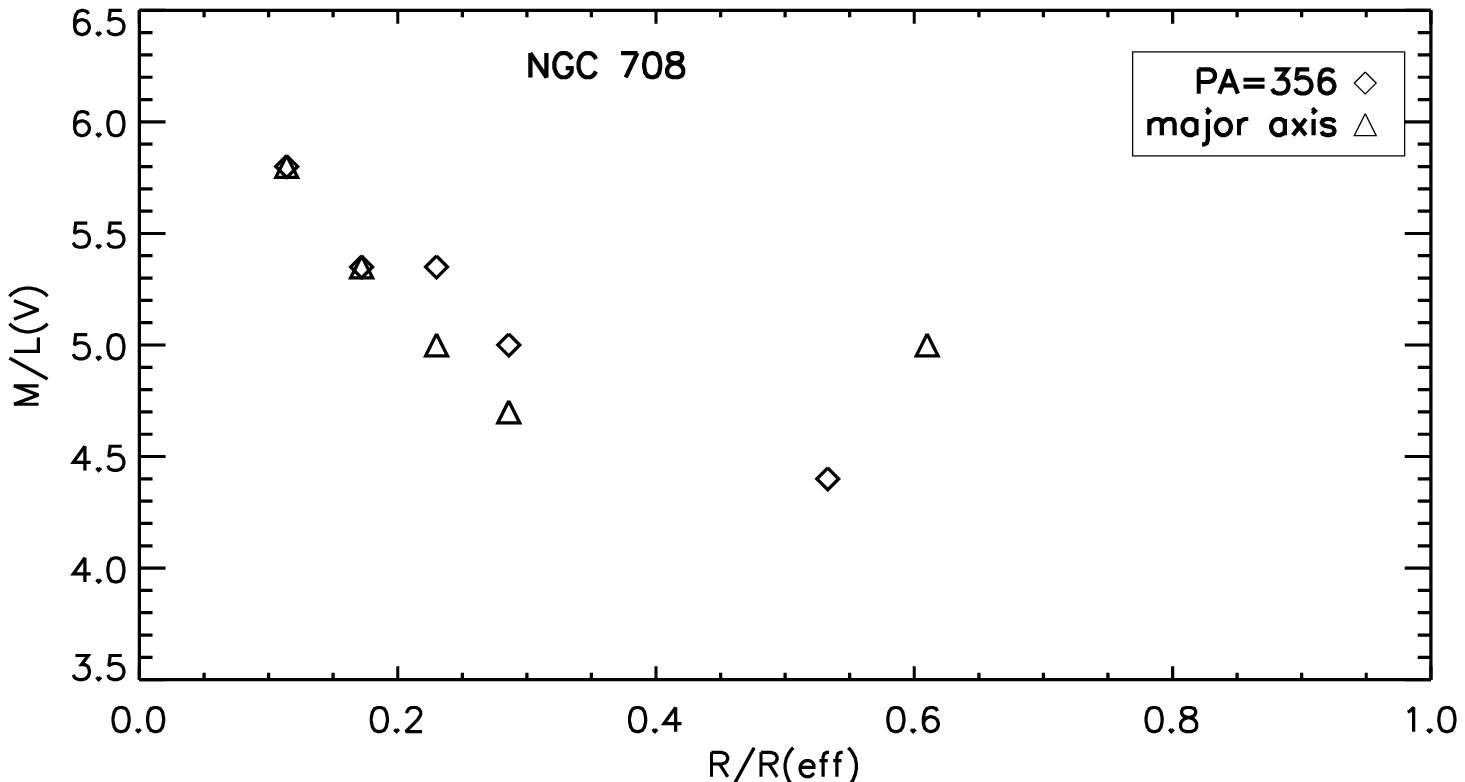}{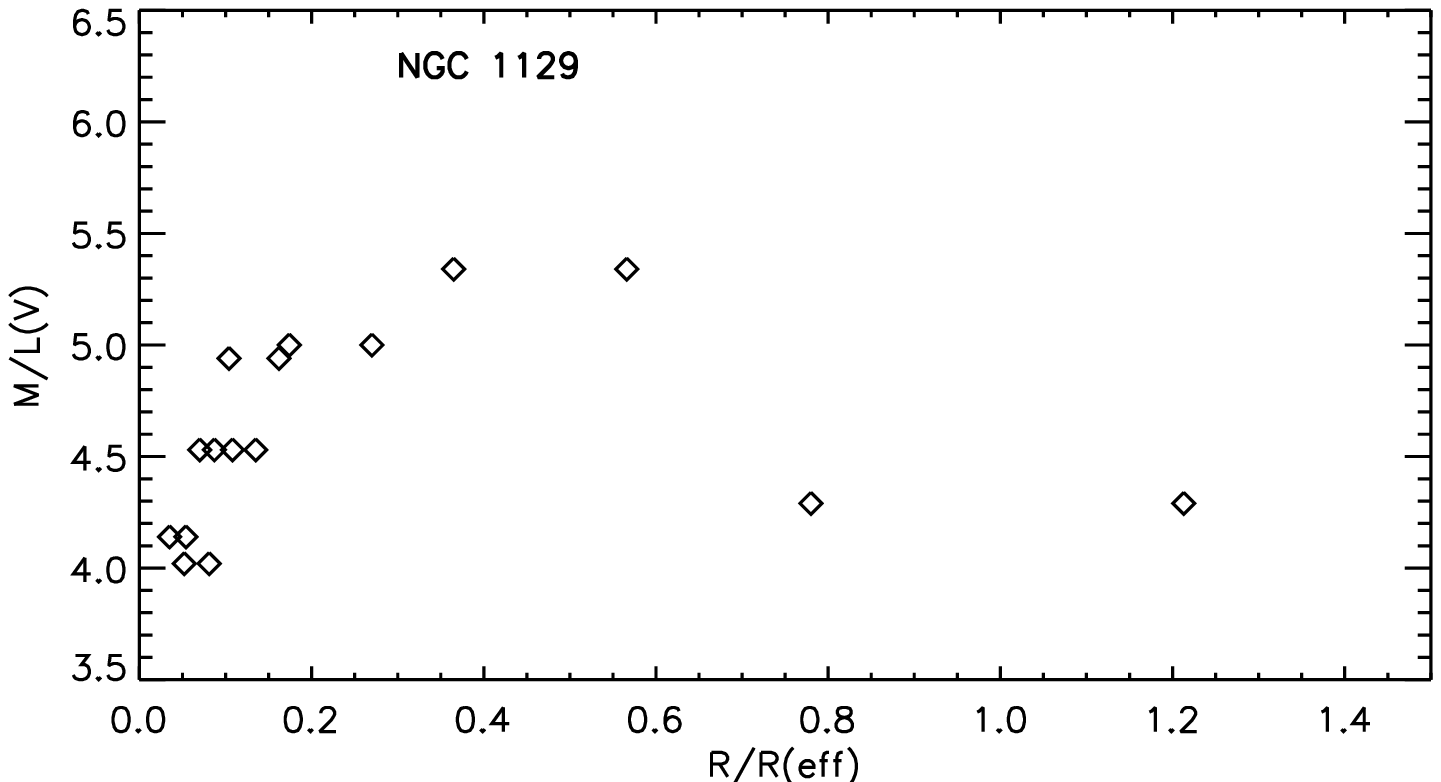}{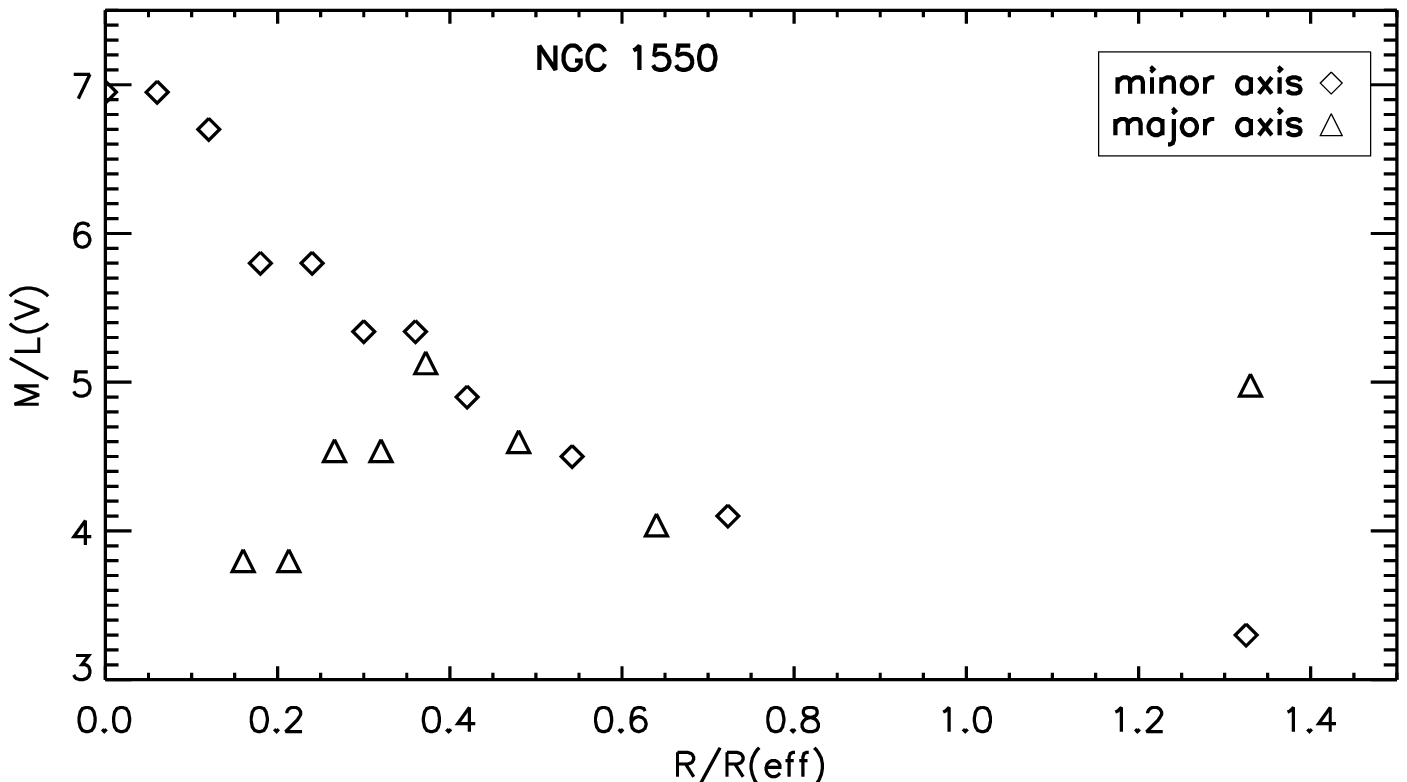}{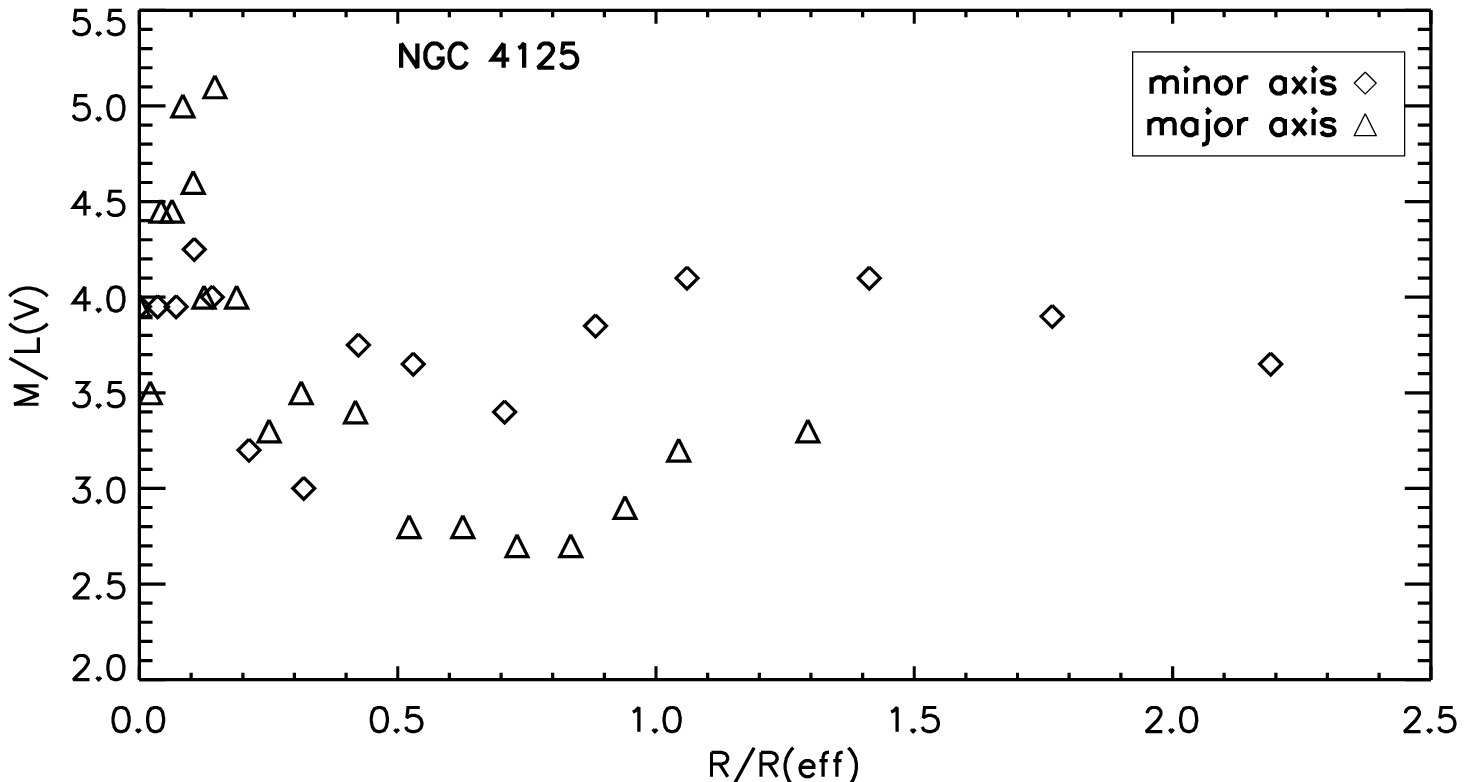}{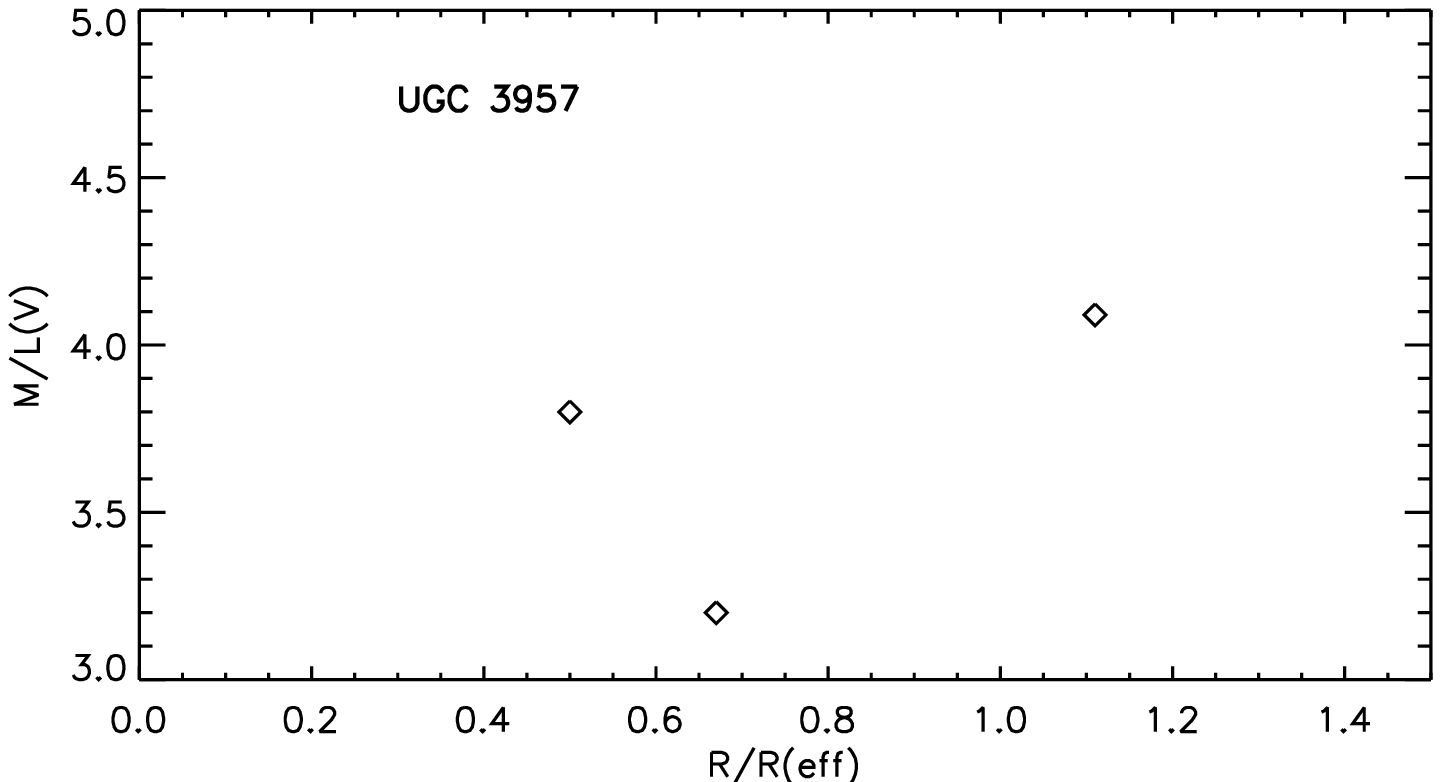}
\caption{The variations of the stellar population mass-to-light ratios along the radius in five elliptical galaxies under consideration; in NGC~1129 both cross-sections  are united since they give coincident dependencies.}
\label{mlv}
\end{figure*}

Radial variations of the stellar population mass-to-light ratio in this case reflect mostly the variations of the metallicity. We have calculated $M/L(V)(R)$ in every galaxy by using the model grid by \citet{maras05}; the [Z/H] and ages found from the Lick indices above have been used to select $M/L(V)$ corresponding to the stellar population properties
at every radius. The radial profile of $M/L(V)$ for every galaxy is shown in Figure~\ref{mlv}. The profiles are presented for the Kroupa IMF; if we prefer the classic Salpeter one, all the $M/L(V)$ values should be increased by a factor of 1.54 (found by confronting $M/L(V)$ for the Kroupa IMF with that for the Salpeter IMF calculated by \citealt{maras05}).

\begin{figure*}
\plotfiveq{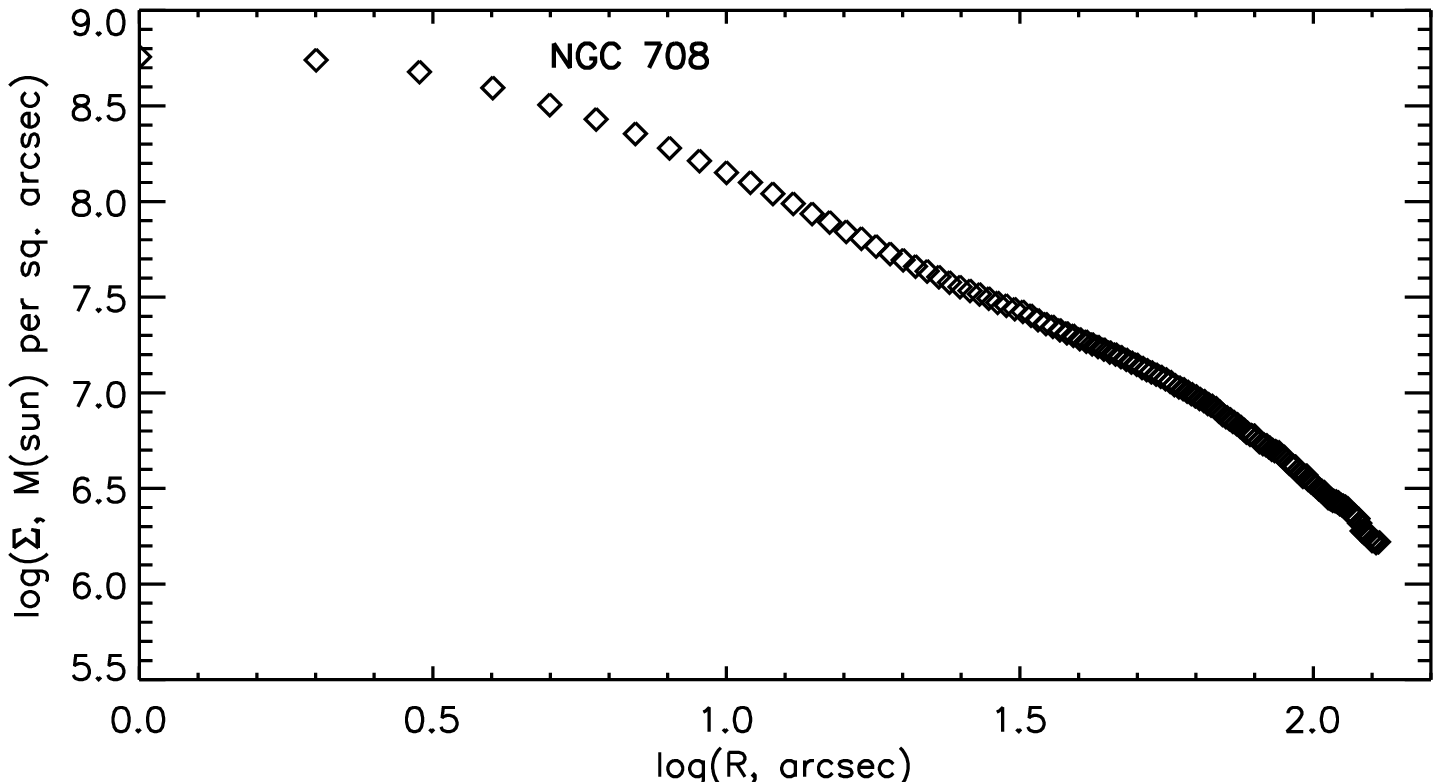}{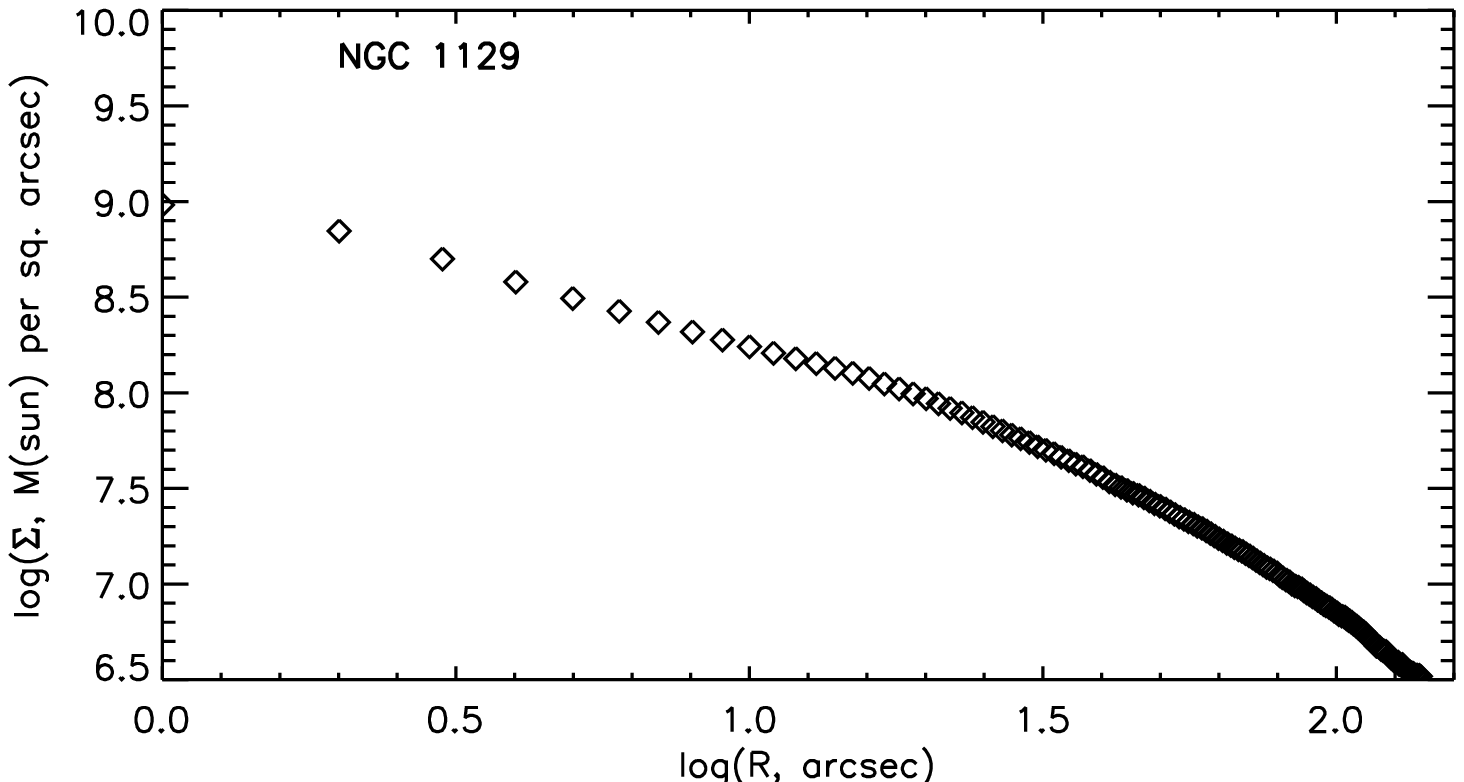}{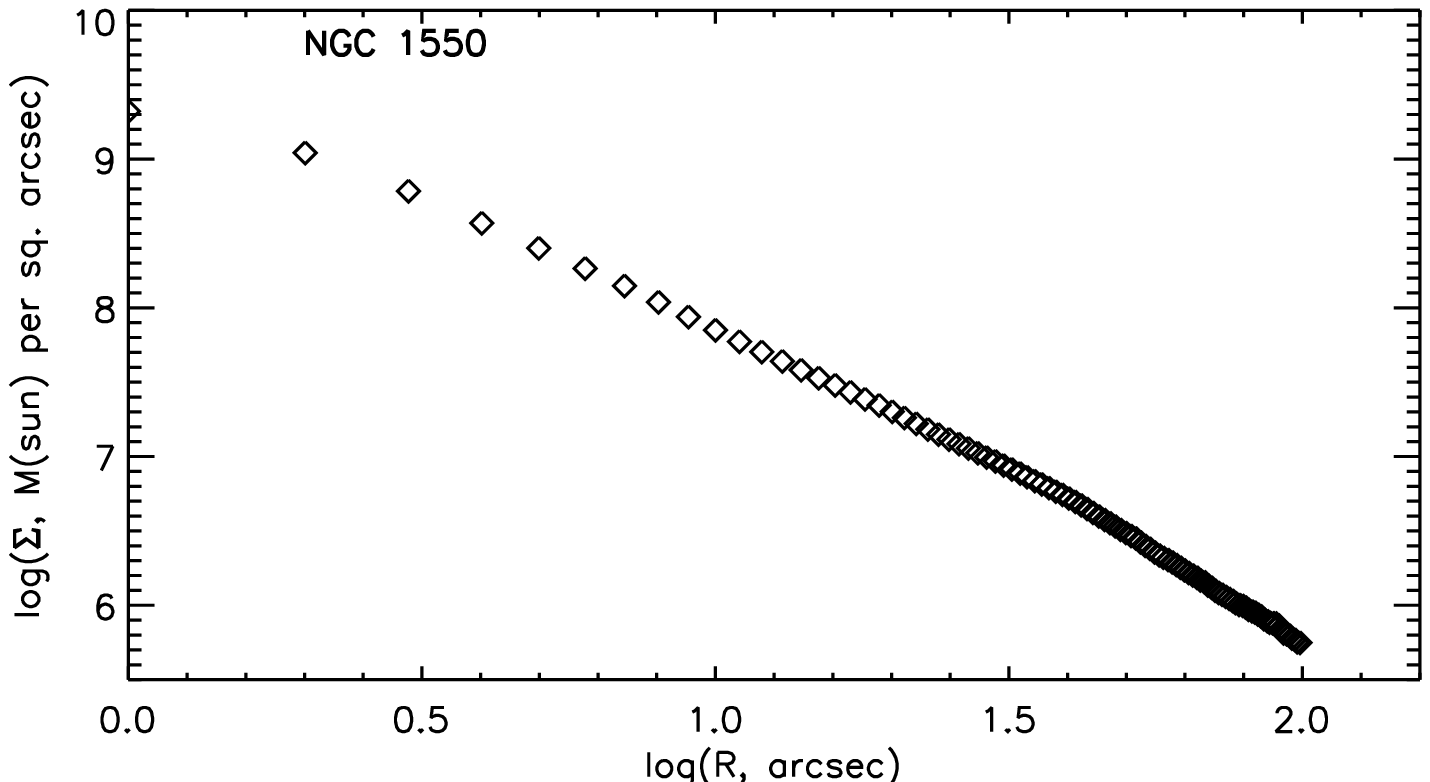}{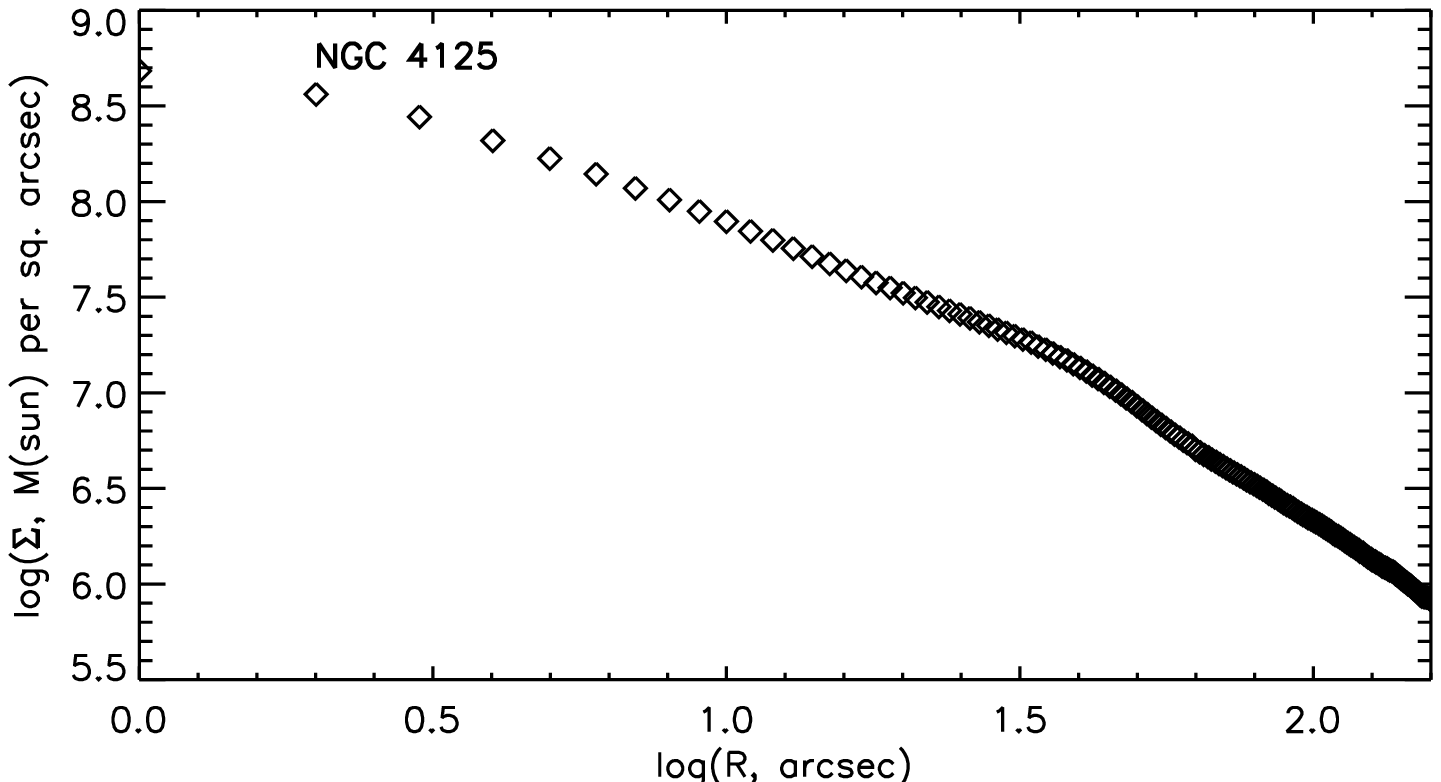}{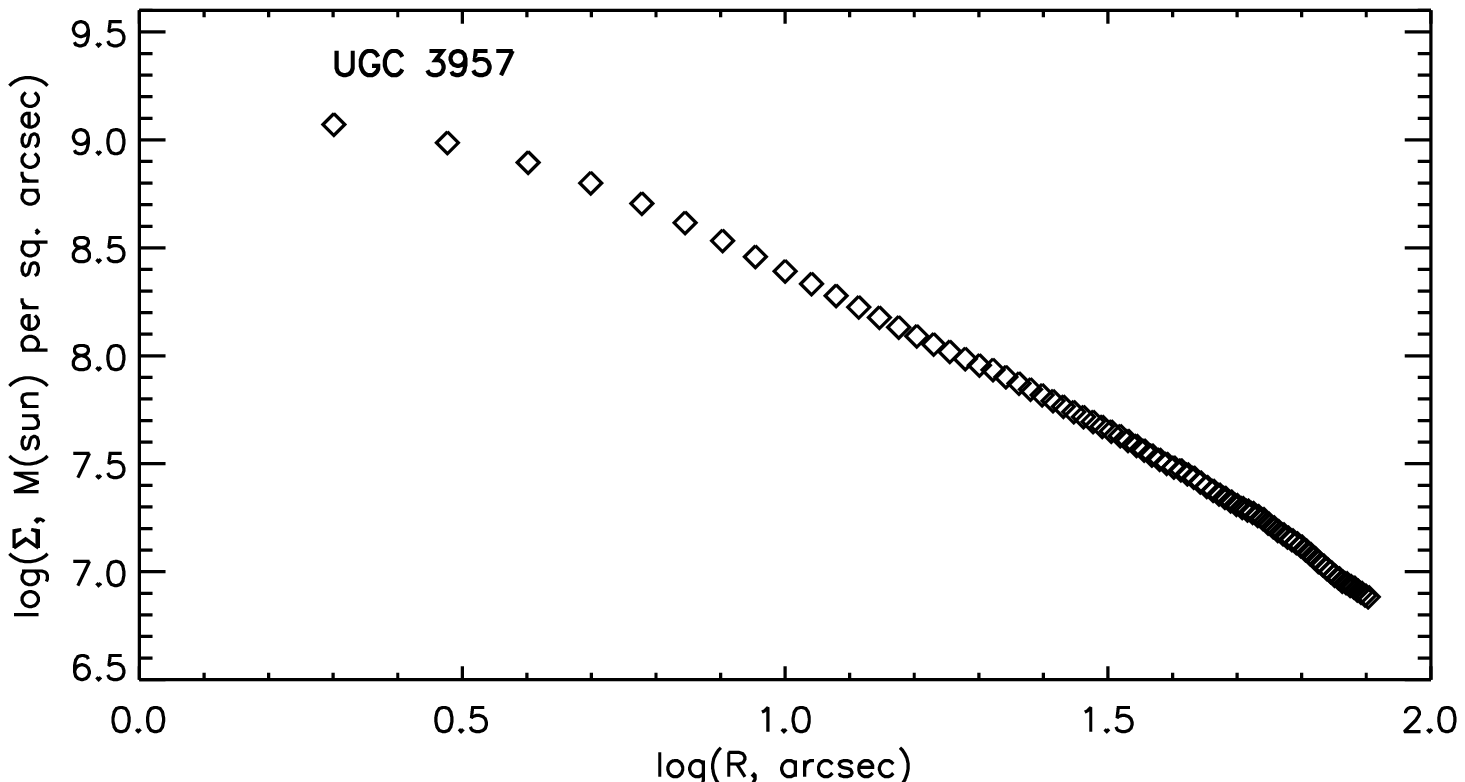}
\caption{The radial profiles of the surface {\bf mass} density along the
radius in five elliptical galaxies under consideration.}
\label{massdens}
\end{figure*}

We have approximated the profiles of Figure~\ref{mlv} by smooth logarithmic or polynomial curves and have used the dependencies derived to transform the surface brightness profiles (this time, the surface brightness profiles obtained from the isophote analysis, with the corresponding azimuthally averaged values of $R_{\rm eff}$) into surface mass density profiles (Figure~\ref{massdens}). Our aim was to estimate, though under very simple assumptions, the stellar mass which is contained within the radii
$R_{\rm sweet}$, to compare it with the dynamical masses derived in previous subsections. The profiles of Figure~\ref{massdens} were then deprojected with the formulae invented by \citet{kholopov}, and after that we have integrated the volume mass density profiles up 
to $R_{\rm sweet}$ under the assumption of spherical symmetry. It is obvious that the assumption of spherical symmetry is very rude for our objects, especially for NGC~708 and NGC~4125, and the fact that the surface mass density profiles are not going to infinity but are cut at arbitrary radii provides only lower limits of the
stellar mass estimates, however some feeling of the dark matter fraction 
within the optical borders of the giant elliptical galaxies can be obtained.

\begin{table*}
\caption{Stellar masses ($\disp V_{c}^{*}=\sqrt{\frac{GM_{*}(<R_{\rm sweet})}{R_{\rm sweet}}}$) and the fraction of dark matter within $R_{\rm sweet}$ for the Kroupa and Salpeter IMFs.}\label{massfrac}
\begin{tabular}{lcccc}
\hline 
Galaxy& \multicolumn{2}{c}{Kroupa IMF} &  \multicolumn{2}{c}{Salpeter IMF} \\
\hline
& $V_{c}^{*}$, km s$^{-1}$ & \%\ of dark matter & $V_{c}^{*}$, km s$^{-1}$ 
& \%\ of dark matter \\
\hline
NGC~0708 & 178 & 77 &  221 &  64 \\
NGC~1129 & 218 & 76 & 270 & 63 \\
NGC~1550 & 190 & 75 & 236 & 62 \\
NGC~4125 & 248 & 56 & 308 & 32 \\
UGC~3957 & 174 & 87 & 216 & 79 \\
\hline
\end{tabular}
\end{table*}

As we can see in Table~\ref{massfrac}, there is a range of dark mass presence among our small sample. In particular, NGC~4125 has the most of all its mass in stars. However, if we refer to the Salpeter IMF, a typical fraction of dark matter within $R_{\rm sweet}$ is 60\%. For the Kroupa IMF the sample averaged fraction is $\sim 75$\%. The comparison between the stellar and dynamical mass estimates measured within $R_{\rm sweet}$ is shown in Figure \ref{analysis_v3} (right panel).

%________________________________________________________________

\section{Discussion}
\label{sec:discussion}

\begin{table*}
\centering
\caption{Ellipticity and effective radius for the sample galaxies. \label{tab:eff} The columns
are: (1) - galaxy name; (2) - ellipticity; (3) - effective radius defined from the de Vaucouleurs fit to the ellipse-averaged surface brightness profile; (4) - effective radius defined from the de Vaucouleurs fit to the long-slit profile.}
\begin{tabular}{lccccccrll}
\hline
Name   & Ellipticity  & $R_{\rm eff}$, arcsec & $R^{slit}_{\rm eff}$, arcsec \\
(1)   & (2) & (3) & (4) \\
\hline

NGC 708        &  0.45 & $63.5 \pm 1.5$ & $58.4 \pm 2.9$  \\

NGC 1129        &  0.22 & $86.7 \pm 2.4$ & $61.7 \pm 6.8$ \\
NGC 1550        &  0.11 & $25.7 \pm 0.7$ & $18.5 \pm 1.1$ \\
NGC 4125        &  0.46 & $56.4 \pm 0.4$  & $54.3 \pm 0.7$ \\
UGC 3957        &  0.1  & $33.9 \pm 0.8$ &  $22.1 \pm 1.0$ \\

\hline
\end{tabular}
\end{table*}

\begin{figure*}
\plottwo{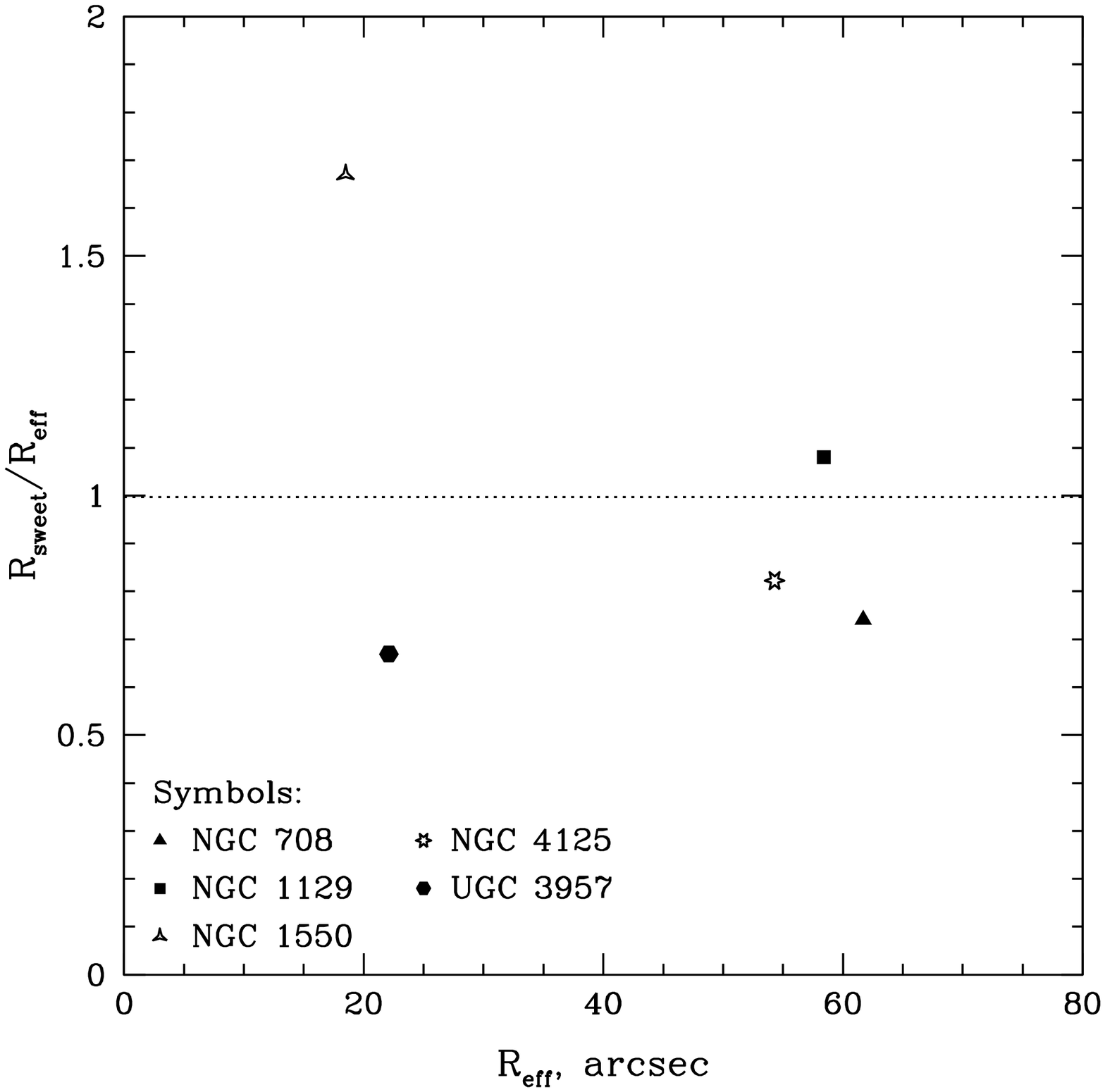}{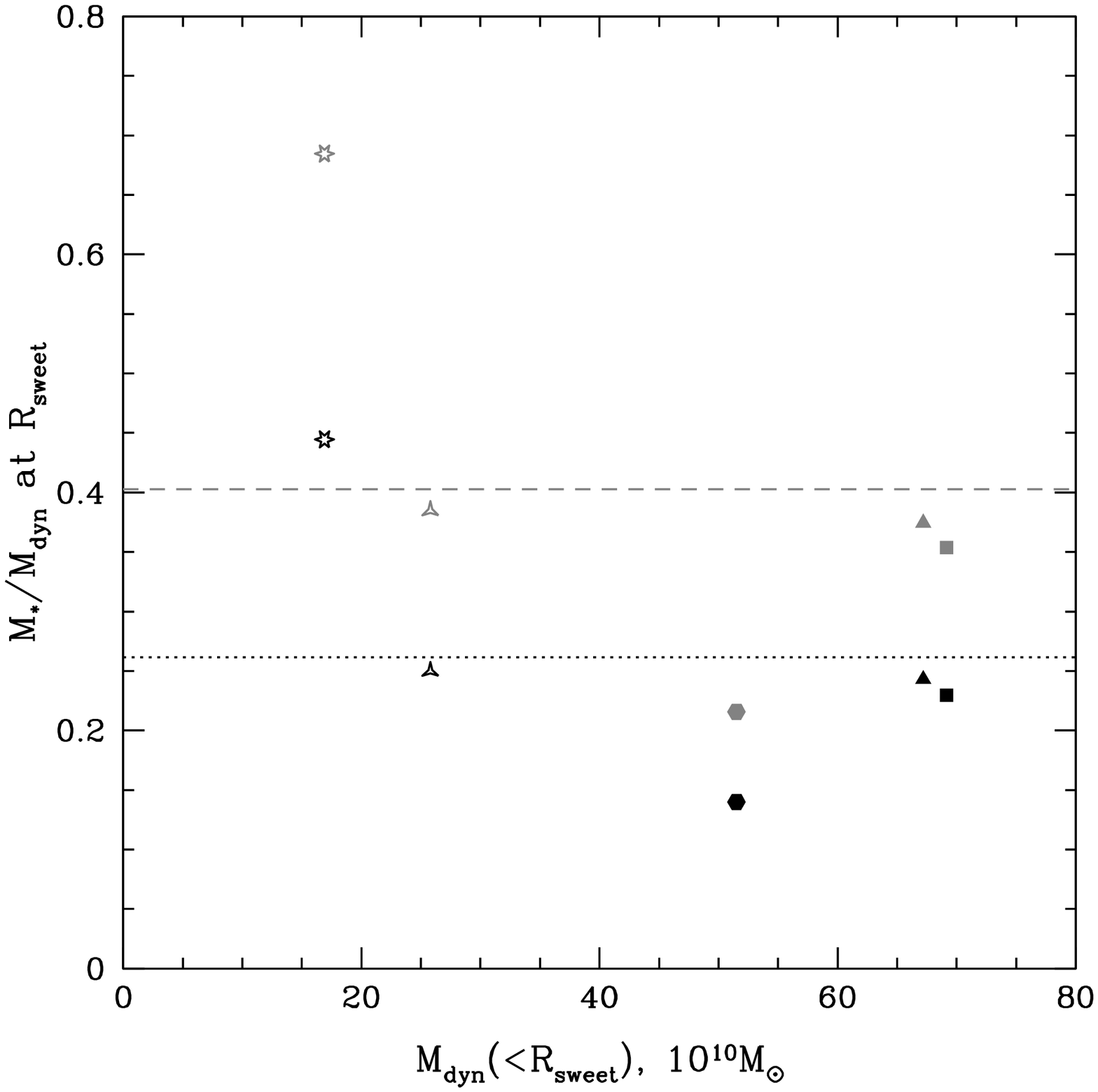}
\caption{Left: the comparison between the sweet radius $\disp R_{\rm sweet}$ and the effective radius $\disp R^{slit}_{\rm eff}$, defined from the de Vaucouleurs fit to the long-slit surface brighness profile. Different symbols denote different galaxies: the solid triangle - NGC 708, the solid square - NGC 1129, the curved triangle - NGC 1550, the star - NGC 4125, the solid hexagon - UGC 3957. The dotted line shows to average ratio $\disp R_{\rm sweet}/R^{slit}_{\rm eff}$.
Right: the comparison between the stellar $\disp M_{*}(<R_{\rm sweet})$ and dynamical $\disp M_{dyn}(<R_{\rm sweet}) = \left[V_c^{iso}\right]^2R_{\rm sweet}/G$ mass at the sweet radius. Black symbols show the ratio $\disp M_{*}/M_{dyn}$ for the Kroupa IMF and grey symbols - for the Salpeter IMF. The black dotted and grey dashed lines indicate the average ratio $\disp M_{*}/M_{dyn}$  for the Kroupa and Salpeter IMFs, respectively. }
\label{analysis_v3}
\end{figure*}

We discuss one simple and fast, but nevertheless reliable method for estimating masses of early-type galaxies from the stellar surface brightness and the line-of-sight velocity dispersion profiles only. The method is based on the ansatz that the relation between the projected velocity dispersion and the circular speed is almost insensitive to the anisotropy of stellar orbits at a characteristic radius  $R_{\rm sweet}$  where derived circular speed profiles for isotropic distribution of stars, pure circular and pure radial stellar orbits are close to each other. $R_{\rm sweet}$ lies close to the radius $R_2$ where the surface brightness $I(R)$ declines as $R^{-2}$, which is in turn not far from the effective radius of the galaxy $R_{\rm eff}$. Although the method allows to estimate mass within some particular radius only (or in radial range where $I(R) \appropto R^{-2}$), it does not require any apriori parametrization of mass or anisotropy profiles and the resulting estimate does not depend significantly on the quality of data. The method has already been tested on a sample of 65 simulated galaxies drawn from cosmological simulations by \cite{2010ApJ...725.23120} in \cite{Lyskova.et.al.2012}. When averaged over the subsample of massive ($\disp \sigma_p(R_{\rm eff})>150$ $\kms$) slowly rotating galaxies the recovered circular speed is almost unbiased ($\disp \overline{\Delta_{opt}}<1 \%$) with modest scatter (RMS = 5.4 $\%$). Note that  in \cite{Lyskova.et.al.2012} the surface brightness and the line-of-sight velocity dispersion profiles were calculated in a set of logarithmic concentric annuli around the center of the simulated galaxy. In this paper we (i) mimic the long-slit observations by computing the profiles along the slits, and also (ii) extend our analysis to rotating elliptical galaxies by considering $\disp V_{\rm rms}(R)=\sqrt{\sigma_p^2(R)+V_{\rm rot}^2(R)}$ instead of $\sigma_p(R)$ in equations (\ref{eq:main}) or (\ref{eq:agd_simple}), where  $\disp V_{\rm rot}(R)$ is the observed rotational velocity. Tests have been performed on the sample of massive simulated galaxies ($\disp \sigma_p(R_{\rm eff}) > 150$ $\kms$, edge-on view) that includes both fast and slow rotators in proportion close to the observed one reported by ATLAS$^{3d}$ team \citep{Emsellem.et.al.2007}. We find that the circular speed recovered from $I(R)$ and $\sigma_p(R)$ measured along the slit aligned with the apparent major axis of the galaxy is on average underestimated by $~4-5 \%$ and the RMS-scatter is about $6 \%$. The bias almost vanishes when $\sigma_p(R)$ in equations (\ref{eq:main}) is substituted  with $\disp V_{\rm rms}(R)=\sqrt{\sigma_p^2(R)+V_{\rm rot}^2(R)}$ and the RMS-scatter remains the same. If profiles are measured along apparent major and minor axes of the galaxy then we can reduce the scatter arising from trixiality of elliptical galaxies. Indeed, in this case the RMS-scatter is reduced down to $5 \%$. 

%Abundance
X-ray circular speed profiles for all galaxies are inferred under an assumption of constant metallicity ($Z=0.5 Z_{\odot}$), although the errorbars are estimated in a conservative way, allowing for the abundance gradients. If the abundance steadily increases to the galaxy center, then the assumption of the flat metallicity profile would lead to overestimated circular velocity. The decreasing at small radii abundance would imply instead that $V_c^X$ is underestimated.
Abundance measurements for low temperature ($\lesssim 1.5$ keV) systems remains among the most important uncertainties in X-ray determination of the circular speed of elliptical galaxies. For hotter systems the impact of abundance is much less severe. For instance, the conservative estimate for NGC0708 suggests that the circular speed may be over-/underestimated as much as 25-30 \% (at $R \lesssim 30-40''$).

The inferred circular speed estimates from optical data and from X-ray for all galaxies in our sample agree with each other remarkably well, especially for NGC 1550 and NGC 1129, indicating the relaxed dynamical state of these galaxies, close to isotropic distribution of stellar orbits within  1-2 effective radii and that hot gas in these objects is in the hydrostatic equilibrium. The only rotating galaxy in our sample,  NGC 4125, is also the only galaxy with non-negligible nonthermal pressure support (at the level of $\approx 36 \%$).  For UGC 3957 both X-ray and optical methods give the same (within errorbars) result at the sweet point and at larger radii we observe that $V_c^X(R) > V_c^{\rm iso} (R)$ what may be interpreted as the radially biased stellar velocity anisotropy. NGC 708 is the most difficult galaxy for interpretation. At $R \lesssim 30 ''$ where optical data are quite reliable $V_c^X(R)$ lies below $V_c^{\rm iso} (R)$. Then at $30'' \gtrsim R \gtrsim 60'' $ these two curves are roughly consistent with each other. At the sweet spot which is located slightly beyond the radial range with available optical data the X-ray mass estimate exceeds the optical one by $\sim 40 \%$ although the reliability of the $V_c^{\rm iso}$ at this radius is under question and errorbars are quite large. The average ratio between the optical $V_c^{\rm iso}$ and $V_c^X$ at the sweet spot is equal to $\disp 0.98 $ with $\approx 0.11$ rms-scatter\footnote{$\disp \overline{x}=\disp \frac{\sum{x}}{N} \pm \frac{RMS}{\sqrt{N}}$, $RMS=\sqrt{\frac{\sum{(x-\overline{x})^2}}{N}}$}. Given the scatter, this result indicates that on average the non-thermal contribution to the total gas pressure is consistent with zero. Two galaxies - NGC 4125 and NGC 0708 - that have the lowest central velocity dispersions and are showing the largest deviation of $\disp \left< \frac{V_{c}^{\rm iso}}{V_{c}^X} \right>$ from the mean value, appear to be especially prone to abundance uncertainties in X-ray analysis. Low temperature of NGC 4125 ($\disp T \approx 0.5$ keV) does not allow to disentangle reliably continuum and emission lines. In its turn, NGC 0708 has higher temperature at the sweet radius ($\approx 1.7$ keV) but shows significant abundance gradients, what leads to large spread in the resulting circular velocity curves. If we exclude these two galaxies,  then the average ratio is $\disp \left< \frac{V_{c}^{\rm iso}}{V_{c}^X} \right> \approx 0.96$ with $RMS \approx 0.03$. This scatter is comparable to the expected value of $5.4 \%$ coming from the analysis of a sample of simulated massive elliptical galaxies without significant rotation (\citealt{Lyskova.et.al.2012}).   

It should be mentioned that for our analysis we deliberately use the surface brightness profiles measured along the slit rather than ellipse-averaged radial profiles. On one hand, $I(R)$ along specific PA could be affected by local inhomogeneities in brightness and the signal-to-noise ratio is smaller compared to azimuthally-averaged profiles. On the other hand, analysing the projected velocity dispersion and surface brightness profiles measured in the same way seems to be more self-consistent and justified.   Moreover, we aim to demonstrate perfomance of our simple mass estimator using the most basic observables, thus on purpose neglecting all possible complications. Apart from using the original surface brightness distribution along the specific PA, we also simplify the analysis by neglecting departures from spherical symmetry (see eq. \ref{eq:SB}-\ref{eq:Vrms}). If we take into account information on ellipse-averaged radial profiles and ellipticity of a given galaxy, we will get the $V_c$-estimate similar to the reported one (within errorbars), although the averaged surface profiles are slightly shallower that original ones pushing $R_{\rm sweet}$ towards larger radii where kinematics is getting less reliable.

The full version of analysis, i.e. equations (\ref{eq:main}), is recommended to use when the projected velocity dispersion profile is reliable over the radial range of interest. As the circular speed $V_c^{\rm rad}(R)$ recovered for pure radial orbits depends on the second derivative $\disp \delta=\frac{d^2\ln[I(R)\sigma^2_p(R)]}{d(\ln R)^2}$, it relies on the quality of $\sigma_p(R)$. If the dispersion profile is noisy and does not decline steeply, then $R_2$ - the radius where $\disp \alpha = d \ln I(R) / d \ln R = 2$ - can be used as the sweet spot.
For our sample of galaxies the average ratio $\disp \left< \frac{V_c^{\rm iso}(R_{\rm sweet})}{V_c^{\rm iso}(R_{2})} \right>$ of circular speed estimates calculated from equations (\ref{eq:main}) at $R_{\rm sweet}$ and at $R_2$ is equal to $1.02$ with $RMS = 0.016$. When the observational data do not allow to use the full analysis, then the circular speed can be estimated using the simplified analysis  (eq. \ref{eq:agd_simple}). The average ratio $\disp \left< \frac{V_c^{\rm iso}(R_{\rm sweet})}{V_c^{iso,s}(R_{2})} \right>$ equals to $1.04$ with $RMS = 0.032$, where $\disp V_c^{iso,s}$ is calculated using equations (\ref{eq:agd_simple}).

%reff
As expected the sweet radius is found to lie close to $R_2$  and also not  far from the effective radius $R_{\rm eff}$ of a galaxy (effective radii used here, $R^{slit}_{\rm eff}$ are listed in Table \ref{tab:eff}). For our sample the average ratio $\disp \left< \frac{R_{\rm sweet}}{R_{2}} \right> \approx 1.06$ with rms-scatter $\disp RMS \approx 15 \%$, while $\disp \left< \frac{R_{\rm sweet}}{R^{slit}_{\rm eff}} \right> \approx 1.0$ with $RMS \approx 36 \%$ scatter. The ratio $\disp  \frac{R_{\rm sweet}}{R^{slit}_{\rm eff}}$ as a function of $\disp R^{slit}_{\rm eff}$ is shown in Figure \ref{analysis_v3} (left panel).  For successful implementation of our simple estimator the surface brighness and kinematic profiles should extend slightly beyond $R_{\rm sweet}$ or $R_2$ as equations (\ref{eq:main}) require differentiation of the observed profiles. For S\'{e}rsic surface brightness distribution $\disp I(R) =I(R_{\rm eff}) \exp\left[-b_{n}\left( (R/R_{\rm eff})^{1/n}-1\right)\right] $, where $\disp b_{n} \simeq 2n-0.324$ (for $\disp 0.5 \leq n \leq 10$, \citealt{Ciotti.1991}), $R_2$ is related to $R_{\rm eff}$ via \citep{Graham.Driver.2005}

\be
\label{eq:R_2_eff}
\disp R_{2} \simeq \left( \frac{2n}{b_n}\right)^n R_{\rm eff} \simeq 1.2R_{\rm eff} ,
\ee

i.e. $R_{2} \simeq 1.2 R_{\rm eff}$. So in terms of effective radii the observed profiles should extent out to $\sim 1.2 - 1.5 R_{\rm eff}$ for reliable mass determination. 
It should be noted that the value of the effective radius strongly depends on a measurement techique. The effective radii $\disp R_{\rm eff}$ and $\disp R^{slit}_{\rm eff}$ for our sample galaxies obtained from the de Vaucouleurs fit to the ellipse-averaged and long-slit surface brightness profiles correspondingly are listed in Table \ref{tab:eff}. For some galaxies $\disp R_{\rm eff}$ and $\disp R^{slit}_{\rm eff}$ are different by a factor of $\sim 1.5$.  The effective radius could vary significantly depending on (i) whether it is measured with or without extrapolation of data, (ii) parametric form of the stellar distribution profile used to fit the data, (iii) radial range used to fit, for instance, the S\'ersic profile, (iv) quality of photometric data \citep[see, e.g.,][]{Kormendy.et.al.2009, Cappellari.et.al.2013}. In contrast with the simple mass estimator proposed by \cite{2010MNRAS.406.1220W} our estimator is not tied to the effective radius. The sweet spot is defined from {\it local} properties of $I(R)$ and $\sigma_p(R)$ or even from $I(R)$ alone.

\section{Conclusion}
\label{sec:conclusion}

We discuss a simple mass estimator that allows one to estimate the circular speed $V_c$ from {\it local} properties of the surface brightness and the line-of-sight kinematics at a characteristic radius where the $V_c$-estimate is largely insensitive to the unknown anisotropy of stellar orbits. Although the method is designed for non-rotating spherical galaxies, we extend it also to mildly rotating axisymmetric and slowly-rotating triaxial ones, substituting $\sigma_p(R)$ in equation \ref{eq:main} or \ref{eq:agd_simple} with $\disp V_{\rm rms}(R)=\sqrt{\sigma_p^2(R)+V_{\rm rot}^2(R)}$, where $\disp V_{\rm rot}(R)$ is the rotational velocity.  Tests on the sample of massive simulated galaxies show that the recovered from $I(R)$ and $\sigma_p(R)$ measured along apparent major and minor axes of the galaxy circular speed is almost unbiased with the RMS-scatter of $\sim 5 \%$. 

We apply the method to M87 and compare our simple mass estimate with circular speed profiles derived from X-rays and the state-of-the-art Schwarzschild modeling, thus revisiting the results of \cite{2008MNRAS.388.1062C,2010MNRAS.404.1165C}. At the sweet radius $R_{\rm sweet}=141''$ we derive $V_c^{\rm opt}=524$ $\kms$, that agrees well with the circular speed obtained in \cite{Murphy.et.al.2011}. After comparing the optical $V_c$-estimate with the X-ray derived one, we conclude that at the sweet radius the non-thermal contribution to the total gas pressure  is $\sim 25 \%$. The true value of the non-thermal contribution in M87 could be even lower, since X-ray data near the sweet radius are affected by the shock \citep{Forman.et.al.2007}.

We observe a sample of five X-ray bright elliptical galaxies with  the 6-m telescope of the SAO RAS and measure the surface brightness, line-of-sight velocity and velocity dispersion distribution of stars up to two effective radii along one or two slits. We apply our simple method to estimate the circular speed and compare it with the circular speed measurements based on the X-ray analysis of Chandra data. We conclude that optical and X-ray $V_c$-estimates agree with each other remarkably well implying the sample averaged non-thermal pressure support of $\disp \sim 4\% \pm 20 \% $ , i.e. to be consistent with zero.

From deep long-slit spectral data obtained with SCORPIO/BTA we derive high-precision Lick indices profiles out to $\sim 2$ effective radii, which in turn used to estimate the radial variations of the stellar population mass-to-light ratios and the dark matter fraction within $R_{\rm sweet}$, typical value of the latter is $\sim 60 \%$ for the Salpeter IMF and $\sim 75 \%$ for the Kroupa IMF.    

%________________________________________________________________

\section{Acknowledgments} 
We are grateful to the referee for very useful comments and suggestions.
NL is grateful to the International Max Planck Research School on Astrophysics (IMPRS) for financial support. NL acknowledges Scott Tremaine and Thorsten Naab for helpful discussions and Jeremy Murphy for providing kinematic profiles for M87.
This research has made use of the NED which is operated by the Jet Propulsion Laboratory, California Institute of Technology, under
contract with the National Aeronautics and Space Administration. This work was partly supported by the  Research Program OFN-17 of the Division of Physics, Russian Academy of Sciences, and by the Program of State Support for Leading Scientific Schools of the Russian Federation (grant no. NSh-6137.2014.2). 
AM is also grateful for the financial support of the `Dynasty' Foundation.
We thank   Azamat Valeev, Timur Fatkhullin and Alexander Vinokurov  for supporting the SCORPIO observations, and esspecially Victor Afanasiev for his great contribution  to spectroscopy at the 6 m telescope.

\label{lastpage}

\end{document}